\documentclass[12pt]{article}

\usepackage{natbib}
\usepackage[T1]{fontenc}
\usepackage{palatino}
\usepackage{graphics}
\usepackage[demo]{graphicx}
\usepackage{booktabs,caption}
\usepackage{threeparttable}
\usepackage{adjustbox}
\usepackage[format=plain, labelsep=period, labelfont=bf, font=footnotesize,singlelinecheck=false,skip=5pt]{caption}
\usepackage{tabularx}
\usepackage{array}
\newcolumntype{H}{>{\setbox0=\hbox\bgroup}c<{\egroup}@{}}
\usepackage{amsthm}
\usepackage{lscape}
\usepackage{appendix}
\usepackage[bookmarks]{hyperref}
\usepackage{subcaption}
\usepackage{epstopdf}
\usepackage{amsmath}

\usepackage{multirow}
\usepackage{setspace}
\usepackage{amssymb}
\usepackage{textcomp}
\usepackage{amsfonts}
\usepackage{soul}
\usepackage{bbm}
\usepackage{multirow}
\usepackage{multicol}
\usepackage{verbatim}
\usepackage{mdwlist}
\usepackage{color,soul}
\usepackage{eurosym}
\usepackage{rotating}
\usepackage{lmodern}
\usepackage{color}
\usepackage{longtable}
\usepackage{etoolbox}

\usepackage{epigraph}
\usepackage{xargs}
\usepackage[left=2cm,right=2cm,bottom=2cm,top=2cm]{geometry}
\usepackage{ifpdf}

\makeatletter

\newcommand*{\@rowstyle}{}
\newcommand*{\rowstyle}[1]{
  \gdef\@rowstyle{#1}%
  \@rowstyle\ignorespaces%
}
\newcolumntype{=}{
  >{\gdef\@rowstyle{}}%
}
\newcolumntype{+}{
  >{\@rowstyle}%
}
\makeatother

\title{A general methodology to measure labour market dynamics }

\author{Davide Fiaschi\thanks{Corresponding author. University of Pisa, Department of Economics, Via Ridolfi 10, 56124 Pisa (Italy), Phone: +39 0502216208, Email: davide.fiaschi@unipi.it. }  \and Cristina Tealdi\thanks{Department of Economics, Heriot-Watt University, EH14 4AS Edinburgh (UK) and IZA Institute of Labor, Phone: +44 0131 4513803, Email: c.tealdi@hw.ac.uk. } \\
}

\date{\today}

\begin{document}

\maketitle

\begin{abstract}
We propose a general methodology to measure labour market dynamics, inspired by the search and matching framework, based on the estimate of the transition rates between labour market states. We show how to estimate instantaneous transition rates starting from discrete time observations provided in longitudinal datasets, allowing for any number of states. We illustrate the potential of such methodology using Italian labour market data. First, we decompose the unemployment rate fluctuations into inflow and outflow driven components; then, we evaluate the impact of the implementation of a labour market reform, which substantially changed the regulations of temporary contracts.
\end{abstract}



\newpage 

\section{Introduction}

The unemployment rate  represents a static and partially informative variable, and yet is the single most important measure considered when evaluating labour market performance and designing optimal welfare policy \citep{layard2005unemployment, barnichon2012ins}.  While considerable progress has been made in the theory of modelling labour market dynamics towards understanding the deep functioning of labour markets \citep{pissarides2000equilibrium, mortensen1994job}, the methodological empirical counterpart, although emerging \citep{shimer2012reassessing, elsby2009ins}, is still lacking  a systematic approach.
To this end, we propose a  methodology to estimate instantaneous time transition rates across labour market states, on which the search and matching framework is built \citep{phelps1968money}, starting from discrete time observations and allowing for any number of states. This methodology, which is a generalization of the method proposed by \cite{shimer2012reassessing}, and it is also based on the work of \cite{israel2001finding}, allows for an easy implementation of inference on transition rates via bootstrap methods, the decomposition of the contribution of inflows and outflows to the change in labour market shares, and the evaluation of labour market policies via forecast.

For a number of different (historical) reasons most of the economic literature considers unemployment indicators as the main proxies of labour market performance \citep{perugini2007labour}. While classical economists, such as Adam Smith, focused on employment, subsequent theoretical evolution led to a clear-cut tendency towards the use of unemployment rates.
However, while the unemployment rate is a useful indicator of particularly low labour market performance, its exclusive use is questionable. In fact the view of the labour market it offers is partial, incomplete and at odd with dynamic approaches typical of macroeconomic frameworks, such as search and matching models, which focus on flows of individuals across different labour market states. Moreover, in countries/regions with significant labour market segmentation and important local labour markets, the role of  aggregate unemployment rates in affecting wage bargaining and dynamics towards labour market equilibria is significantly reduced.

This raises the issue of considering additional variables, e.g., inflows and outflows, as complementary indicators of labour market performance \citep{valli1970programmazione,garibaldi2002anatomy}. Two countries might in fact have similar unemployment rates, while hiding very
different labour markets, specifically in terms of workers turnover and flows \citep{ Blanchard_Portugal_2001}. There is, therefore, a need to first, identify relevant workers' states in high frequency panel datasets, in which the longitudinal feature is essential to follow the individuals' career paths. Second, it is crucial to adopt an appropriate and empirically implementable methodological approach to analyse such dynamics and interpret them to properly evaluate the performance of the labour markets.

This paper proposes a general approach to estimate instantaneous transition rates starting from discrete time transitions between different labour market states based on the work of \citet{israel2001finding}.  Our approach has two main advantages: it is flexible to include any number of labour market states; and inference via bootstrap and forecasting are easily implementable. Our methodology advances the knowledge on how to measure the cyclicality of transition rates and on the decomposition of the unemployment fluctuations over the business cycle into inflow and outflow driven components \citep{shimer2012reassessing, silva2013ins}. In addition, we discuss how the estimates of the transition rates and their forecast could be used to assess the changes of the labour market dynamics for guiding policy evaluation. 

We apply this methodology to the Italian labour market by using longitudinal quarterly labour force data for the period 2013-2020. First, we illustrate how the decomposition of the unemployment rate fluctuations points to a major role played by the transitions from and to inactivity. We then report transition rates between five labour market states and the corresponding shares before and after the implementation of a labour market reform, which significantly modified the regulations of temporary contracts. Overall, we find strong support for the utilisation of labour market flows as well as labour market stocks for the assessment of labour market dynamics.

This paper directly relates to the search and matching  literature, which is the theoretical framework at the basis of the flow approach to the labour market \citep{mortensen1970theory, phelps1968money}. From a  methodological point of view,  approaches to measure flows in and out of unemployment, using publicly available data have been proposed by \cite{elsby2009ins, darby1986ins, fujita2009cyclicality}.  These studies differ along two main dimensions: (i) how many labour market states are included and (ii) the "time aggregation", as flows in and out of
each state are taking place in continuous time while data are collected in discrete times. \cite{shimer2012reassessing} uses US monthly statistics on employment, unemployment and short-term unemployment to compute the probabilities that an employed worker becomes unemployed and that an unemployed worker finds a job, i.e, the job finding and the job exit rates. His approach is based on two key assumptions: (i) individuals only move between employment and unemployment and (ii) individuals are homogeneous.\footnote{In any period all unemployed individuals have the same job finding probability and all employed individuals have the same job exit probability.}   This methodology however is implementable when the number of labour market states is small. As \citet[p.133]{shimer2012reassessing} claims, when expanding the number of labour market states, `the theory here is more cumbersome, the data limitations are more serious, and the data analysis is more involved'. Using the same  methodology, \cite{elsby2015importance} discusses the importance of including inactivity, while \cite{silva2013ins, fontaine2020labour}  and \citet{borowczyk2020ins} expand the number of labour market states by including temporary employment,  public employment and part-time employment, respectively. \cite{elsby2009ins}  account for the  time aggregation bias using a discrete-time variant of the \cite{shimer2012reassessing}  procedure, based on the fact that the US Current Population Survey (CPS) uses the week as its reference period.
Finally, some papers have followed applying such methodologies in different countries \citep{petrongolo2008ins, smith2011ins, hertweck2015ins, baussola2014transitions, gomes2012labour}, while \citet{gomes2015importance} discusses the importance of the data frequency. 

The paper is organized as follows. Section \ref{methodology} explains in detail the proposed methodology, which is based on a view of the labour market in terms of flows,  as in the search and matching framework. Section \ref{sec:empirics} applies the methodology using Italian data to illustrate its main advantages and Section \ref{sec:concludingRemarks} concludes the paper. The appendices collect the technical material.

\section{Methodology}\label{methodology}

As discussed in the introduction, we propose a dynamic approach to the labour market, based on observed transitions of working age individuals between different labour market states (e.g. unemployed, employed, inactive, etc.), compared to the traditional static approach, based on stock variables, such as unemployment rates and labour market participation.
Specifically, we take a microeconomic perspective by tracking movements of individuals across labour market states and discuss under which conditions these transition rates can summarize the whole dynamics of the labour market. The theoretical (economic) basis of the proposed methodology is the search and matching model, which supplies a meaningful economic interpretation to estimated (instantaneous) transition rates. We discuss how this methodology is sufficiently flexible to deal with `realistic and empirically implementable' scenarios in the labour market \citep[p.3]{pissarides2000equilibrium}.

We first illustrate a simplified version of labour market dynamics with three labour market states (employed, unemployed and inactive) and map the corresponding transition rates into a standard continuous time search and matching framework (Section \ref{sec:transitionBetweenStatus}). Next, we discuss the general framework and characterize the short-run dynamics, the equilibrium distribution, how to deal with the presence of seasonality, and the decomposition of the the contribution of individual transition rates to labour market dynamics (Section \ref{sec:transitionK}). Finally, we describe the methodology to estimate the instantaneous transition rates starting from discrete time observations (Section \ref{sec:empiricalImplementation}).

\subsection{Transitions between labour market states \label{sec:transitionBetweenStatus}}

The seminal papers that proposed the search and matching framework \citep{mortensen1970theory, phelps1968money} introduced a flow approach to the labour market. 
The core idea on which the search and matching model is built is that trade in the labour market is a decentralized economic activity, in which the process for both individuals and firms to find each other, agree on the wage and start producing is costly and time consuming \citep{pissarides2000equilibrium}. While similar to other macroeconomic models, the search and matching framework features optimizing agents, rational expectations, and equilibrium outcomes, the real emphasis in the model is placed on the flows between labour market states.

Consider a simple model in which a mass of employees $L$ can belong to one of the three labour market states available in the economy: employed ($e$),  unemployed ($u$) or inactive ($n$); hence:
\begin{equation}\label{LF}
	e+u+n=L.
\end{equation}
 Assume that individuals can freely move across the labour market states at some instantaneous transition rates. Specifically, unemployed  individuals become employed at rate $\alpha$, while employed worker become unemployed at rate $\lambda$. Inactive individuals join the unemployment pool at rate $\varphi_u$ and find a job at rate $\varphi_e$. Finally, employed and unemployed individuals become inactive at rate $\mu$ and $\gamma$, respectively. On the basis of these rates, assuming no entry and exit from the working age population, the number of employees who are employed, unemployed and inactive evolve according to the following system of differential equations:
\begin{eqnarray}
	\begin{cases}
	\label{dote} \dot{e}=-(\lambda +\mu) e +\alpha u +\varphi_e n; \\
	\label{dotu} \dot{u}=-(\alpha +\gamma) u +\lambda e  +\varphi_u n; \text{ and}\\
	\label{dotn} \dot{n}=-(\varphi_u  +\varphi_e) u +\mu e  +\gamma u.
	\end{cases}
\end{eqnarray}
The observed dynamics are the outcome of inflows and outflows of individuals between the three labour market states and are fully described by the set of transition rates. The economic theory has provided several explanations for these transitions (see, e.g., \citealp{pissarides2000equilibrium, garibaldi2005equilibrium, mortensen1994job}). For example,  $\alpha$ is directly related to \textit{job creation}, i.e. the process by which a firm and a worker meet and agree to form a match, which results in the worker's flow from unemployment to employment; while $\lambda$ is directly related to \textit{job destruction}, which takes place in case of a separation between a firm and a worker, which results in the worker's flow from employment into unemployment. 

Consider the dynamics of the shares of individuals on the working age population. In particular, the system of differential equations (\ref{dotn}), assuming $L$ to be constant, can be written as:
\begin{eqnarray}
	\begin{cases}
	\label{eq:dotpie} \dot{\pi_e}=-(\lambda +\mu) \pi_e +\alpha \pi_u +\varphi_e \pi_n; \\
	\label{eq:dotpiu} \dot{\pi_u}=-(\alpha +\gamma) \pi_u +\lambda \pi_e  +\varphi_u \pi_n; \text{ and}\\
	\label{eq:dotpin} \dot{\pi_n}=-(\varphi_u  +\varphi_e) \pi_u +\mu \pi_e  +\gamma \pi_u.
	\end{cases}
\end{eqnarray}

where $\pi_e \equiv e/L$, $\pi_u \equiv u/L$, and $\pi_n \equiv n/L$, while Equation (\ref{LF}) becomes:
\begin{equation}
\pi_e + \pi_u + \pi_n =1.
\label{eq:piLF}
\end{equation}

From Equations (\ref{eq:dotpin})-(\ref{eq:piLF}), by setting $\dot{\pi_e}=\dot{\pi_u}=\dot{\pi_n}=0$, we can then calculate the steady-state equilibrium values of the labour market shares of the working age population as:
\begin{eqnarray}
		\begin{cases}
\label{eqe} \pi_e^{EQ}=\dfrac{\varphi_e(\varphi_u+\alpha)+ \alpha\varphi_u }{(\gamma +\varphi_u ) (\lambda+\mu)+ \alpha(\varphi_u+\mu)+\varphi_e(\lambda +\varphi_u+\alpha)}; \\
\label{equ} \pi_u^{EQ}=\dfrac{\varphi_u (\lambda+\mu)+ \varphi_e\lambda  }{(\gamma +\varphi_u ) (\lambda+\mu)+ \alpha(\varphi_u+\mu)+\varphi_e(\lambda +\varphi_u+\alpha)}; \text{ and}\\
\label{eqn}	\pi_n^{EQ}=\dfrac{\alpha \mu+\gamma(\lambda+\mu) }{(\gamma +\varphi_u ) (\lambda+\mu)+ \alpha(\varphi_u+\mu)+\varphi_e(\lambda +\varphi_u+\alpha)},
		\end{cases}
\end{eqnarray}
where $\pi_e^{EQ}$, $\pi_u^{EQ}$ and $\pi_n^{EQ}$ are the equilibrium share of employed individuals, of unemployed individuals, and of inactive individuals respectively.

The system of equations (\ref{eqn}) highlights that different combinations of transition rates can lead to the same equilibrium shares of employed, unemployed and inactive individuals. In particular, the same equilibrium unemployment share can be the result of either high job finding and job destruction rates or of the opposite scenario, of low job finding and job destruction rates. \cite{Blanchard_Portugal_2001} report, for instance, that similar average unemployment rates for Portugal and the USA for the period 1983-1995 are the result of low (Portugal) versus high (USA) job finding and job destruction rates.\footnote{The unemployment rate is defined as the ratio between the number of individuals who are unemployed and the number of individuals in the labour force, which according to our definition would read as 	$\dfrac{\pi_u^{EQ}}{\pi_u^{EQ}+\pi_e^{EQ}} = \dfrac{\varphi_u (\lambda+\mu)+ \varphi_e\lambda}{\varphi_u (\lambda+\mu)+ \varphi_e\lambda+\varphi_e(\varphi_u+\alpha)+ \alpha\varphi_u}$.}
Moreover, a comparison between observed employment and unemployment shares of the working age population and their equilibrium values is particularly appealing as it is suggestive of the direction the labour market is moving to. In this respect, it is also possible to compute the expected speed of convergence to the long-run equilibrium \citealp[p. 184]{cox1977theory}.

\subsection{Transitions with K states}\label{sec:transitionK}

In a more general setting with $K$ labour market states, the system of equations (\ref{eq:dotpin}) can be expressed as follows:
\begin{equation}
\dot{\boldsymbol{\pi}} = \boldsymbol{\pi}\mathbf{Q},
\label{eq:dynamicsStocks}
\end{equation} 
where $\boldsymbol{\pi}$ is a $ 1\times K $ vector collecting the shares of individuals in the working age population in different $K$ states, and $\mathbf{Q}$ is a  $K \times K$ matrix, whose elements are the instantaneous transition rates between different states, with the constraint that:
\begin{equation}
\boldsymbol{\pi}\mathbf{1}^T  = 1,
\label{eq:constantWorkingAgePop}
\end{equation}
where $\mathbf{1}$ is a  $1 \times K $ vector of ones; Equation (\ref{eq:constantWorkingAgePop}) simply states that the shares of working age individuals in the $K$ labour market states sum to one.

The matrix of (instantaneous) transition rates $\mathbf{Q}$ is assumed to satisfy the following conditions:
\begin{eqnarray} 
		\begin{cases}
q_{ii}  \leq 0  \; \forall i; \\ 
q_{ij}  \geq 0  \; \forall i,j; \; \text{and} \\
\sum_{j=1}^{K} q_{ij}  =  0 \; \forall i,
	\end{cases}
\label{eq:hypothesisOnQ}
\end{eqnarray}
  which amounts to assume that the process governing the labour market dynamics is \textit{conservative} \citep[p. 180]{cox1977theory}, i.e., there are no entries and exits from/to the working age population and, hence, the working age population is constant. Under general conditions (i.e., finite $K$), the matrix $\mathbf{Q}$ represents a continuous time \textit{honest} Markov process with discrete states \citep[p. 182]{cox1977theory}, i.e.:\footnote{The definition of the exponential matrix is the following:
\[
\exp\left(\mathbf{Q}t\right) = \sum_{r=0}^{\infty} Q^r \dfrac{t^r}{r!} = \mathbf{I} + \mathbf{Q}t + \left(\mathbf{Q}t\right)^2/2! + \left(\mathbf{Q}t\right)^3/3! + \cdots,
\]
where $\mathbf{I}$ is the  $K \times K$ identity matrix.}
\begin{equation}
\mathbf{P}(t) = \exp\left(\mathbf{Q}t\right),
\label{eq:MarkovMatrixContinuosTime}
\end{equation} 
where $\mathbf{P}(t)$ is the matrix collecting the transition probabilities from period $0$ to period $t$, with $\mathbf{Q}^0=\mathbf{I}$. When $\mathbf{Q}$ is constant over time, the general solution to Equation (\ref{eq:dynamicsStocks}) is \citep[p. 129]{hirsch2012differential}:
\begin{equation}
\boldsymbol{\pi}(t) = \boldsymbol{\pi}(0)\exp\left(\mathbf{Q}t\right),
\label{eq:GeneralSolution}
\end{equation}
where $\boldsymbol{\pi}(0)$ is the  $1 \times K$ vector which collects the shares at time 0. A non-trivial equilibrium is characterized by $\dot{\boldsymbol{\pi}} = 0$, i.e., $\boldsymbol{\pi} \mathbf{Q}=0$. Solving Equation (\ref{eq:GeneralSolution}), using Equation (\ref{eq:constantWorkingAgePop}), we get that the equilibrium distribution of $\boldsymbol{\pi}$, $\boldsymbol{\pi}^{EQ}$, reads as:\footnote{The proof uses $ \boldsymbol{\pi} \mathbf{1}^T \mathbf{1} = \mathbf{1}$.}
\begin{equation}
\boldsymbol{\pi}^{EQ} = \mathbf{1} \left(\mathbf{1}^T\mathbf{1} - \mathbf{Q}\right)^{-1},
\label{eq:equilibriumMassProbabilityDistributionStocks}
\end{equation}
where $\boldsymbol{\pi}^{EQ}$ is a $1 \times K$ row-vector whose elements are non-negative and sum to 1.
Finally, the convergence to equilibrium is exponential and the speed of convergence is measured by the eigenvalues of the  $\mathbf{Q}$ matrix \citep[p. 110]{hirsch2012differential}.

\subsubsection{Determinants of the changes in the labour market shares}\label{sec:AnalysDeterminantsChanges}

\cite{pissarides1986unemployment} and \cite{shimer2012reassessing} propose a methodology for decomposing the contribution of inflows and outflows to the changes in the labour market shares over time. This approach  is based on the construction of counterfactual shares computed by setting all but one transition rates to their average/trends/long-run/reference values and allowing only the element $(s,r)$ of the matrix $\mathbf{Q}$, denoted as $q_{sr}$, to vary over time. The counterfactual matrix of transition rates $\mathbf{Q}^{CF}$ is therefore constructed as:
\begin{equation}
\begin{cases}
q^{CF}_{ij}\left(t\right) = \bar{q}_{ij}(t) \; \forall (i,j)\neq (s,r) \text{ and } i \neq j; \\
q^{CF}_{sr}\left(t\right) = q_{sr}\left(t\right);  \text{ and} \\
q^{CF}_{ii}\left(t\right) = - \sum_{j=1,j \neq i}^{K} q^{CF}_{ij}\left(t\right),
\end{cases}
\label{eq:counterfactualQ}
\end{equation}
where $\bar{q}_{ij}(t)$ can be set to its initial value (\citealp{pissarides1986unemployment}) or to its average value over the observed period (\citealp{shimer2012reassessing}), or to the trend component of the $q_{ij}(t)$ series (\citealp{silva2013ins}).
The counterfactual matrix $\mathbf{Q}^{CF}$ is then used to compute the counterfactual matrix of shares $\boldsymbol{\pi}^{CF}$ per each period to be compared with the observed matrix of shares $\boldsymbol{\pi}$. To effectively compute $\boldsymbol{\pi}^{CF}$, \cite{shimer2012reassessing} takes as reference the equilibrium shares, i.e.:
\begin{equation}
\boldsymbol{\pi}^{CF}\left(t\right) = \mathbf{1} \left[\mathbf{1}^T\mathbf{1} - \mathbf{Q}^{CF}\left(t\right)\right]^{-1},
\label{eq:counterfactualEquilibriumShare}
\end{equation}
as equilibrium and observed shares are highly correlated in his sample ($\rho=0.98$). This high correlation however is not always guaranteed. An alternative way, which does not rely on this hypothesis but on the milder conjecture that one-period ahead forecasts are very close to the observed shares, computes the counterfactual matrix of shares period by period as follows:
\begin{equation}
\boldsymbol{\pi}^{CF}(t) = \boldsymbol{\pi}(t-1)\exp\left(\mathbf{Q}^{CF}\right).
\label{eq:counterfactualActualShare}
\end{equation}
Counterfactual and observed shares are then compared in different ways depending on the objective of the analysis. Since \cite{pissarides1986unemployment}'s goal is to explain the continuous increase in unemployment in Britain between the late 60s and early 80s, he makes a direct comparison between $\boldsymbol{\pi}$ and $\boldsymbol{\pi}^{CF}$. \cite{shimer2012reassessing}, instead, being interested in the determinants of the unemployment volatility, extracts the cyclical component of both shares, denoted by $\tilde{\boldsymbol{\pi}}$ and $\tilde{\boldsymbol{\pi}}^{CF}$, and computes the contribution of the $(s,r)$ element  of the  $\mathbf{Q}$ matrix to the fluctuation of the  $k$ element of $\boldsymbol{\pi}$ as:
\begin{equation}
\dfrac{\mathrm{Cov}\left(\tilde{\boldsymbol{\pi}}_k,\tilde{\boldsymbol{\pi}}^{CF}_k\right)}{\mathrm{Var} \left(\tilde{\boldsymbol{\pi}}_k\right)},
\label{eq:contributionCyclicalComponent}
\end{equation} 
where $\mathrm{Cov}\left(\tilde{\boldsymbol{\pi}}_k,\tilde{\boldsymbol{\pi}}^{CF}_k\right)$ is the covariance between the $k$ element of $\tilde{\boldsymbol{\pi}}$ and $\tilde{\boldsymbol{\pi}}^{CF}$ and $\mathrm{Var} \left(\tilde{\boldsymbol{\pi}}_k\right)$ is the variance of $\tilde{\boldsymbol{\pi}}$.

\subsubsection{Seasonality}
So far we have assumed the matrix $\mathbf{Q}$ to be constant, however it might not always be the case. We could in fact observe seasonality in transition rates \citep{shimer2012reassessing}, for instance directly related to seasonal fluctuations of employment in specific sectors, such as tourism and agriculture. In this scenario, from Equation (\ref{eq:GeneralSolution}) we can define an ``annual'' $\mathbf{Q}_a$ based on a set of $\tau$ ``seasonal'' $\mathbf{Q}$s, i.e.:
\begin{eqnarray}
\boldsymbol{\pi}(t) &=& \notag \boldsymbol{\pi}(t-\tau) \exp\left({\mathbf{Q}(t-\tau)}\right) \dots \exp\left({\mathbf{Q}(t-1)}\right) = \boldsymbol{\pi}(t-\tau) \exp\left({\sum_{s=0}^{\tau-1} \mathbf{Q}(t-\tau+s)}\right)=\\ &=& \boldsymbol{\pi}(t-\tau) \exp\left(
\mathbf{Q}_a \right),
\label{eq:annualQ}
\end{eqnarray}
where $\mathbf{Q}_a$ is the sum of the seasonal $\mathbf{Q}$s and $\tau$ is chosen according to the seasonality, e.g., monthly ($\tau=12$) or quarterly ($\tau=4$).

\subsection{Empirical implementation}\label{sec:empiricalImplementation}

Since observations on the labour market states of individuals are available at discrete time, a direct estimation of $\mathbf{Q}$ is not feasible. To circumvent this issue, we first estimate $\mathbf{P}$ in discrete time and then estimate $\mathbf{Q}$ using Equation (\ref{eq:MarkovMatrixContinuosTime}). \citet[p. 92]{anderson1957statistical} show that each element $p_{ij}$ of the matrix $\mathbf{P}$ can be estimated by maximum likelihood as follows:
\begin{equation}
\hat{p}_{ij} = \dfrac{m_{ij}(t)}{m_{i}(t)},
\end{equation}
where $m_{ij}(t)$ is the number of individuals in period $t$ in state $i$ moving in period $t+1$ in state $j$ and $m_{i}(t)$ is the total number of individuals in period $t$ in state $i$.
From the estimate of $\mathbf{P}$, we then get an estimate of $\mathbf{Q}$ using Equation (\ref{eq:MarkovMatrixContinuosTime}), under the conditions discussed by \cite{israel2001finding}. In particular, they argue that under mild conditions, the matrix $\tilde{\mathbf{Q}}$, which is defined by the following geometric infinite series:
\begin{equation}
\tilde{\mathbf{Q}} =  \sum_{r=1}^{\infty} -(-1)^r\dfrac{\left(\mathbf{P} - \mathbf{I}\right)^r}{r} = \left(\mathbf{P} - \mathbf{I}\right)-\dfrac{\left(\mathbf{P} - \mathbf{I}\right)^2}{2} + \dfrac{\left(\mathbf{P} - \mathbf{I}\right)^3}{3} - \dfrac{\left(\mathbf{P} - \mathbf{I}\right)^4}{4} + \cdots
\label{eq:FromPtoQ}
\end{equation}
is such that $\exp\left(\mathbf{\tilde{Q}}\right)=\mathbf{P}$ and its rows sum to zero (Theorem 2 in \citealp{israel2001finding}). A potential drawback of using Equation (\ref{eq:FromPtoQ}) is that it does not ensure that $\mathbf{\tilde{Q}}$ is a ``valid'' $\mathbf{Q}$, i.e., $\mathbf{\tilde{Q}}$ satisfies all Conditions (\ref{eq:hypothesisOnQ}). Specifically, there is no guarantee that all off-diagonal entries of matrix $\mathbf{\tilde{Q}}$ are non-negative.\footnote{In Section 3, \cite{israel2001finding} propose two methods to circumvent this issue. The first is to set $q_{ij}=\max\left(\tilde{q}_{ij},0\right)$ for $i \neq j$ and $q_{ii}=  (\tilde{q}_{ii} +  \sum_{j \neq i }\min\left(\tilde{q}_{ij},0\right)$, i.e. to set to zero all negative off-diagonal elements and change the diagonal elements to make sure the sum of each row is equal to zero. The second method sets the negative off-diagonal values to zero and spans the difference on all positive entries to assure that the sum of each row is equal to zero.}
Finally, \citet[p. 97]{zahl1955markov} shows that the properties of the maximum likelihood estimate of $\mathbf{P}$ are inherited by $\mathbf{Q}$. In Section \ref{sec:empirics} we will use bootstrap as an alternative robust approach to inference (see Appendix \ref{app:bootstrapProcedures}).

\section{Empirical applications}\label{sec:empirics}

In this section we discuss two applications of our methodology using data from Italy. The first application aims at identifying the determinants of the labour market shares' volatility, specifically of unemployment rate (Section \ref{sec:DecompositionVolatilityShares}), while the second focuses on the implications of  labour market policies. Before moving into the applications, we first provide a short overview of the features of the Italian labour market (Section \ref{sec:italianLabourMarket}), then we describe the data used in the analysis (Section \ref{sec:dataset}) and finally we provide evidence of the goodness of our estimates (Section \ref{sec:goodnessEstimates}).

\subsection{The Italian labour market}\label{sec:italianLabourMarket}

Following important labour market reforms in the 1990s and early 2000s, labor market outcomes have improved substantially in Italy: employment and labor force participation rates have increased, and the unemployment rate dropped. But despite these improvements, the Italian labour market is still under-performing compared to those in most other European countries \citep{OECDreport2019}. Specifically, the participation rate is still substantially below that in most other European countries, the unemployment rate is higher, and the shares of temporary employment and self-employment are significantly higher compared to the EU average (Table \ref{SEandTC}).  The fast growing share of temporary employment led to the implementation of several reforms over the years with the goal to facilitate the transitions of individuals from temporary to permanent employment, while reducing the unemployment (and inactivity) rate and the growing share of self-employment\footnote{The category of para-subordinate workers in Italy, i.e. individuals who are legally self-employed but who are often “economically dependent” on a single employer, is relatively large. These  workers are disadvantaged relative to employees in terms of  the welfare provisions that they are entitled to receive \citep{Raitano2018}.} \citep{boeri2019tale, di2019heterogeneous}.\footnote{Specifically, in March 2014 a labour market reform (\textit{Decreto Poletti}) increased the flexibility of temporary contracts; in March 2015 the \textit{Jobs Act} changed the regulations of the open-ended contract, by introducing firing costs increasing with tenure; and, finally, in July 2018 the \textit{Decreto Dignità} increased the rigidity of temporary contracts.} This evidence provides support for considering five labour market states when applying our methodology: inactive, unemployed, temporary employed, permanent employed and self-employed.\footnote{Age, gender and education would be further interesting dimensions to explore, but are outside the scope of this paper.}

\begin{table}[t]
	\centering
	\caption{Labour market characteristics for a select sample of European countries. } 
	\label{SEandTC}
	\scriptsize
	\begin{tabular}{@{\extracolsep{-1pt}}lcccc} 
		\toprule
		\\[-1.8ex]
		\textbf{Country}&\textbf{Self-employment}&\textbf{Temporary-employment}&\textbf{Unemployment}&\textbf{Labour force participation}\\
		&(\% total employment)&(\% dependent employment)&(\% labour force)&(\% working age)\\
		\hline
		\\[-1.8ex]
		Greece&31.9&12.5&17.5&68.4\\
		\textbf{Italy}&	\textbf{22.7}&	\textbf{17.0}&	\textbf{10.2}&	\textbf{65.7}\\
		Portugal&16.9&20.8&6.7&75.5\\
		Spain&15.7&26.3&14.2&75.0\\
		United Kingdom&15.6&5.2&4.0&78.8\\
		Ireland&14.4&9.8&4.5&73.1\\
		Belgium&14.3&10.9&5.4&69.0\\
		France&12.1&16.4&8.5&71.7\\
		Germany&9.6&12.0&3.2&79.2\\\\[-1.8ex]
		\hline \\[-1.8ex]
		\textbf{EU average}&	\textbf{15.3}&	\textbf{13.2}&	\textbf{6.4}&	\textbf{74.2}\\
		\bottomrule\\[-1.8ex]
		\multicolumn{2}{l}{\textit{Source}: OECD, 2019.}
	\end{tabular} 
	\global\let\\=\restorecr
\end{table}

\subsection{Data \label{sec:dataset}}

We use Italian quarterly longitudinal labour force data as provided by the Italian Institute of Statistics (ISTAT) for the period 2013 (quarter I) to 2020 (quarter III).\footnote{Data for the period 2013 (quarter I) to 2020 (quarter III) are available upon request at: https://www.istat.it/it/archivio/185540.} The Italian Labour Force Survey (LFS) follows a simple rotating sample design where households participate for two consecutive quarters, exit for the following two quarters, and come back in the sample for other two consecutive quarters. As a result, 50\% of the households, interviewed in a quarter, are re-interviewed after three months, 50\% after twelve months, 25\% after nine and fifteen months. This rotation scheme allows to obtain 3 months longitudinal data, which include almost 50\% of the original sample.

The longitudinal feature of these data is essential for achieving a complete picture of significant economic phenomena of labour market mobility. Per each individual who has been interviewed we observe a large number of individual and labour market characteristics at the time of the interview and three months before. Taking into account the structure of this database, we compute the labour market flows by calculating the quarter-on-quarter transitions made by individuals between different labour market states. Specifically, we estimate the gross flows using a five-state model (permanent employed, temporary employed, self-employed, unemployed, and inactive). The drawback of these data is the point-in-time measurement of the worker's labour market state, which fails to capture transitions within the period (quarter). For instance, if an employed worker becomes unemployed and finds a new job within a quarter, we do not observe those transitions in our data. However, from Section \ref{sec:empiricalImplementation} we know that, assuming constant (instantaneous) transition rates within the quarter, the latter ($\mathbf{Q}$) can be estimated using the transitions at quarterly frequency  ($\mathbf{P}$).

On average approximately 70.000 individuals are interviewed each quarter, of which 45.000 are part of the working age population. The average quarterly inflow of younger individuals in the working age population is 0.3\%, while the average quarterly outflow of older individuals from the working age population is 0.4\%,  backing our hypothesis of a (almost) constant working age population within quarters.

\subsection{Goodness of estimates}\label{sec:goodnessEstimates}

\begin{figure}[htbp]
	\centering
	\caption{Observed versus fitted labour market shares.}
	\label{fig:ObsVsFittedShares}
	\begin{subfigure}[b]{0.32\textwidth}
		\centering
		\includegraphics[width=0.9\linewidth]{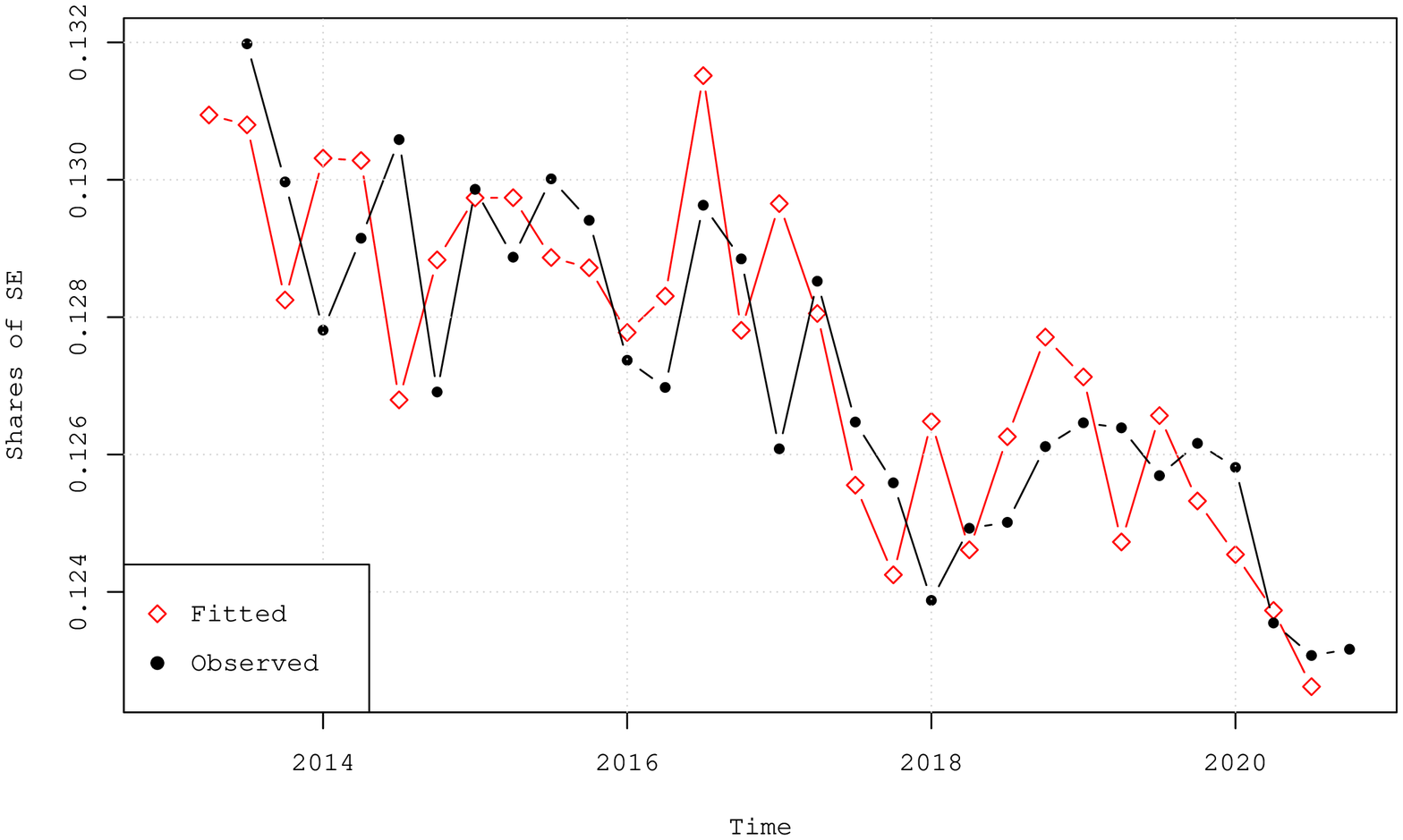}
		\caption{Self-employed.}
		\label{tot_unem}
	\end{subfigure}
	\begin{subfigure}[b]{0.32\textwidth}
		\centering
		\includegraphics[width=0.9\linewidth]{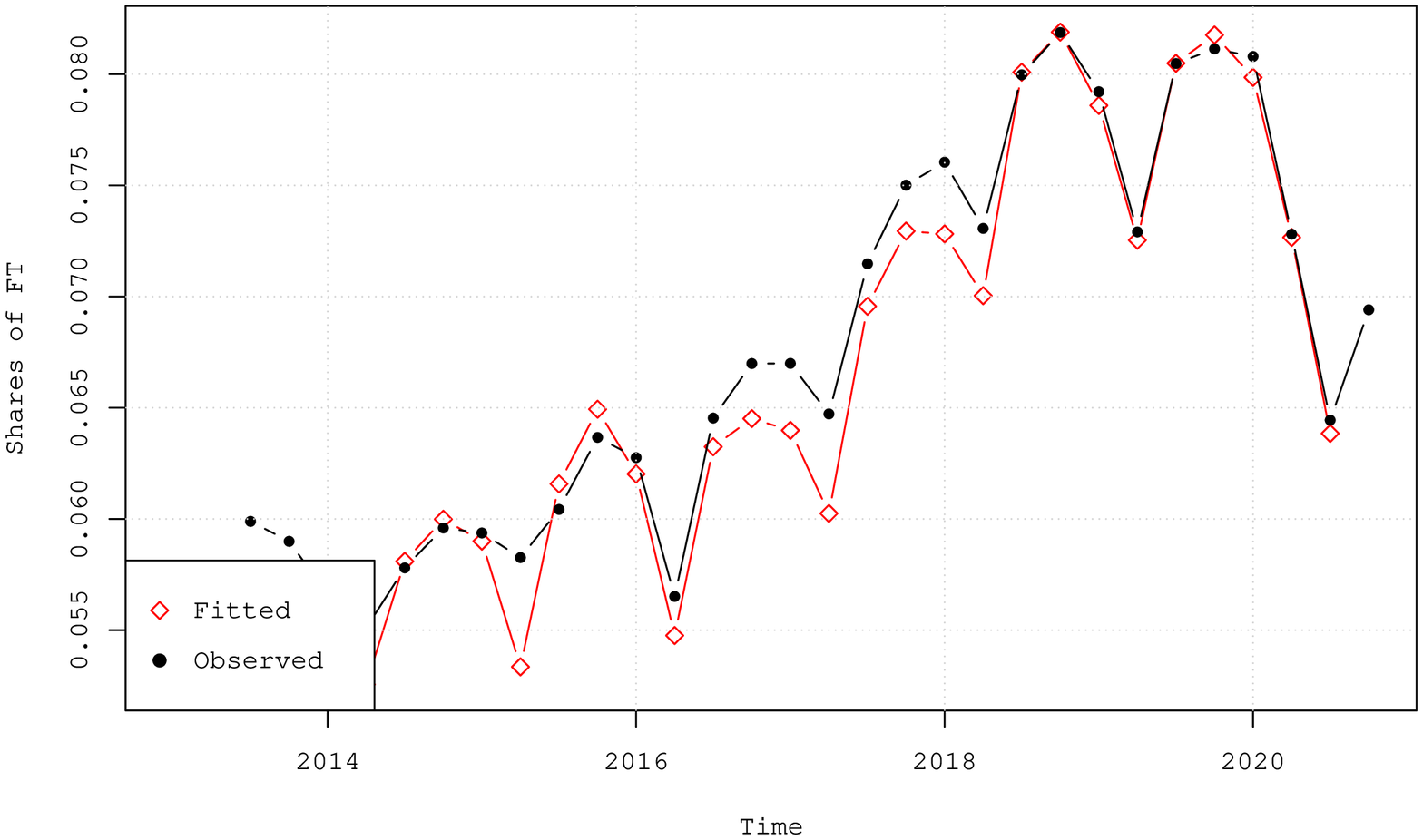}
		\caption{Temporary employed.}
		\label{tot_unem}
	\end{subfigure}
	\begin{subfigure}[b]{0.32\textwidth}
		\centering
		\includegraphics[width=0.9\linewidth]{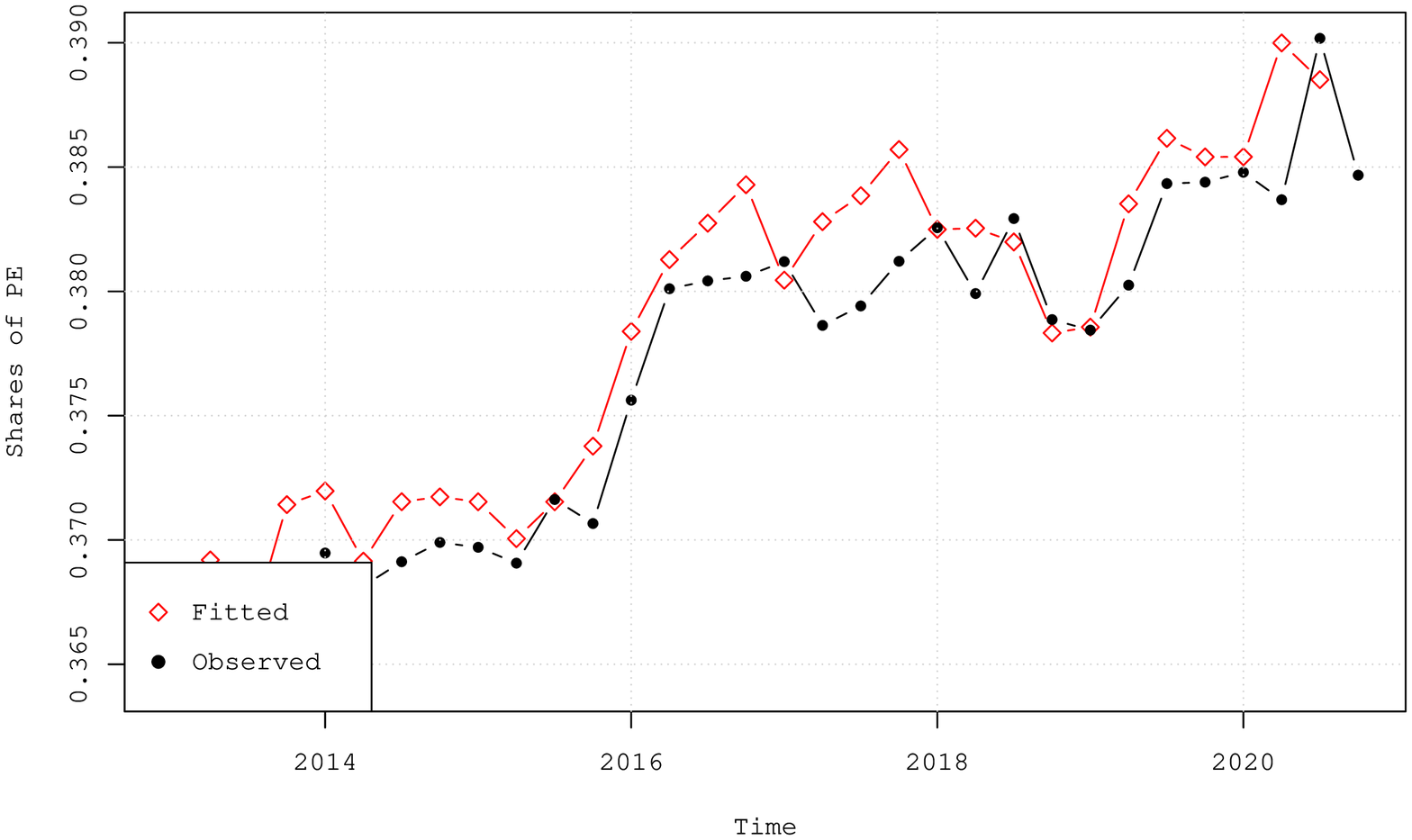}
		\caption{Permanent employed.}
		\label{tot_unem}
	\end{subfigure}\\
	\begin{subfigure}[b]{0.32\textwidth}
		\centering
		\includegraphics[width=0.9\linewidth]{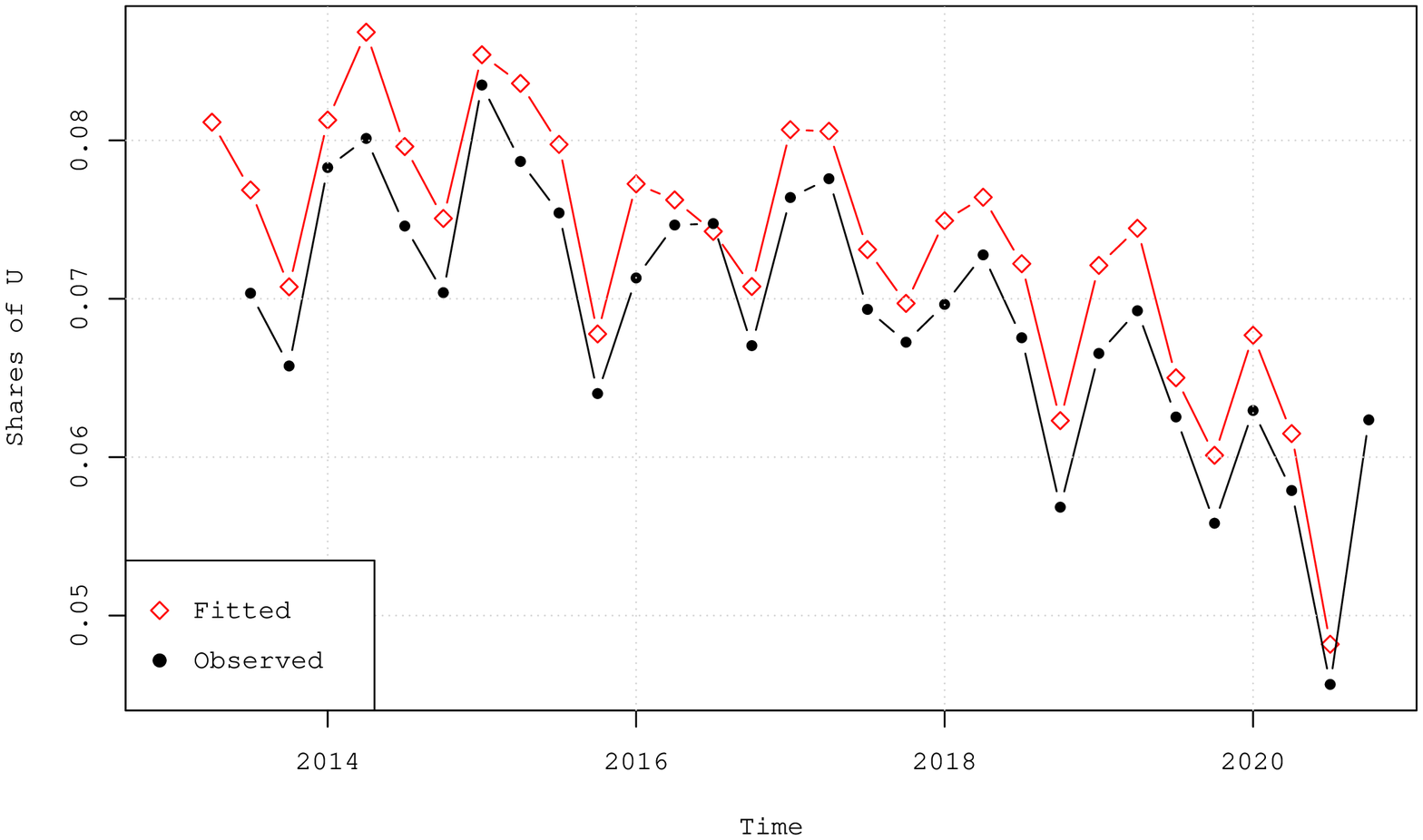}
		\caption{Unemployed.}
		\label{tot_unem}
	\end{subfigure}
	\begin{subfigure}[b]{0.32\textwidth}
		\centering
		\includegraphics[width=0.9\linewidth]{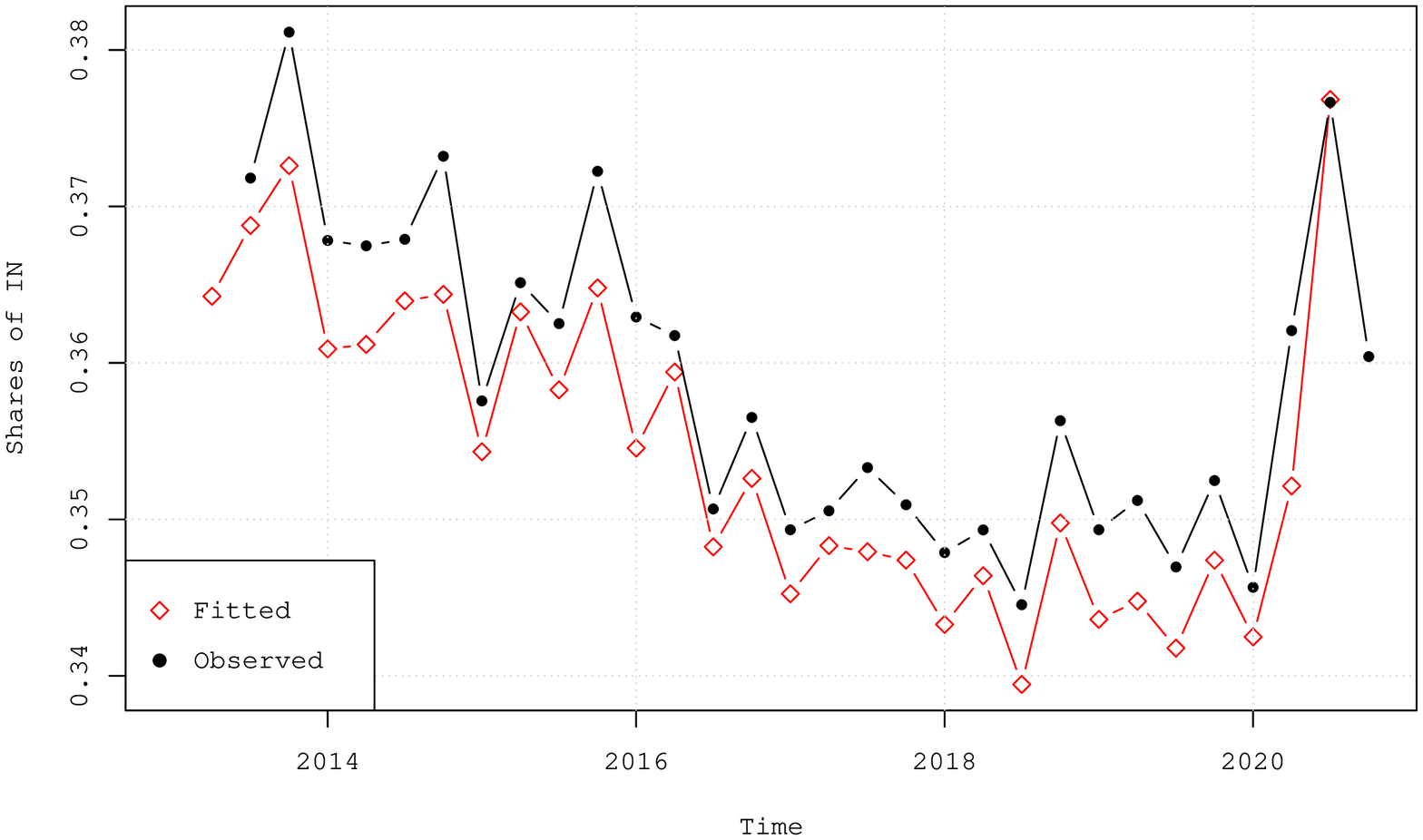}
		\caption{Inactive.}
		\label{tot_unem}
	\end{subfigure}
	\caption*{ \scriptsize{\textit{Note}: the black line reports the observed share of individuals in each labour market state, while the red line reports the fitted share, computed using Equation \ref{eq:GeneralSolution} period by period.}}
\end{figure}

In order to assess how well our methodology works, we compare the observed labour market shares with the fitted ones, i.e., the ones computed using Equation (\ref{eq:GeneralSolution}) in each quarter.\footnote{Data and codes are available at $https://people.unipi.it/davide_fiaschi/ricerca/$.} Specifically, starting from the observed shares at quarter $t-1$ we use the estimated $\mathbf{Q}$ to compute the fitted shares at quarter $t$. Figure \ref{fig:ObsVsFittedShares} reports the observed and fitted shares for all the five labour market shares. A visual inspection suggests that both shares follow fairly similar paths throughout the period; the high correlation between the observed and the fitted shares for all labour market shares confirms this impression.\footnote{The correlations are 0.755 for self-employed, 0.986 for temporary employed, 0.949 for permanent employed, 0.982 for unemployed and 0.973 for inactive.} This evidence therefore supports the validity of our methodology in estimating the instantaneous transition rates, i.e., matrix $\mathbf{Q}$.

\subsection{Contributions to fluctuations in labour market  shares}\label{sec:DecompositionVolatilityShares}

\begin{figure}[htbp]
	\centering
	\caption{Observed and hypothetical unemployment rates.}
	\label{fig:sensibilityUnemploymentRateToTransitionRate}
		\begin{subfigure}[b]{0.32\textwidth}
		\centering
		\includegraphics[width=0.9\linewidth]{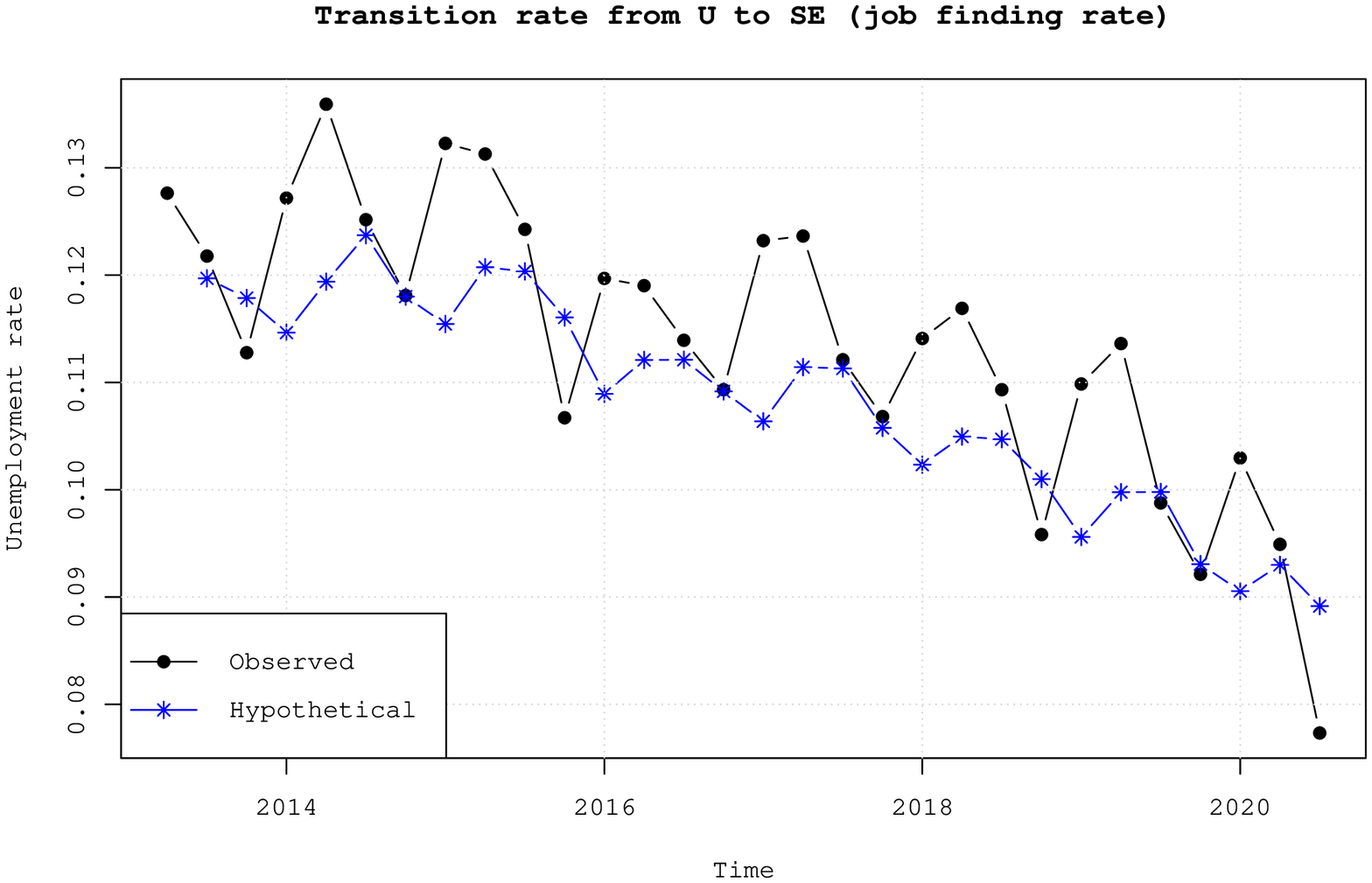}
		\caption{Transitions from U to SE.}
		\label{tot_unem}
	\end{subfigure}
	\begin{subfigure}[b]{0.32\textwidth}
		\centering
		\includegraphics[width=0.9\linewidth]{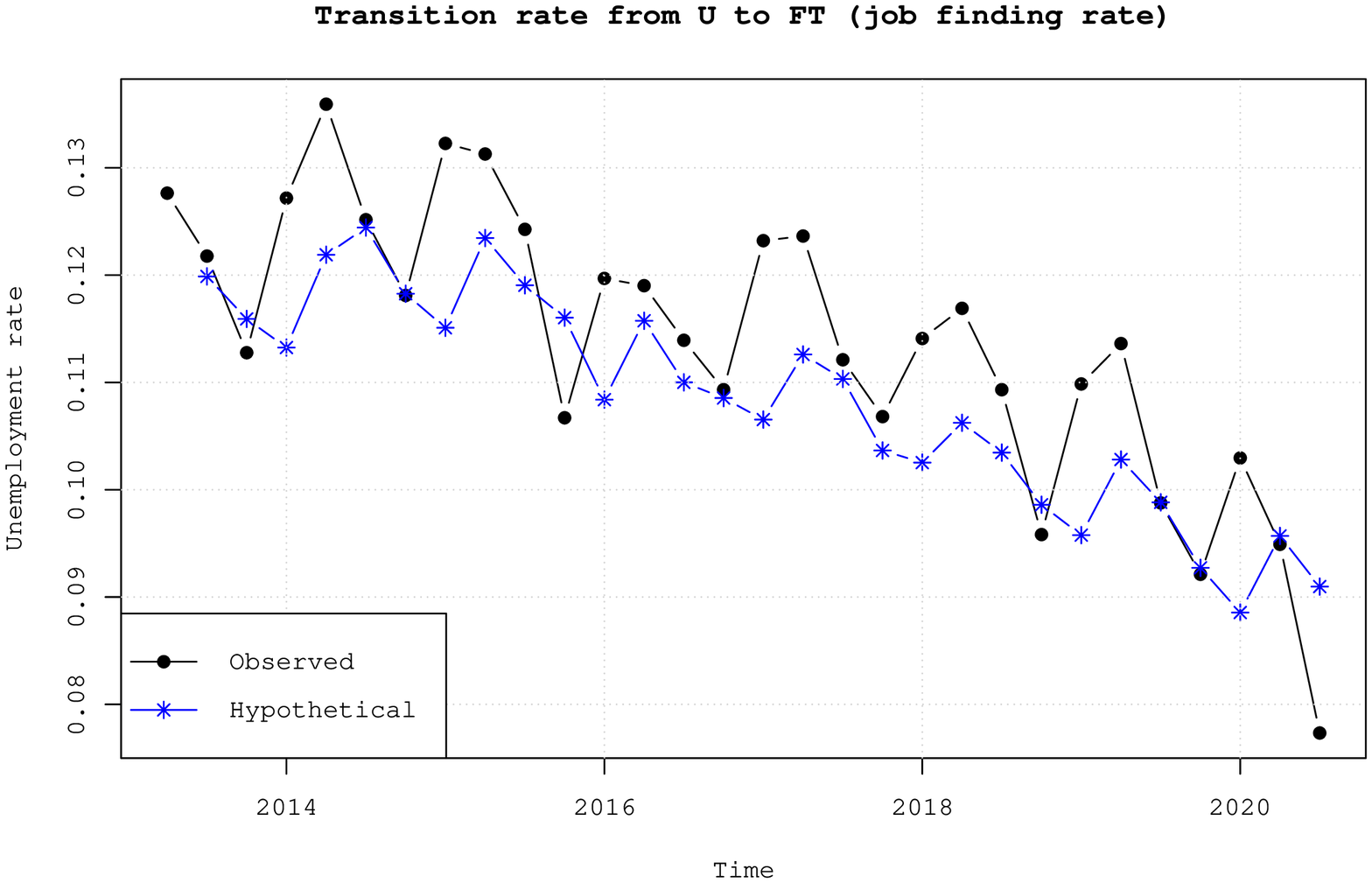}
		\caption{Transitions from U to FT.}
		\label{tot_unem}
	\end{subfigure}
	\begin{subfigure}[b]{0.32\textwidth}
		\centering
		\includegraphics[width=0.9\linewidth]{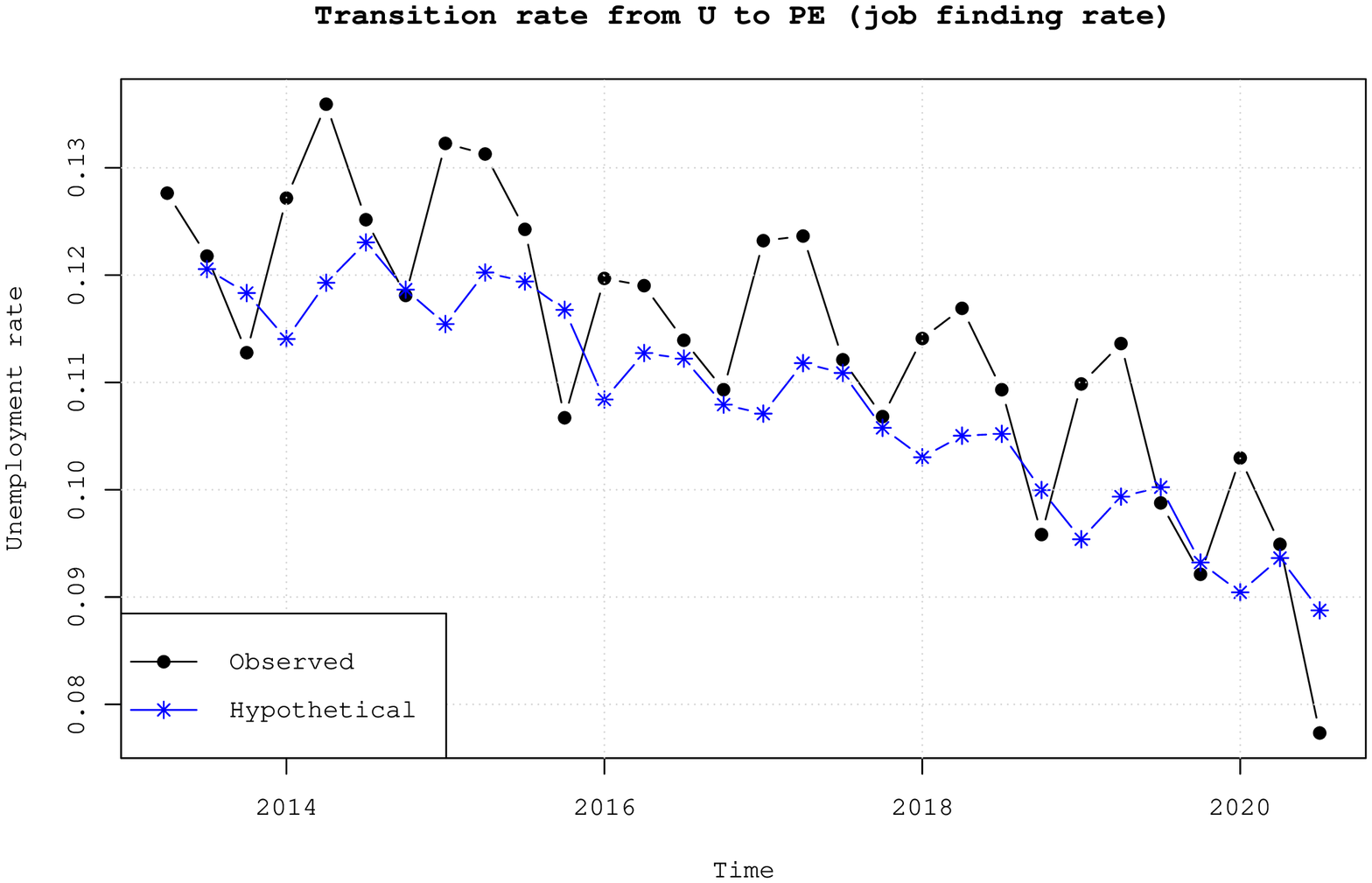}
		\caption{Transitions from U to PE.}
		\label{tot_unem}
	\end{subfigure}\\
	\begin{subfigure}[b]{0.32\textwidth}
		\centering
		\includegraphics[width=0.9\linewidth]{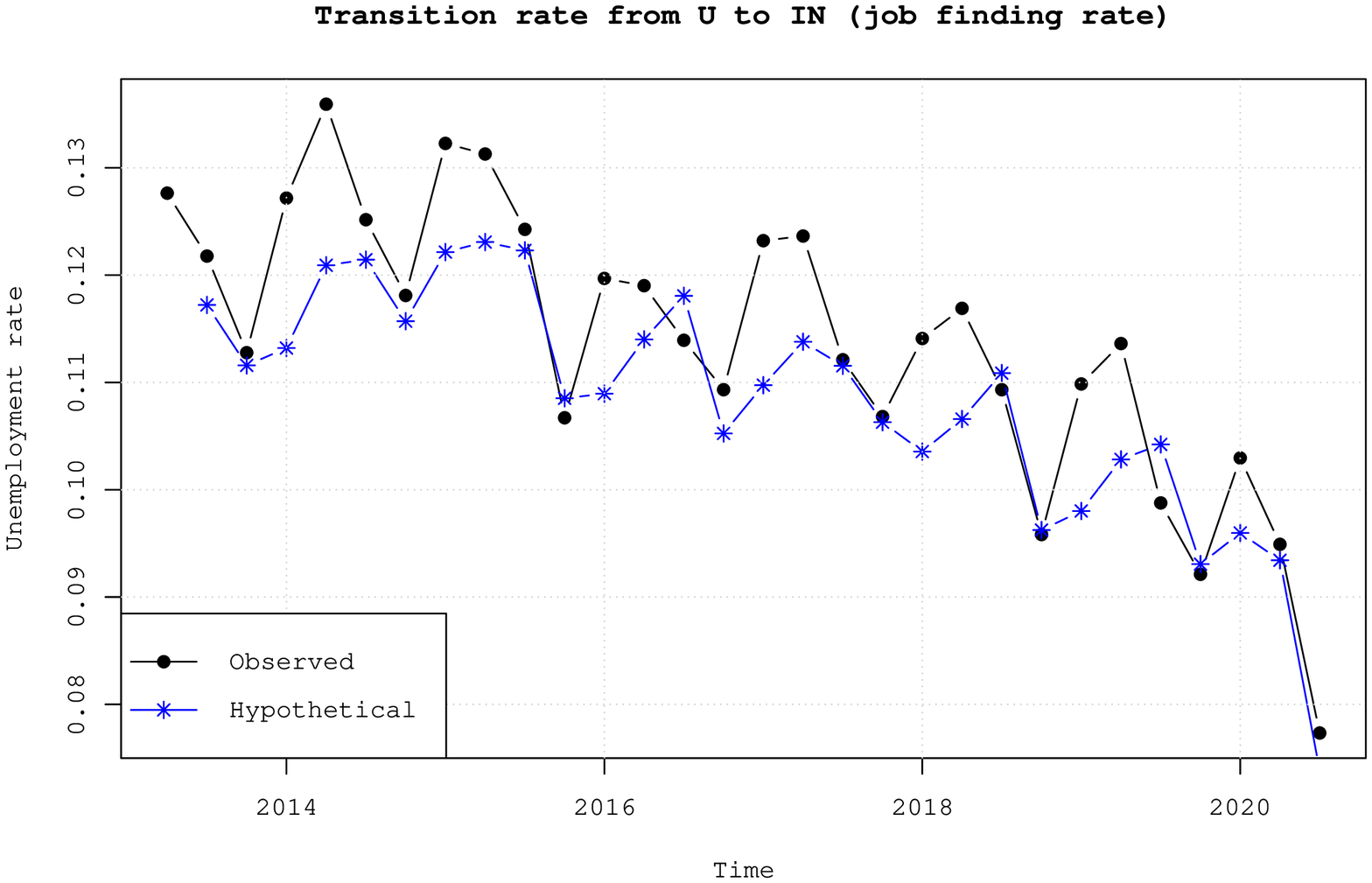}
		\caption{Transitions from U to PE.}
		\label{tot_unem}
	\end{subfigure}
	\begin{subfigure}[b]{0.32\textwidth}
		\centering
		\includegraphics[width=0.9\linewidth]{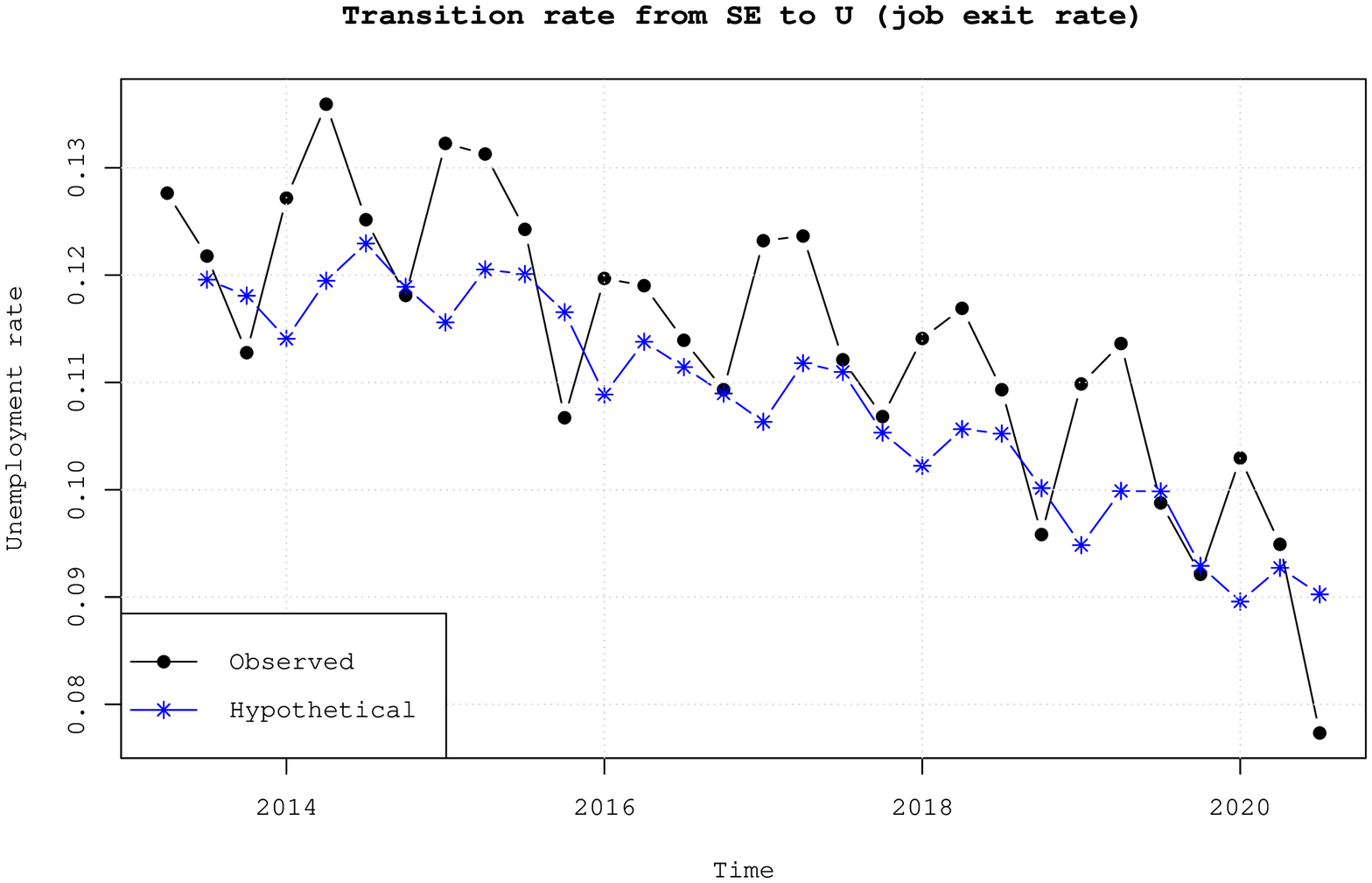}
		\caption{Transitions from SE to U.}
		\label{tot_unem}
	\end{subfigure}
		\begin{subfigure}[b]{0.32\textwidth}
			\centering
			\includegraphics[width=0.9\linewidth]{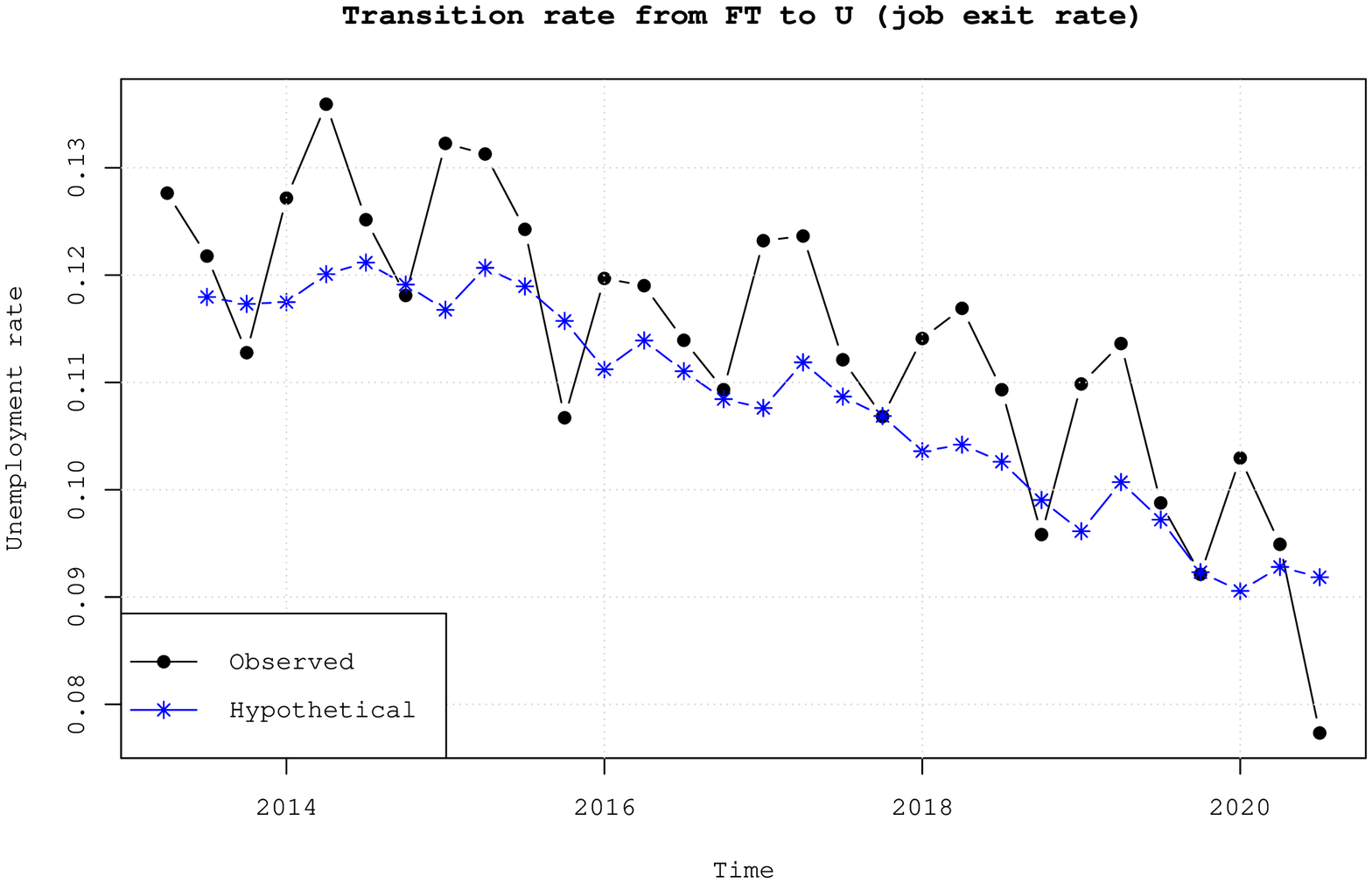}
			\caption{Transitions from FT to U.}
			\label{tot_unem}
		\end{subfigure}
		\begin{subfigure}[b]{0.32\textwidth}
			\centering
			\includegraphics[width=0.9\linewidth]{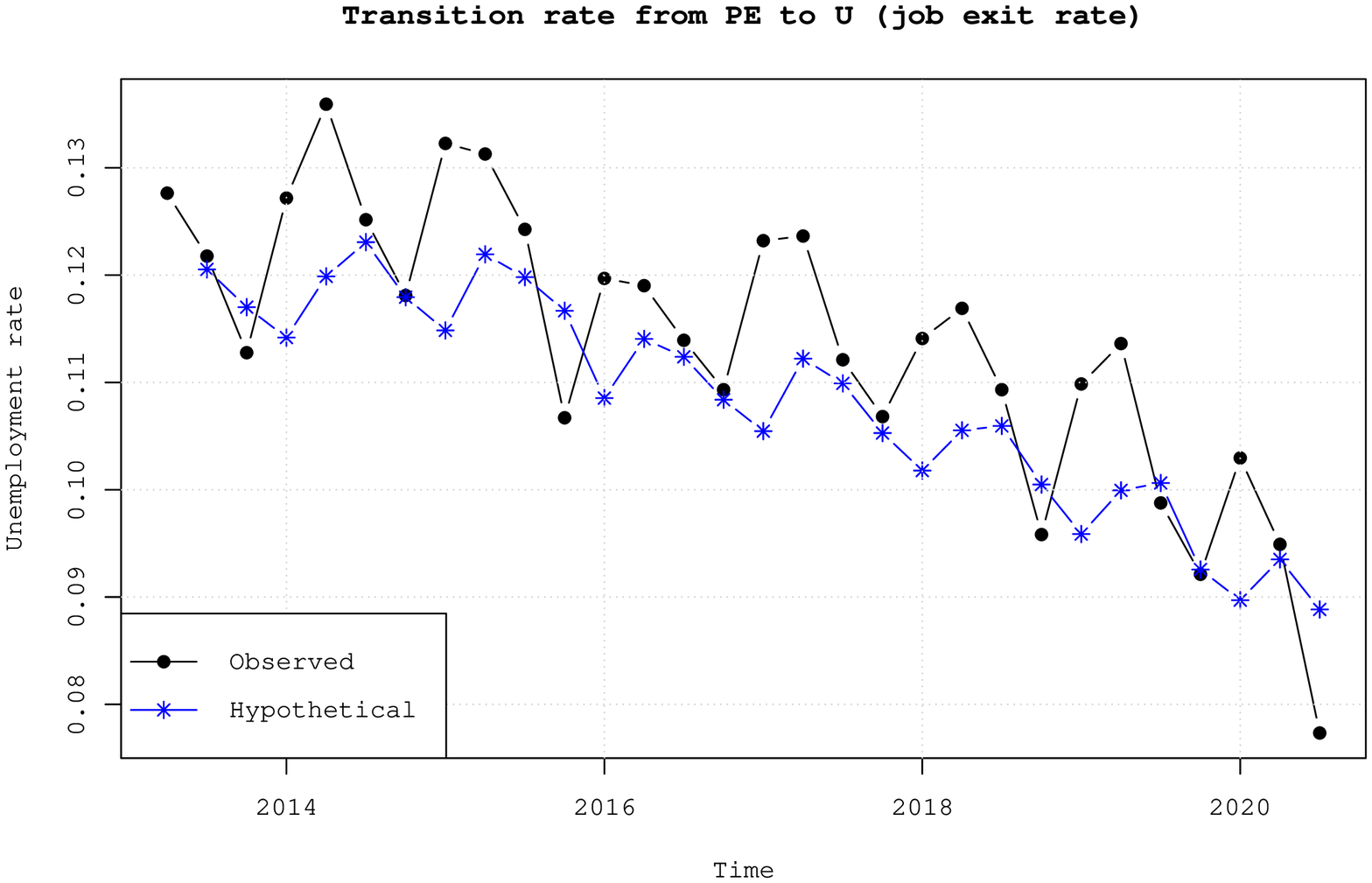}
			\caption{Transitions from PE to U.}
			\label{tot_unem}
		\end{subfigure}
		\begin{subfigure}[b]{0.32\textwidth}
			\centering
			\includegraphics[width=0.9\linewidth]{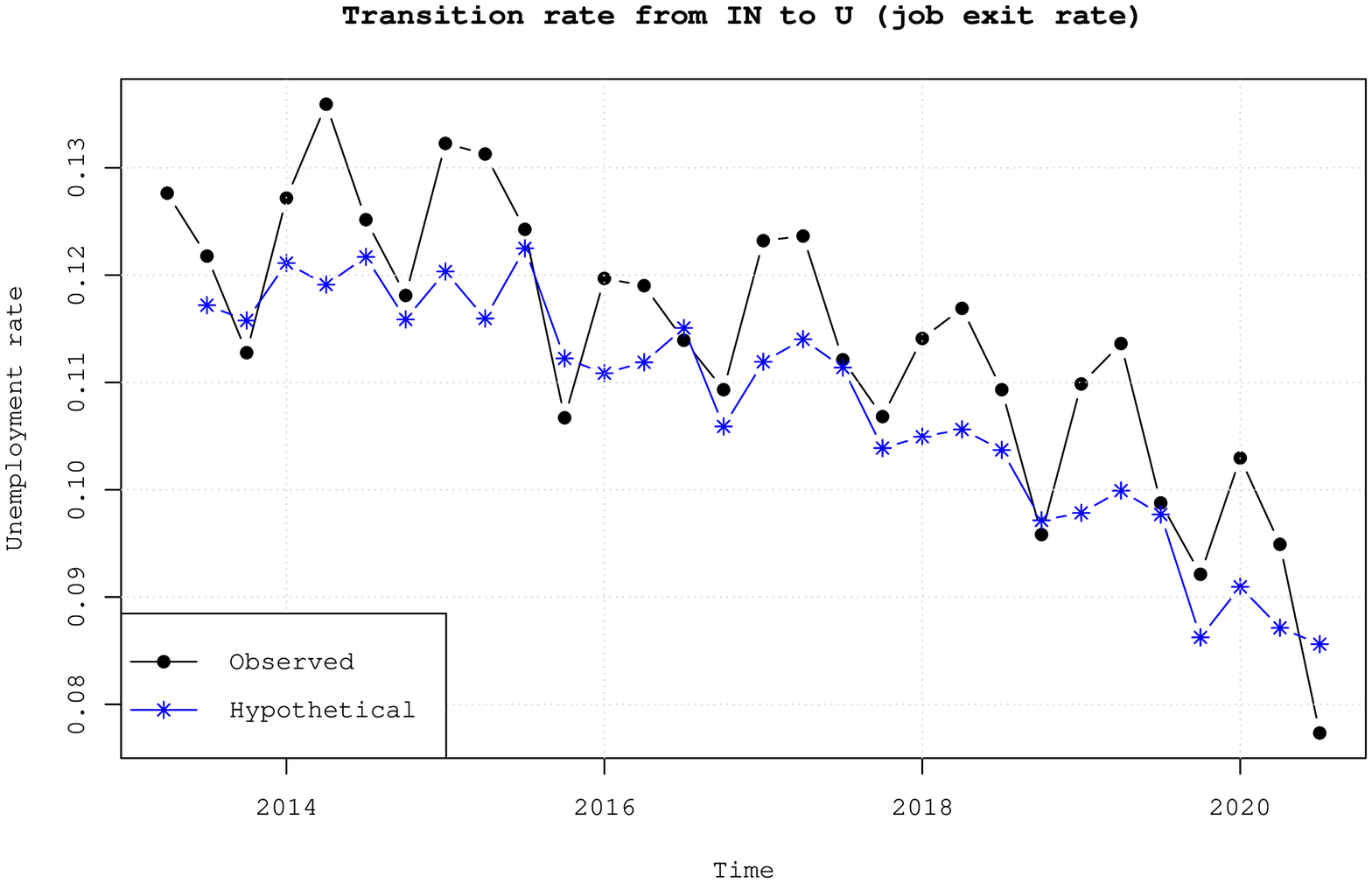}
			\caption{Transitions from IN to U.}
			\label{tot_unem}
		\end{subfigure}
	\caption*{ \scriptsize{Note: the black line reports the observed share of unemployed in the working age population, while the blue line reports the counterfactual share, computed using the methodology described in Section \ref{sec:AnalysDeterminantsChanges} (Equation (\ref{eq:counterfactualQ})).}}
\end{figure}

The aim of this section is to illustrate an application of the methodology discussed in Section \ref{sec:DecompositionVolatilityShares}.
Given that, as shown in Section \ref{sec:goodnessEstimates}, the fitted shares are a very good approximation of the observed shares, we use this strong relationship to distinguish between the contribution of the fluctuations in the inflows and outflows from each labour market state to the fluctuations in the fitted shares. We compute the hypothetical (counterfactual) shares using the methodology explained in Section \ref{sec:AnalysDeterminantsChanges} (Equation (\ref{eq:counterfactualQ})). Specifically, we allow movements over time in just one transition rate and assume that the remaining rates are fixed at their trend values calculated via a Hodrick-Prescott filter as in \cite{silva2013ins}.
We report in Figure \ref{fig:sensibilityUnemploymentRateToTransitionRate} the resulting time series for the unemployment rate\footnote{The resulting time series for all other labour market shares are available upon request.}, with the observed unemployment rate for comparison. 

In Table \ref{tab:decompositionUnemploymentVolatility} we report the decomposition of the contribution of each of the transition rates to the volatility of the unemployment rate. Fluctuations in the transition rate from unemployed to inactive account for about half of the movement in the unemployment rate, while the transitions from inactive to unemployed account for approximately one quarter. The third and fourth most important factors are the transitions from unemployed to temporary employed and viceversa, respectively. This suggests that unemployed workers are weakly attached to the labor force as their transitions to and from inactivity play a major role in determining the volatility of the unemployment rate.

\begin{table}[htbp]
	\footnotesize
	\centering
	\caption{Decomposition of unemployment share volatility.}
	\label{tab:decompositionUnemploymentVolatility}
	\begin{tabular}{lrrrrrrrrr}
		\hline
		\hline
		 \\[-1.8ex]
		& U to SE & U to FT & U to  PE & U to IN & SE to U & FT to U & PE to U & IN to U\\ 
		\hline
		 \\[-1.8ex]
	& 0.0067 &  0.1365&   0.0084&   0.4663 &  0.0022 &  0.0777 &  0.0371 &  0.2789\\
		\hline
		\hline
	\end{tabular}
	\caption*{\scriptsize{\textit{Note}: The entries are computed as the covariance of
			the fitted unemployment share and the counterfactual unemployment share for different time periods. All series are detrended using an HP filter with smoothing parameter equal to $1600$.}}
\end{table}

\subsection{The 2018 labour market reform: \textit{Decreto Dignità} \label{sec:DecretoDignità}}

We now apply the methodology discussed above to analyse labour market changes due to the implementation of a new policy. In particular,  we estimate transitions rates and  labour market shares in the period 2013-2020 and discuss their patterns in light of an important labour market reform approved in 2018 (\textit{Decreto Dignità}),\footnote{Decree July 12, 2018, n. 87 converted into Law August 9, 2018, n. 96.} which significantly changed the regulations of temporary contracts.  Although we do not claim any causal relationship between the reform and the changes in the transition rates and the labour market shares, we believe this  to be an interesting setting on which to test our methodology and prove the importance of considering both stock and flow variables when evaluating the  performance of the labour market. 

The \textit{Decreto Dignità}, which was approved in July 2018, significantly increased the rigidity of the temporary contract legislation with the goal of reducing job instability, defined by the time spent by individuals in temporary employment. Specifically, the reform reduced the maximum length  of temporary contracts from 36 to 24 months. It also introduced the restriction that any temporary contract longer than 12 months could be utilized only in three circumstances: (i) to replace a worker, (ii) for temporary reasons, outside the regular business and (iii) in case of a temporary and unforeseeable increase in business. If the contract was not justified by any of these reasons, the contract would be transformed into a permanent one. The number of extensions within the 24 months were reduced from 5 to 4 and any renewal of the contract would need to be justified by any of the three reasons listed above. The reform also increased the contributions payable by employers on each renewal of a temporary contract.\footnote{Prior to the reform coming into force, this contribution was set at 1.4\% of taxable salary for social security purposes and applied to all temporary contracts. With the reform it increased by 0.5\%. Moreover, the reform increased the firing costs associated with permanent contracts in case of unfair dismissals.} Clearly, the reform made the utilization of the temporary contract more difficult, more costly, it restricted the circumstances in which it could be utilized and it reduced the possibility of renewals/extensions.
As a result, some economists debated on the possible consequences of such reform, fearing that the increase in labour costs would lead to a decrease in labour demand and therefore an increase in unemployment, as predicted by the economic theory.\footnote{https://www.lastampa.it/politica/2018/07/19/news/decreto-dignita-l-audizione-di-tito-boeri-la-stima-di-8-mila-posti-persi-e-ottimistica-1.34032792.}

\subsubsection{Pre and post \textit{Decreto Dignità} \label{sec:prePostDecretoDignità}}

We take  quarter III of 2018, which includes the date of the approval of the \textit{Decreto Dignità} (July 12, 2018), as the time when the dynamics of the Italian labour market is expected to change. We will test whether the reform has affected the transitions rates between labour market states and the equilibrium shares in two ways. First,  we consider two quarters, one immediately before the reform, i.e. quarter II of 2018, and one at the end of the observation period, just before the onset of the Covid-19 pandemic, i.e., quarter IV of 2019. For both quarters considered, we compute estimates based on the average of four quarters (the actual quarter and the three quarters before) and make inference using a ``pairwise'' comparison between the elements of the $\mathbf{Q}$ matrix and the equilibrium shares, using bootstrapping methods. Then, in Section \ref{sec:evaluationByCounterfactual}, we compare the estimated transition rates with counterfactual transition rates, i.e., transition rates that we would have observed if the reform would have not been implemented.  

\paragraph{Labour market shares}
We look at the five annual labour market shares, i.e., permanent and temporary employed, self-employed, unemployed and inactive in quarter II of 2018, immediately before the implementation of the reform and in quarter IV of 2019, at the end of our observation period.\footnote{The annual shares are computed according to Equation (\ref{eq:annualQ}).} We find no change in the share of self-employed, permanent employed and inactive (Table \ref{tab:annualShares}). We instead find an increase in the share of temporary employed (from 7.1\% to 7.8\%) and a decrease in the share of unemployed (from 7.4\% to 6.8\%). 
Looking at the shares of individuals in the five different states before and after the reform, we identify some important changes, but we are not able to understand why these changes are happening, i.e., how the individuals move across states. Is the share of unemployment lower after the reform because more unemployed individuals move to employment or because they move to inactivity? This information is rather important to evaluate the impact of the reform on individuals and firms' choices.

\begin{table}[htbp]
	\footnotesize
	\centering
	\caption{Observed average annual labour market shares.}
	\label{tab:annualShares}
	\begin{tabular}{l|rrrrr}
		\hline
		\hline
		& SE & FT & PE & U & IN \\ 
		\hline
		2018 quarter II & 0.125 & 0.071 & 0.384 & 0.074 & 0.346 \\ 
		2019  quarter IV & 0.126 & 0.078 & 0.383 & 0.068 & 0.344 \\ 
		\hline
		\hline
	\end{tabular}
	\caption*{\scriptsize{\textit{Note}: The annual shares are computed as average ratio of the number of individuals in each state and the working age population. SE refers to self-employment, FT to temporary employment, PE to permanent employment, U to unemployment and IN to inactivity.}}
\end{table}

\paragraph{Transition rates}\label{sec:transRates2Q}
We then compute the estimates of the annual matrix  $\mathbf{Q}$ in quarter II of 2018, and in quarter IV of 2019 (see Equation (\ref{eq:annualQ})) and we perform a pairwise comparison of each element of the two matrices (Table \ref{fig:EstimatesQ2Quarters}). We find that the transition rates from self-employed to unemployed have declined significantly between the two quarters (from 0.05 to 0.037),\footnote{Remember that the diagonal elements of the   $\mathbf{Q}$ matrix are negative and the elements on the same row sum up to zero. A diagonal element which becomes less negative is interpreted as a reduced outflow from that state, and therefore as an increase in the persistence in that state.} while the transition rates from self-employed to self-employed have increased. We also find that the transition rates from temporary employed to temporary employed have significantly decreased, while from temporary employed to permanent employed have significantly increased (from 0.257 to 0.390). We also observe significant increases in the transition rates from permanent employed to inactive (from 0.046 to 0.051), but a decline in the transition rates from permanent employed to unemployed (from 0.034 to 0.027). Transition rate from unemployed to inactive have increased (from 2.371 to 2.561), while transition rates from unemployed to unemployed have decreased. Finally, we find that the transition rates from inactive to unemployed have decreased (from 0.471 to 0.440), while transition rates from inactive to inactive have significantly increased.

\begin{table}[htbp]
	\footnotesize
	\caption{Estimates of annual $\mathbf{Q}$ matrices in two different quarters.}
	\label{fig:EstimatesQ2Quarters}
	\centering
\subfloat[2018 quarter II]{
\begin{tabular}{ =l | +l +l +l +l +l}
  \hline
   \hline
 & SE & FT & PE & U & IN \\ 
  \hline
SE & \textbf{-0.22} & 0.031 & 0.056 & \textbf{0.050} & 0.084 \\ 
\rowstyle{\small}
   & (0.007) & (0.004) & (0.004) & (0.005) & (0.005) \\ 
  FT & 0.026 & \textbf{-1.053} & \textbf{0.257} & 0.419 & 0.351 \\ 
  \rowstyle{\small}
   & (0.004) & (0.022) & (0.011) & (0.02) & (0.016) \\ 
  PE & 0.020 & 0.037 & -0.137 & \textbf{0.034} & \textbf{0.046} \\ 
  \rowstyle{\small}
   & (0.001) & (0.002) & (0.003) & (0.002) & (0.002) \\ 
  U & \textbf{0.109} & 0.659 & 0.109 & \textbf{-3.248} & \textbf{2.371} \\ 
  \rowstyle{\small}
   & (0.009) & (0.024) & (0.01) & (0.05) & (0.043) \\ 
  IN & 0.025 & 0.091 & 0.031 & \textbf{0.471} & \textbf{-0.618} \\ 
  \rowstyle{\small}
   & (0.002) & (0.004) & (0.002) & (0.009) & (0.009) \\ 
   \hline
    \hline
\end{tabular}}
\vspace{0.5cm}
\subfloat[2019 quarter IV.]{
	\begin{tabular}{ =l | +l +l +l +l +l}
		\hline
		\hline
		& SE & FT & PE & U & IN \\ 
		\hline
		SE & \textbf{-0.192} & 0.026 & 0.048 & \textbf{0.037} & 0.081 \\ 
		\rowstyle{\small}
		& (0.007) & (0.003) & (0.004) & (0.004) & (0.005) \\ 
		FT & 0.030 & \textbf{-1.160} & \textbf{0.390} & 0.414 & 0.326 \\ 
		\rowstyle{\small}
		& (0.004) & (0.023) & (0.014) & (0.019) & (0.014) \\ 
		PE & 0.021 & 0.039 & -0.139 & \textbf{0.027} & \textbf{0.051} \\ 
		\rowstyle{\small}
		& (0.001) & (0.002) & (0.003) & (0.002) & (0.002) \\ 
		U & \textbf{0.088} & 0.670 & 0.105 & \textbf{-3.424} & \textbf{2.561} \\ 
		\rowstyle{\small}
		& (0.009) & (0.026) & (0.011) & (0.054) & (0.047) \\ 
		IN & 0.024 & 0.092 & 0.028 & \textbf{0.440} & \textbf{-0.584} \\ 
		\rowstyle{\small}
		& (0.002) & (0.004) & (0.002) & (0.009) & (0.009) \\ 
		\hline
		\hline
	\end{tabular}}
	\caption*{\scriptsize{\textit{Note}: Standard errors are reported in parenthesis. They are calculated using 1000 bootstrap samplings. Differences in the pairwise estimates which are statistically significant at 5\% level are reported in bold (the corresponding p-values are reported in the Appendix \ref{tab:pvaluesestimates}). SE refers to self-employment, FT to temporary employment, PE to permanent employment, U to unemployment and IN to inactivity.}}
\end{table}

\paragraph{Equilibrium labour market shares}

We then compute the equilibrium labour market shares in the two quarters considered (Table \ref{tab:equlibriumShares2Q}). We find that the shares of temporary employed and unemployed are significantly lower after the reform (from 9.1\% to 7.8\% and from 7\% to 6\%, respectively), as observed when we analysed the dynamics of actual labour market shares, but in addition we also observe that in equilibrium the share of permanent employed is significantly higher (from 35.7\% to 38.1\%). 

\begin{table}[htbp]
	\footnotesize
\centering
\caption{Equilibrium labour market shares.}
\label{tab:equlibriumShares2Q}
\begin{tabular}{l|rrrrr}
  \hline
    \hline
 & SE & FT & PE & U & IN \\ 
  \hline
   2018 II & 0.119 & \textbf{0.091} & \textbf{0.357} & \textbf{0.070} & 0.363 \\ 
    \rowstyle{\small} & (0.005) & (0.003) & (0.009) & (0.001) & (0.006) \\ 
   2019 IV & 0.126 & \textbf{0.078} & \textbf{0.381} & \textbf{0.060} & 0.356 \\ 
    \rowstyle{\small} & (0.006) & (0.002) & (0.009) & (0.001) & (0.007) \\ 
   \hline
     \hline
\end{tabular}
 \caption*{\scriptsize{\textit{Note}: The equilibrium shares are calculated on the basis of the annual $\mathbf{Q}$ matrices (Equation (\ref{eq:annualQ})). The shares are computed as the ratio of the number of individuals in each state and the working age population. Standard errors are reported in parenthesis. Differences in the coefficients in the two quarters which are statistically significant at 5\% level are reported in bold (the corresponding p-values are reported in the Appendix \ref{tab:pvaluesestimates}). SE refers to self-employment, FT to temporary employment, PE to permanent employment, U to unemployment and IN to inactivity.}}
\end{table}

\subsubsection{Counterfactual evaluation \label{sec:evaluationByCounterfactual}}

In this section, we use data from before the reform to forecast labour market shares and transition rates after the reform was implemented, which we interpret as counterfactual, and compare these forecasts to the observed ones \citep{blanchard2013growth,pathak2020well}. We compute the forecasts using the methodology described in the Appendix \ref{app:counterfactualViaARIMA}.\footnote{This forecasting methodology has also been used by \citet{barnichon2012ins} and \citet{barnichon2016forecasting}. In the counterfactual evaluation we ignore the constraints listed in the system of equations \ref{eq:hypothesisOnQ} and the fact that the elements of the matrix $\mathbf{Q}$ are estimated variables.}  

\paragraph{Labour market shares}

\begin{figure}[htbp]
	\centering
	\caption{Observed shares of individuals in different labour market states (\% of working age population).}
	\label{fig:ObsMasses}
		\begin{subfigure}[b]{0.3\textwidth}
		\centering
		\includegraphics[width=0.9\linewidth]{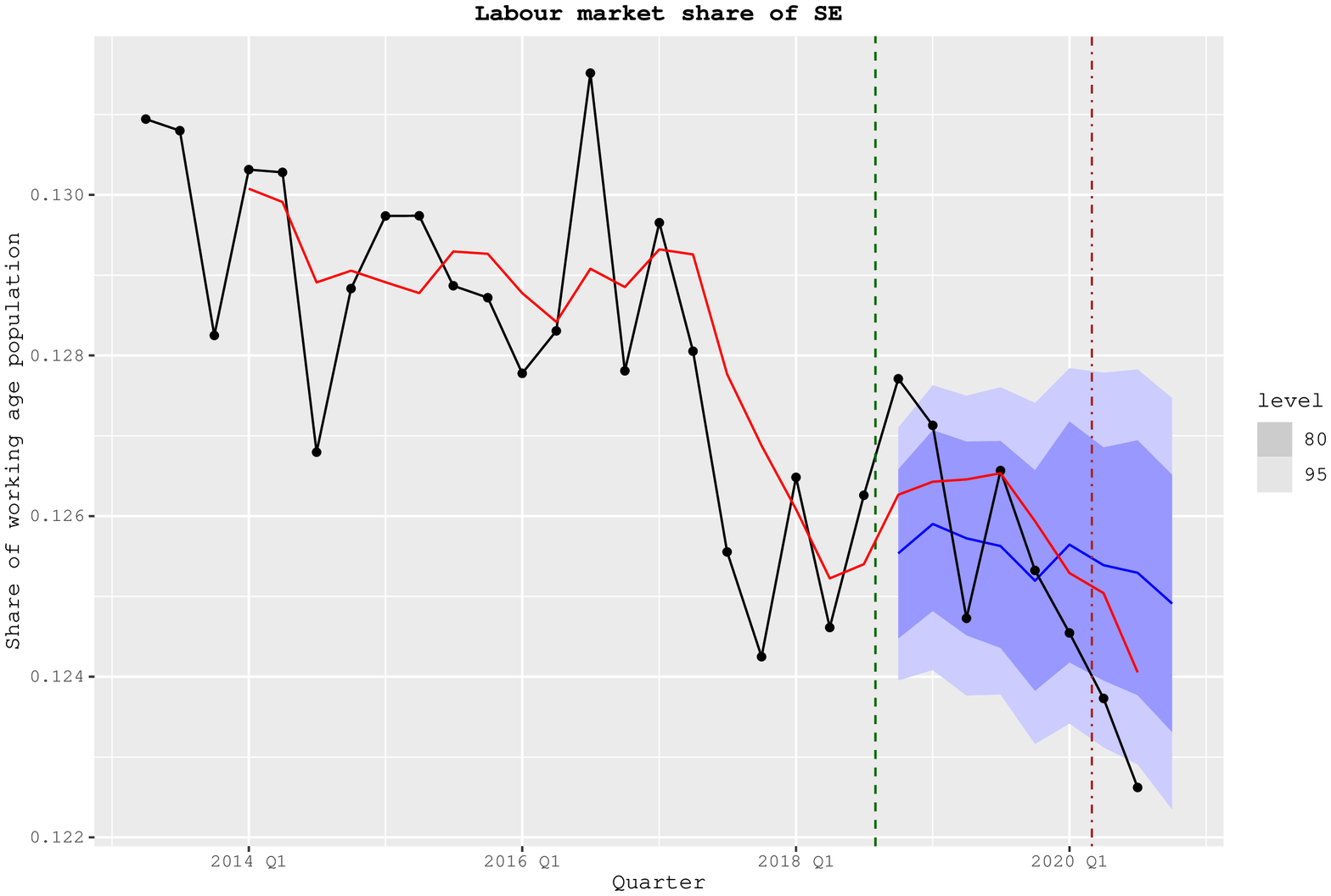}
		\caption{Self-employed.}
		\label{tot_unem}
	\end{subfigure}
	\begin{subfigure}[b]{0.3\textwidth}
		\centering
		\includegraphics[width=0.9\linewidth]{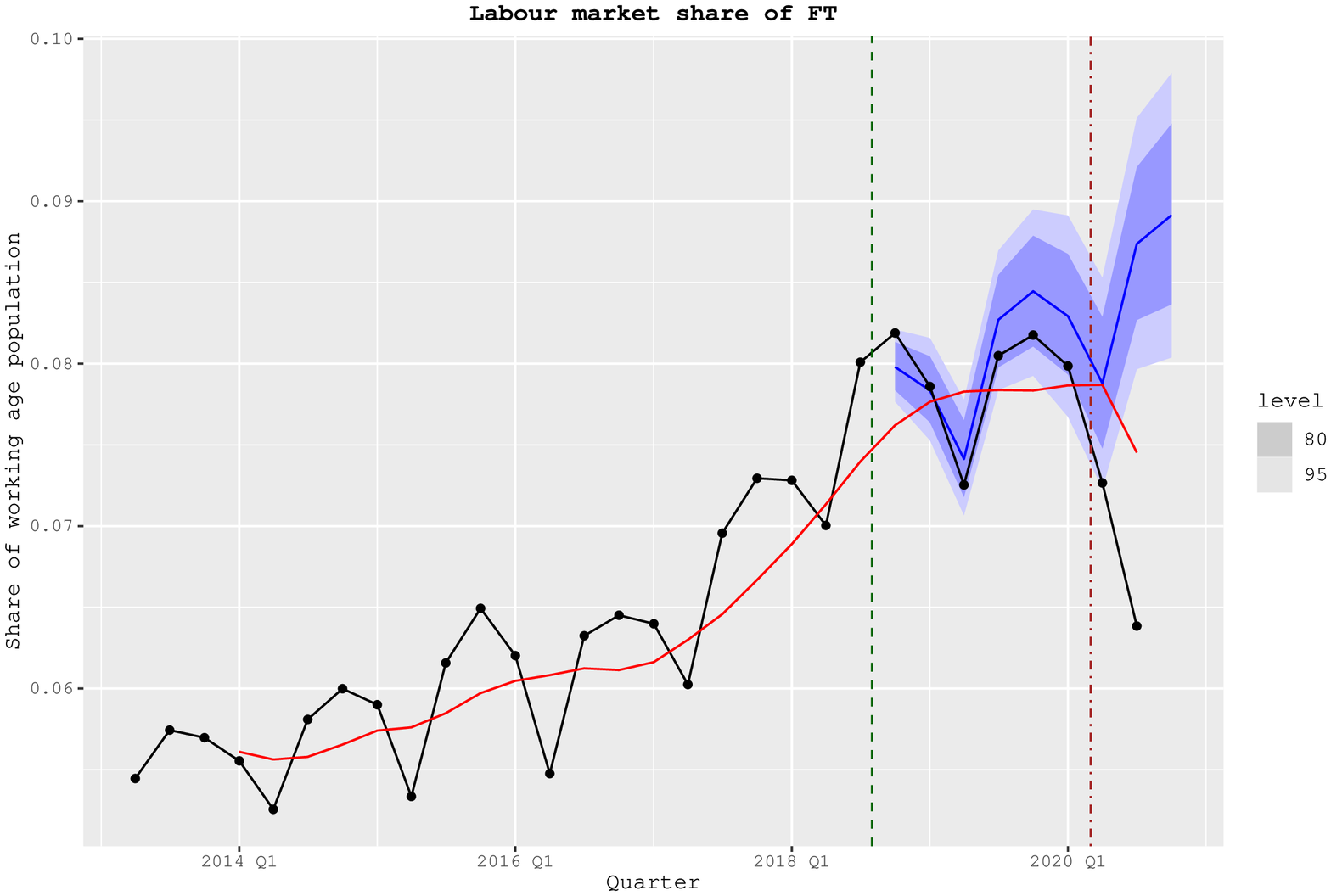}
		\caption{Temporary employed.}
		\label{tot_unem}
	\end{subfigure}
	\begin{subfigure}[b]{0.3\textwidth}
		\centering
		\includegraphics[width=0.9\linewidth]{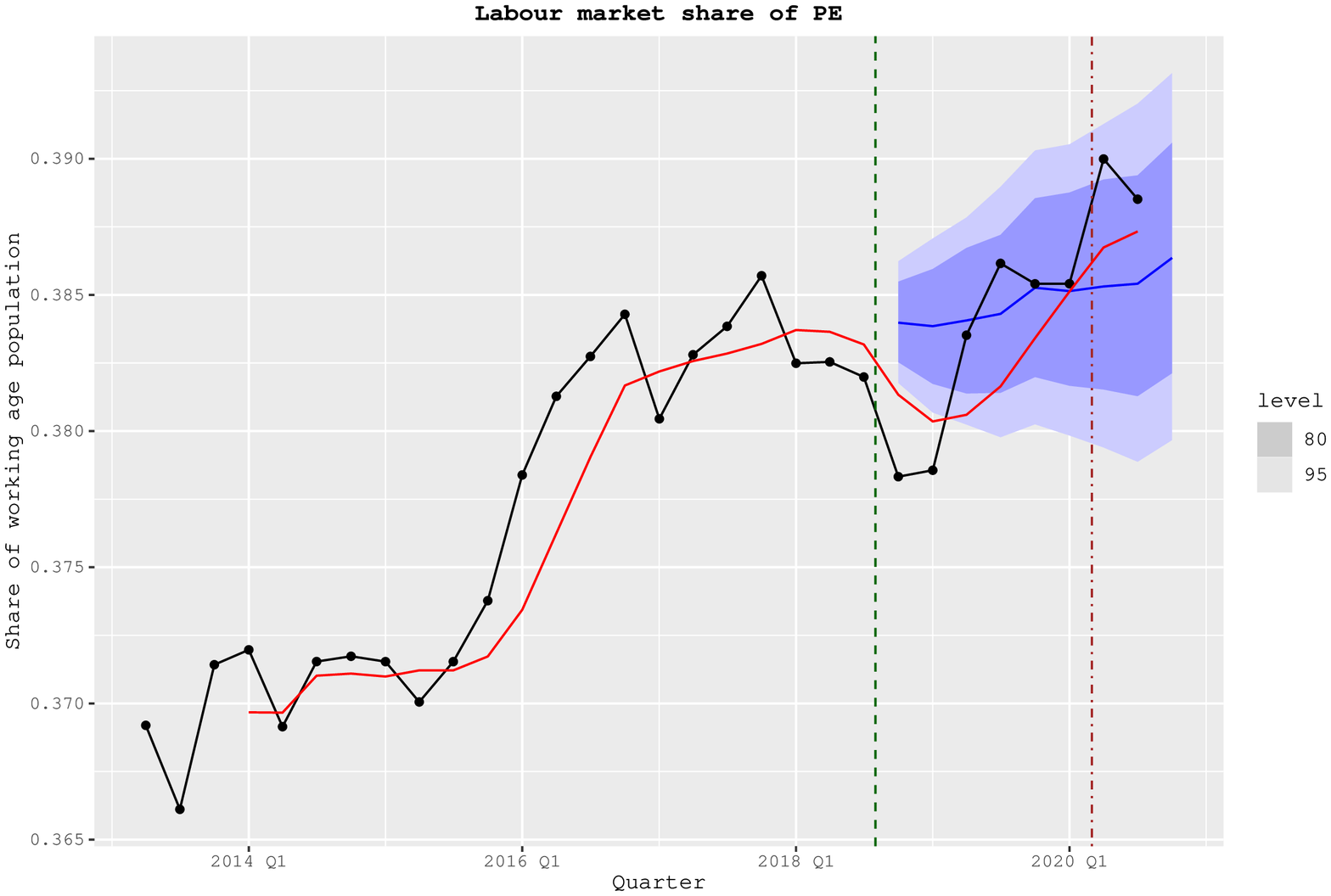}
		\caption{Permanent employed.}
		\label{tot_unem}
	\end{subfigure}\\
	\begin{subfigure}[b]{0.3\textwidth}
		\centering
		\includegraphics[width=0.9\linewidth]{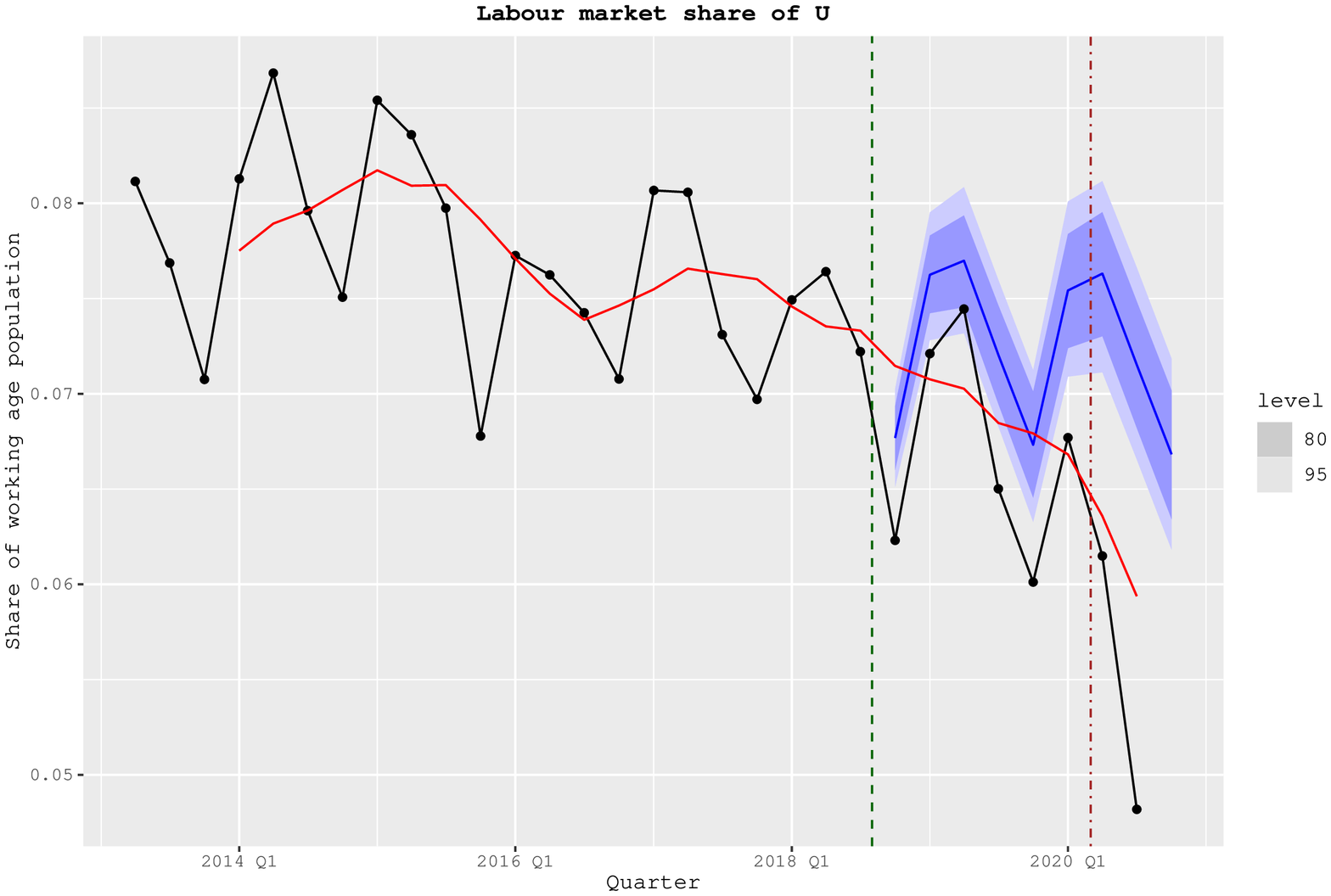}
		\caption{Unemployed.}
		\label{tot_unem}
	\end{subfigure}
	\begin{subfigure}[b]{0.3\textwidth}
		\centering
		\includegraphics[width=0.9\linewidth]{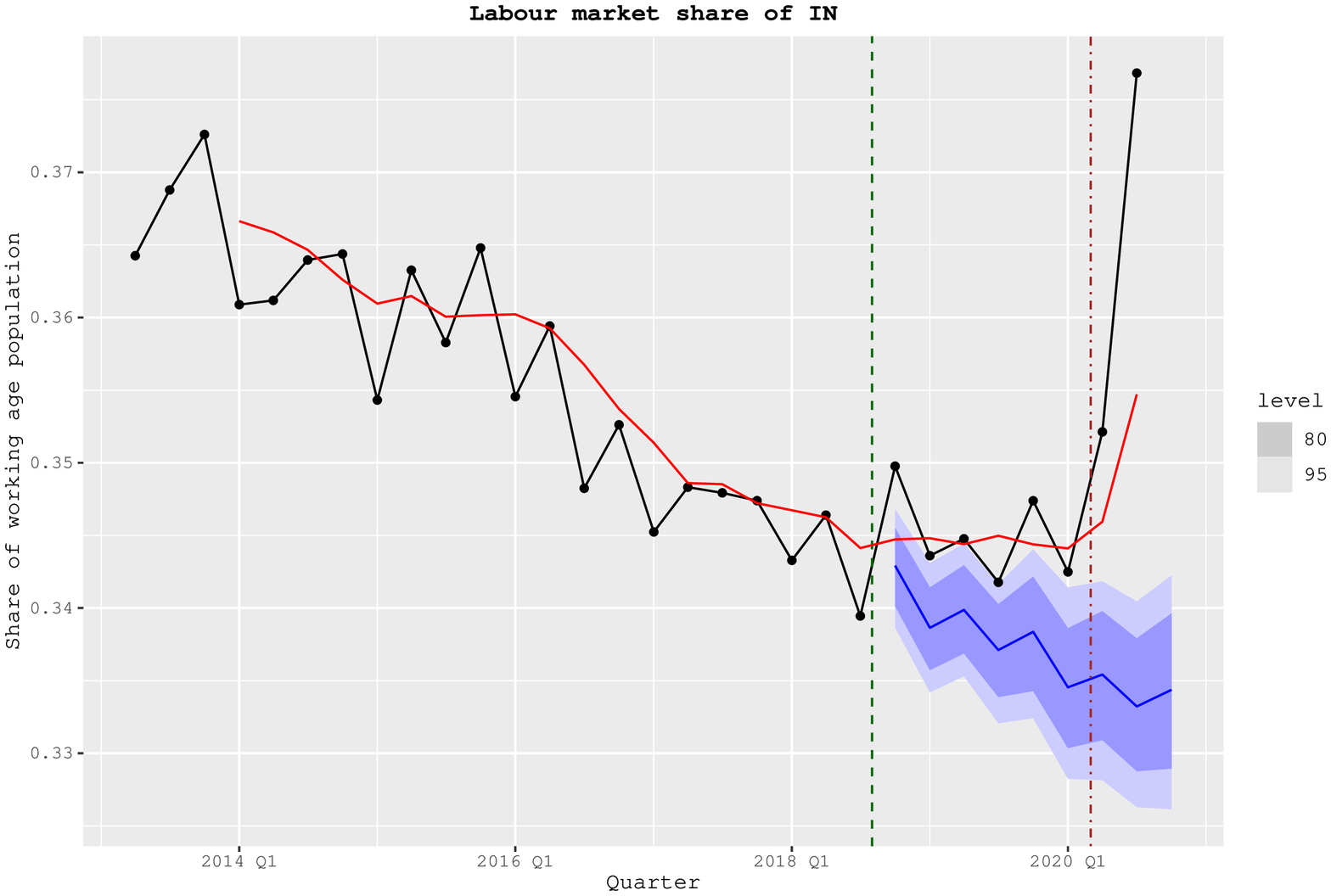}
		\caption{Inactive individuals.}
		\label{tot_unem}
	\end{subfigure}
	\caption*{ \scriptsize{Note: the black line reports the observed share of individuals in each labour market state, while the blue line reports the counterfactual share, with standard errors (purple area) at 80\% and 95\%. The red line is the annual observed share calculated according to Equation (\ref{eq:annualQ}), given $\tau=4$. The vertical green line represents the \textit{Decreto Dignita'} reform implemented in August 2018. Finally, the vertical red dotted line in March 2020 represents the beginning of the Covid-19 lockdown.}}
\end{figure}

We look at the evolution of the five labour market shares in each quarter from 2013 (quarter I) to 2020 (quarter I) and we compare them with the counterfactual shares.\footnote{It is interesting to notice how the forecasts replicate the seasonal fluctuations of labour market shares which are more affected by the seasonality, such as temporary employed, unemployed and self-employed, while showing smoother patterns for shares such as inactive and permanent employed, which are less subject to seasonal fluctuations.} These values are reported in Figure \ref{fig:ObsMasses}. The observed share of individuals hired on a temporary contract has been stable after the reform, compared to a growing trend before the reform. Compared to the counterfactual share, the observed one is significantly lower: the share of individuals who are temporary employed would have reached approximately 8.5\% of the working age population if no reform would have been implemented against the observed average of 7.8\%. Similarly, the share of unemployed individuals is significantly lower after the reform: the difference between the estimated and the counterfactual unemployment share is approximately 1 percentage point. Finally, the share of inactive individuals is significantly higher after the implementation of the reform, compared to the counterfactual estimate. While if there was no reform, the decreasing trend of the share of inactive individuals would have continued to reach 33.5\% of the working age population, with the reform the share is approximately equal to 34.5\%.  We do not pick up instead any relevant difference between the observed and the counterfactual estimates of the shares of permanent and self-employed individuals.

\paragraph{Transition rates}

\begin{landscape}
\begin{figure}[htbp]
	\vspace{-1cm}
	\centering
	\caption{Transition rates across labour market states.}
	\label{fig:transIntensities1}
		\begin{subfigure}[b]{0.25\textwidth}
		\centering
		\includegraphics[width=\linewidth]{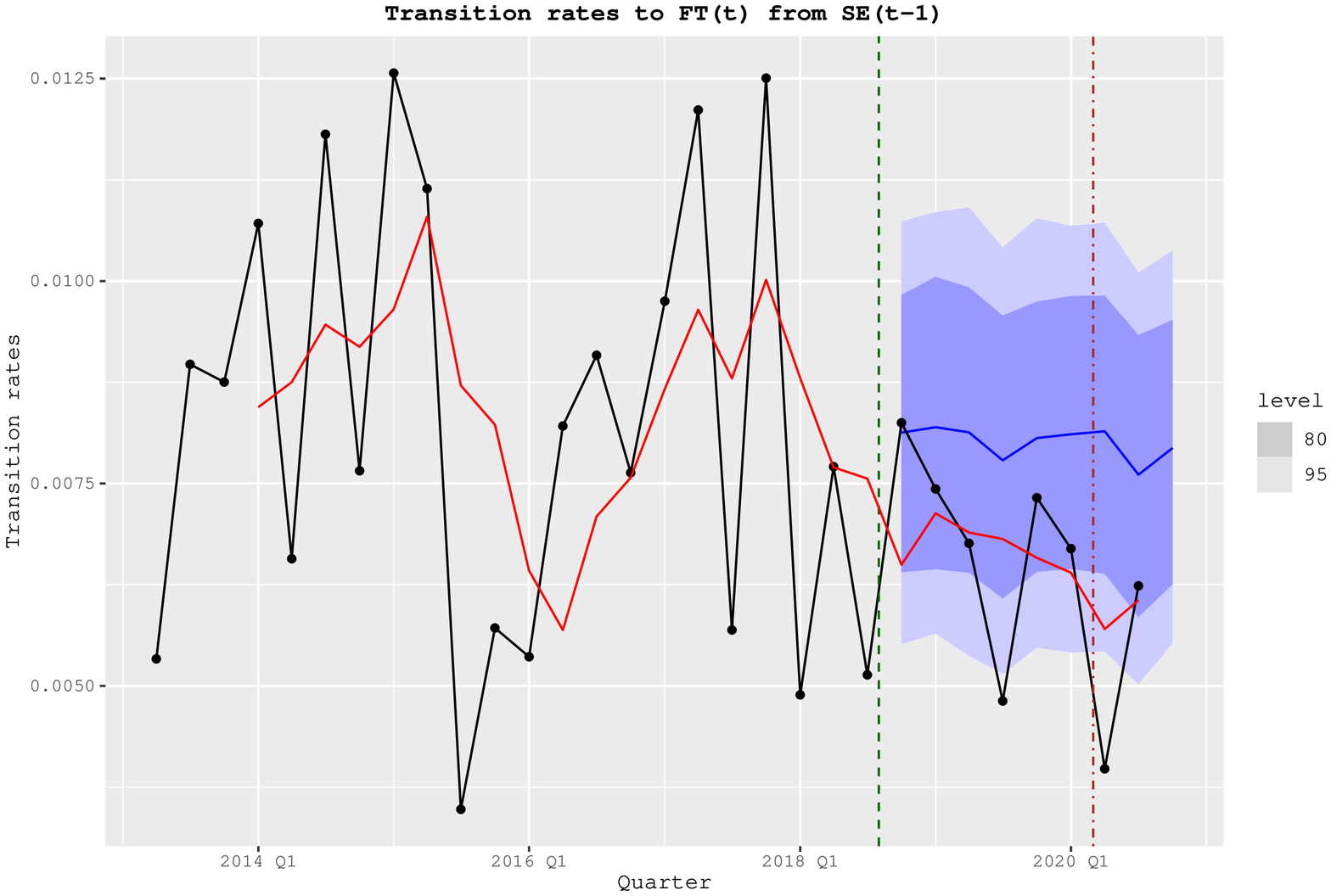}
		\caption{From SE to FT.}
		\label{tot_unem}
	\end{subfigure}
	\begin{subfigure}[b]{0.25\textwidth}
		\centering
		\includegraphics[width=\linewidth]{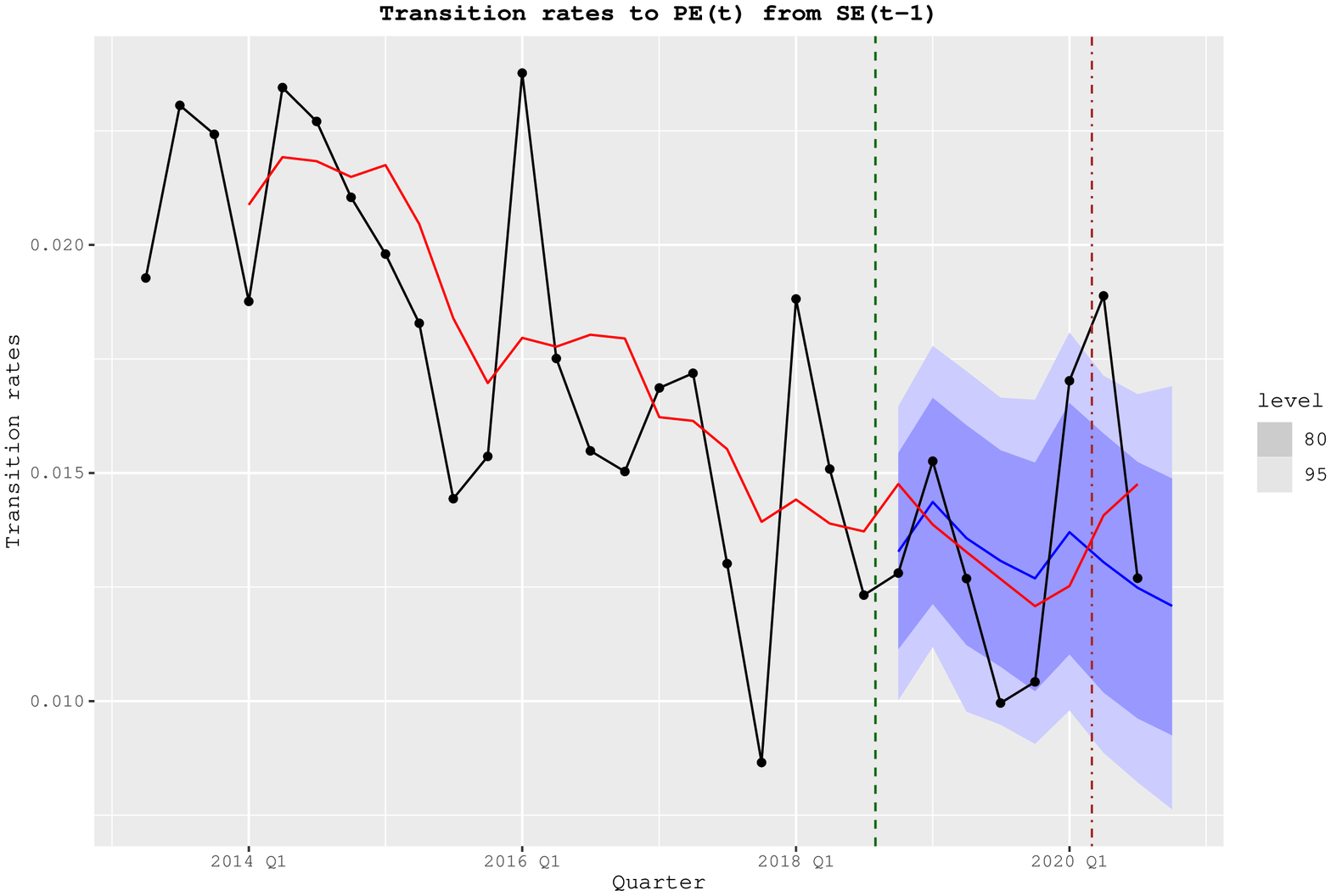}
		\caption{From SE to PE.}
		\label{tot_unem}
	\end{subfigure}
	\begin{subfigure}[b]{0.25\textwidth}
		\centering
		\includegraphics[width=\linewidth]{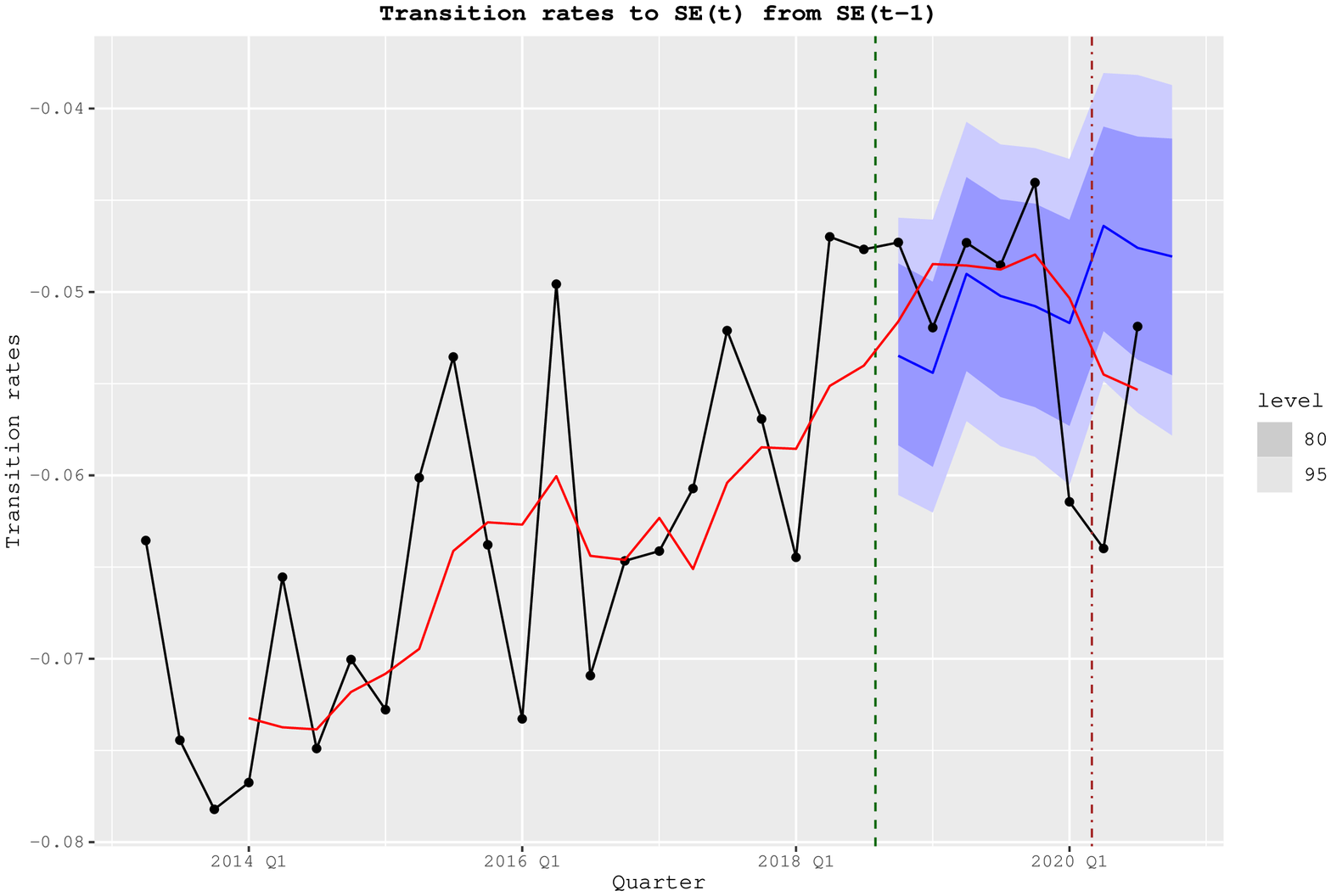}
		\caption{From SE to SE.}
		\label{tot_unem}
	\end{subfigure}
	\begin{subfigure}[b]{0.25\textwidth}
		\centering
		\includegraphics[width=\linewidth]{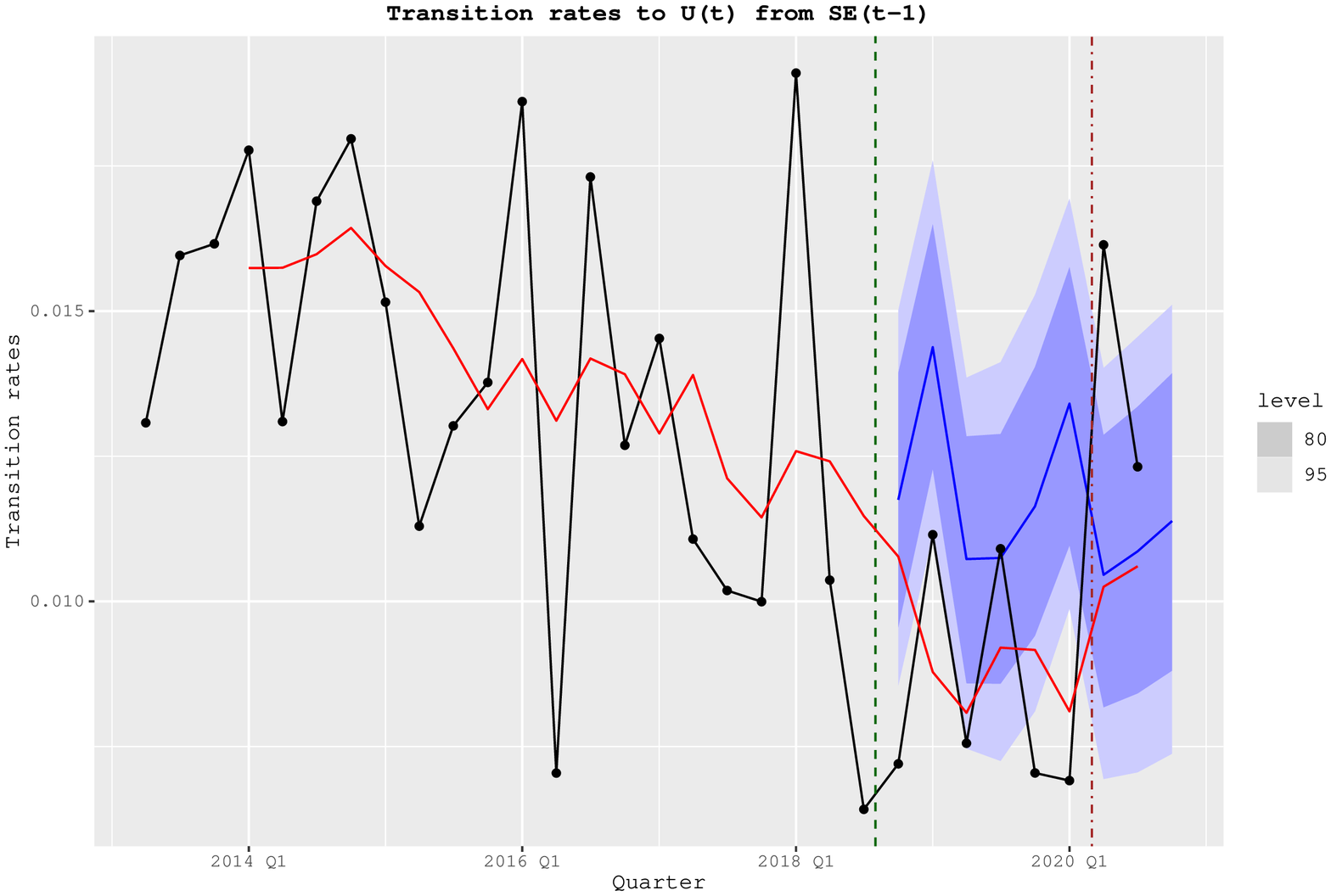}
		\caption{From SE to U.}
		\label{tot_unem}
	\end{subfigure}
	\begin{subfigure}[b]{0.25\textwidth}
		\centering
		\includegraphics[width=\linewidth]{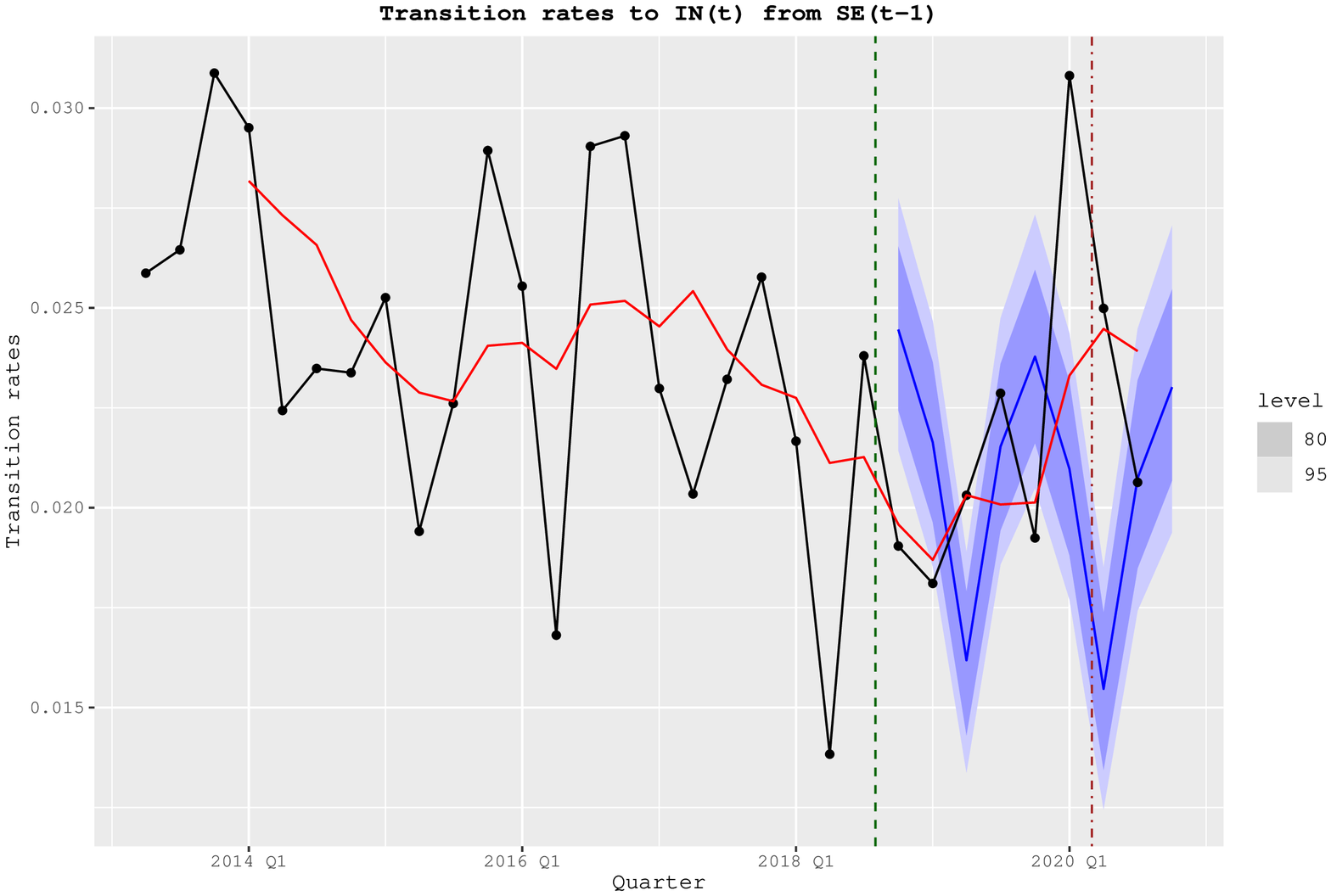}
		\caption{From SE to IN.}
		\label{tot_unem}
	\end{subfigure}
	\begin{subfigure}[b]{0.25\textwidth}
			\centering
			\includegraphics[width=\linewidth]{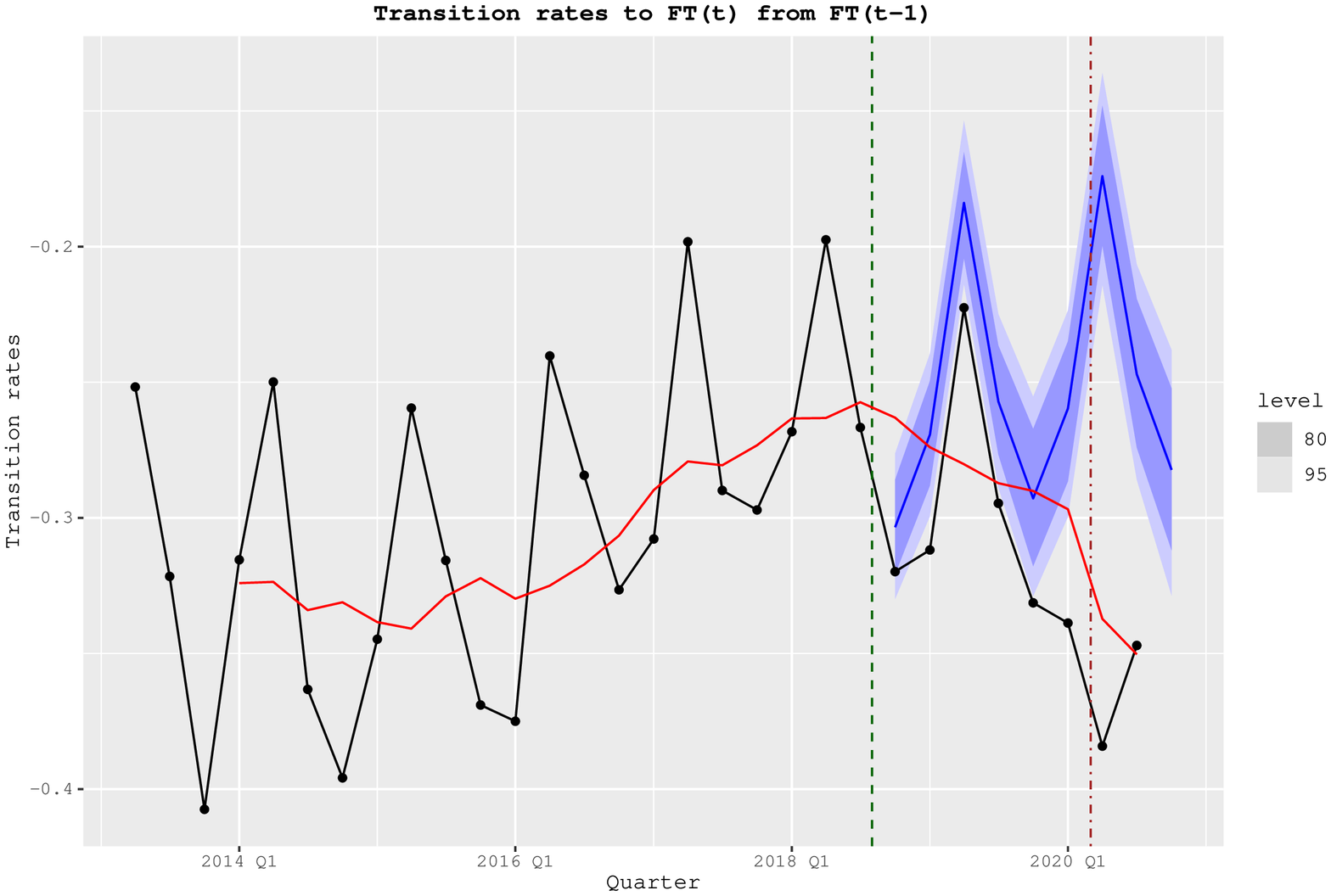}
			\caption{From FT to FT.}
			\label{tot_unem}
	\end{subfigure}
	\begin{subfigure}[b]{0.25\textwidth}
			\centering
			\includegraphics[width=\linewidth]{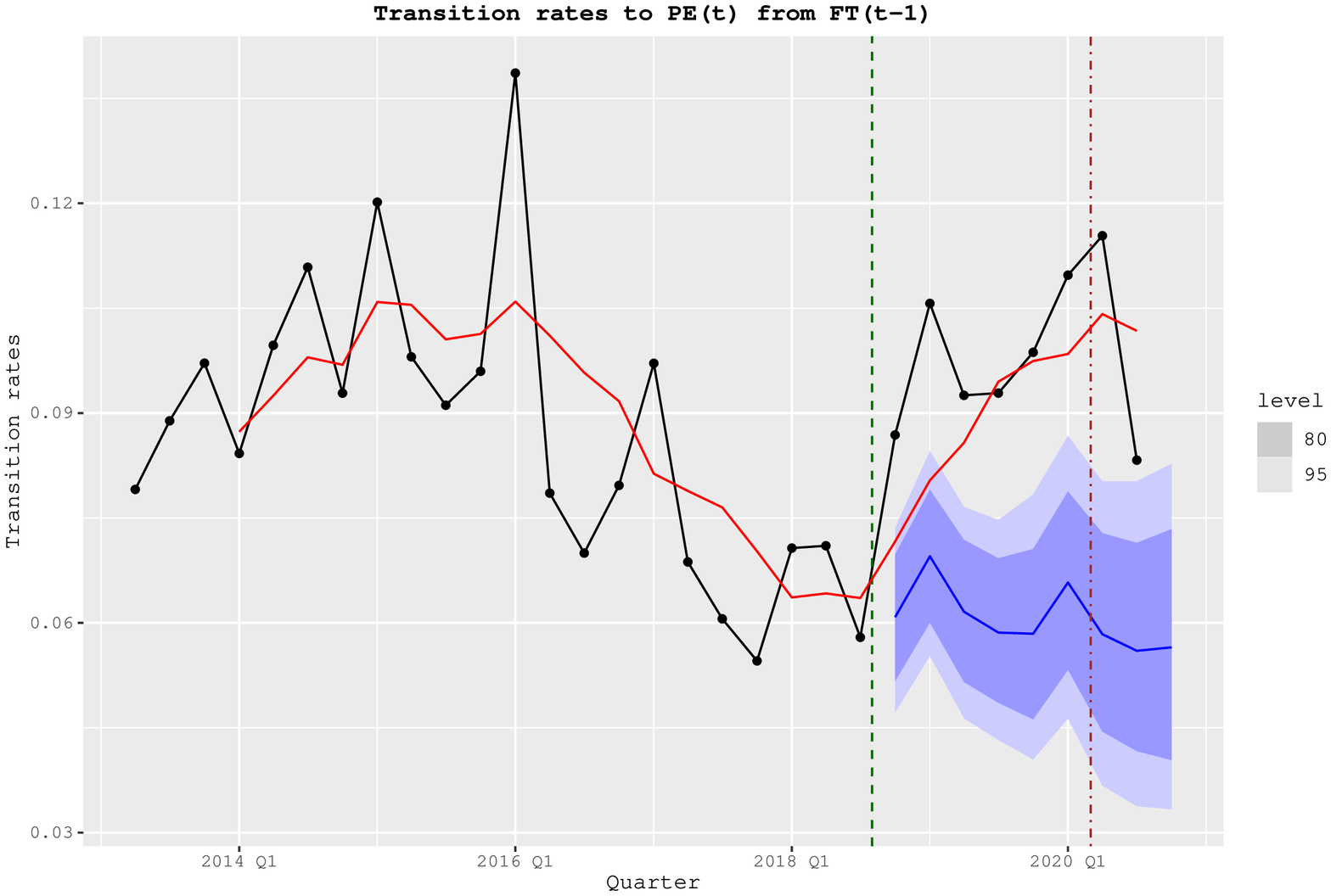}
			\caption{From FT to PE.}
			\label{tot_unem}
	\end{subfigure}
	\begin{subfigure}[b]{0.25\textwidth}
			\centering
			\includegraphics[width=\linewidth]{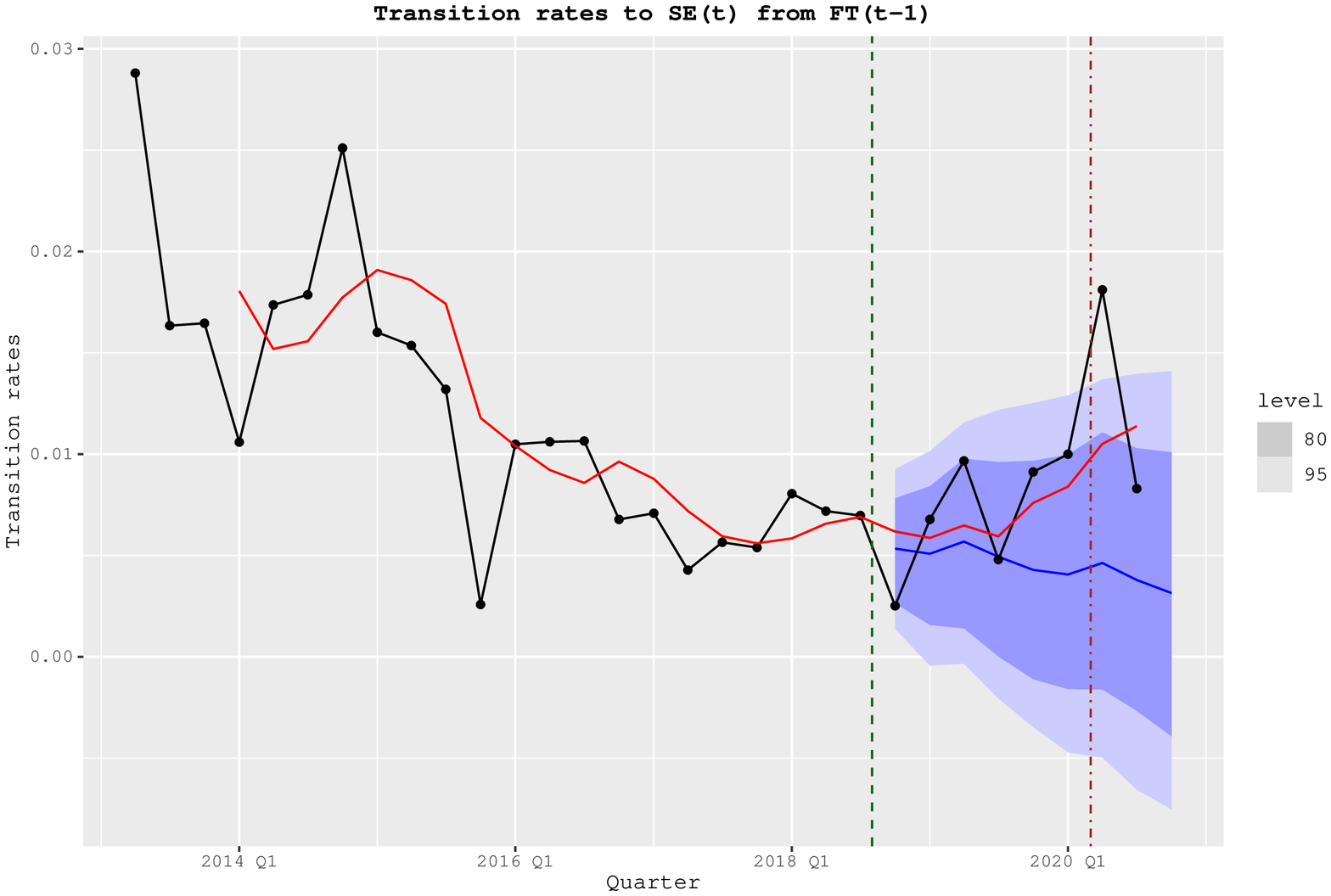}
			\caption{From FT to SE.}
			\label{tot_unem}
	\end{subfigure}
	\begin{subfigure}[b]{0.25\textwidth}
			\centering
			\includegraphics[width=\linewidth]{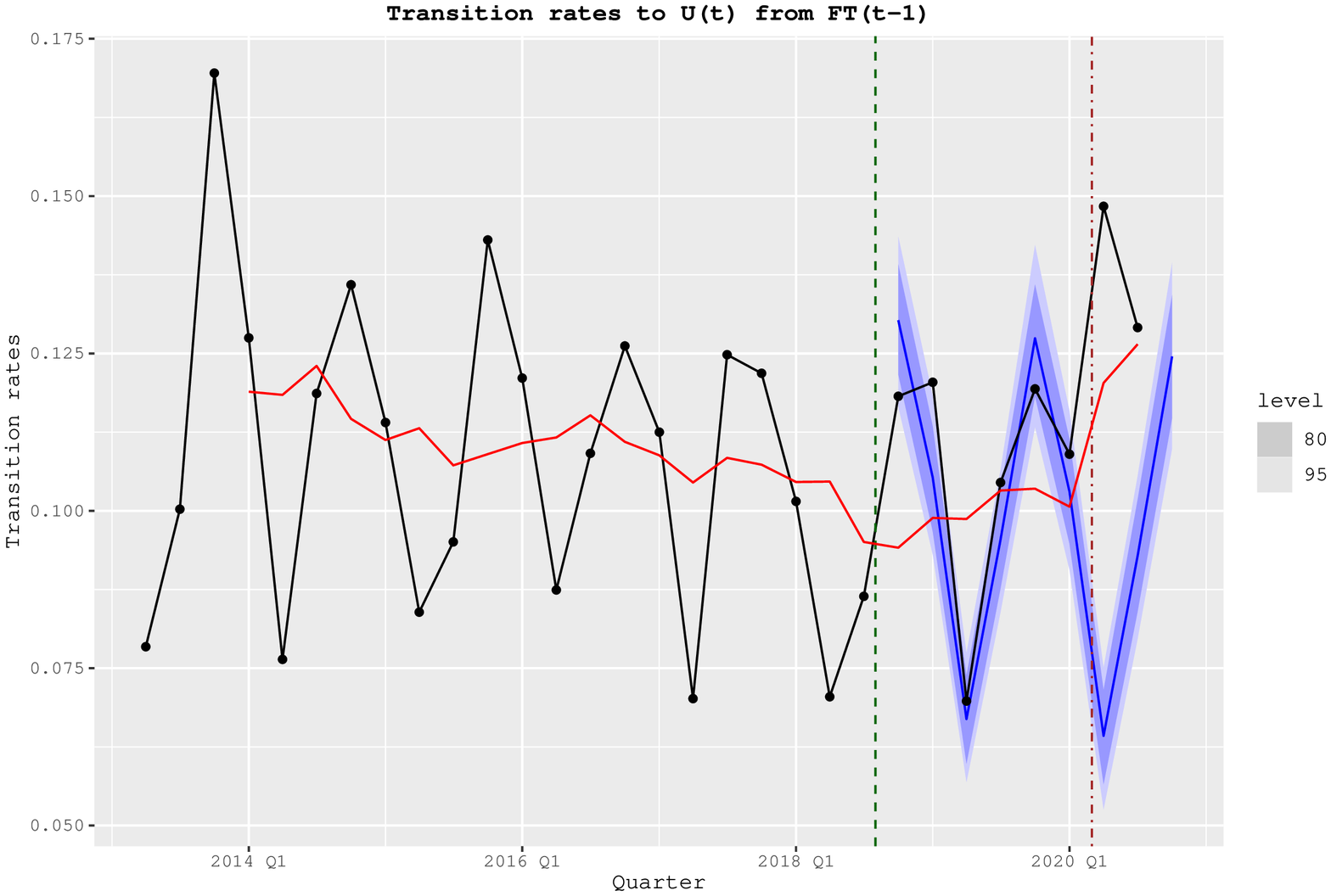}
			\caption{From FT to U.}
			\label{tot_unem}
	\end{subfigure}
	\begin{subfigure}[b]{0.25\textwidth}
	\centering
	\includegraphics[width=\linewidth]{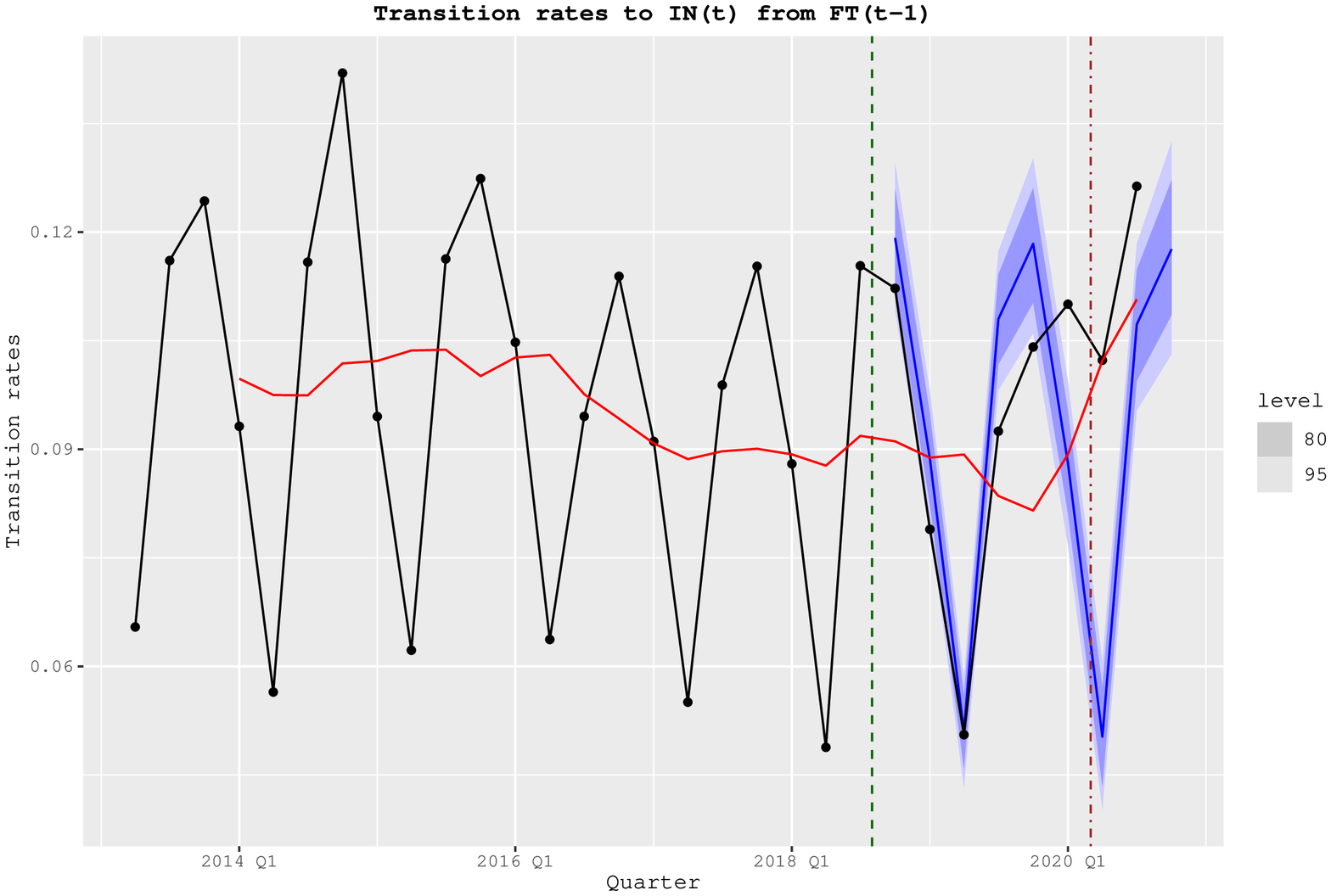}
	\caption{From FT to IN.}
	\label{tot_unem}
	\end{subfigure}\\
	\begin{subfigure}[b]{0.25\textwidth}
		\centering
		\includegraphics[width=\linewidth]{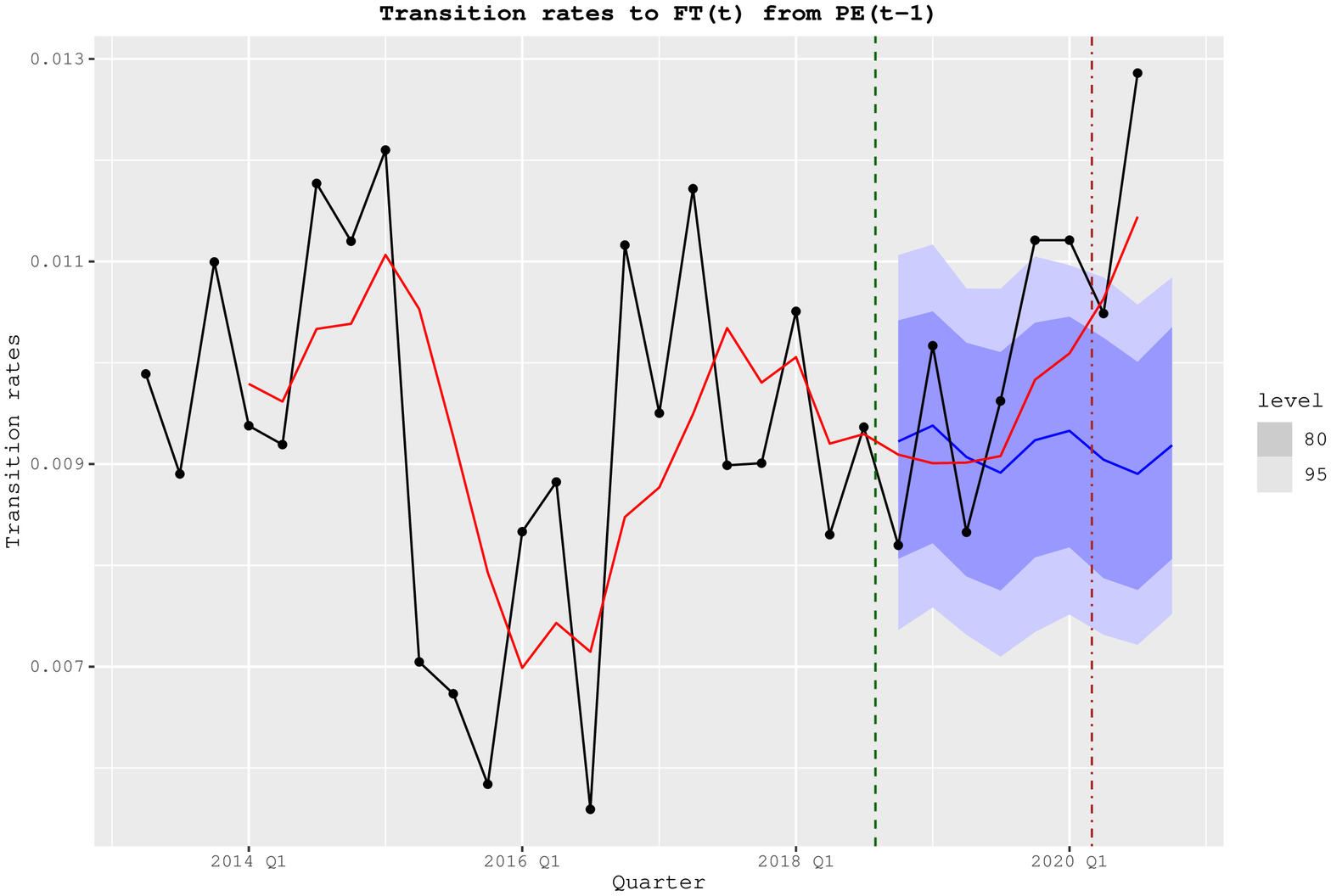}
		\caption{From PE to FT.}
		\label{tot_unem}
	\end{subfigure}
	\begin{subfigure}[b]{0.25\textwidth}
		\centering
		\includegraphics[width=\linewidth]{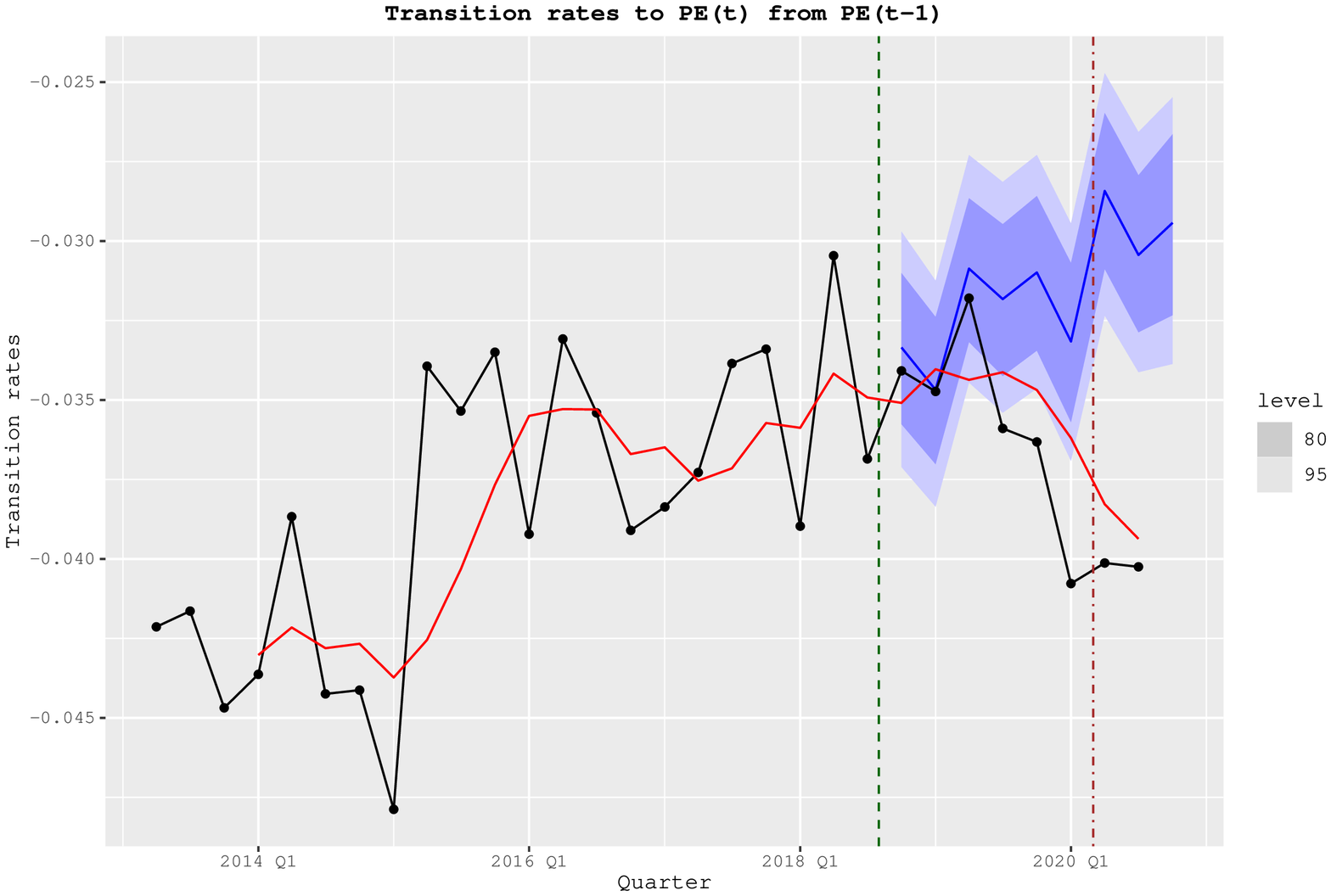}
		\caption{From PE to PE.}
		\label{tot_unem}
	\end{subfigure}
	\begin{subfigure}[b]{0.25\textwidth}
		\centering
		\includegraphics[width=0.9\linewidth]{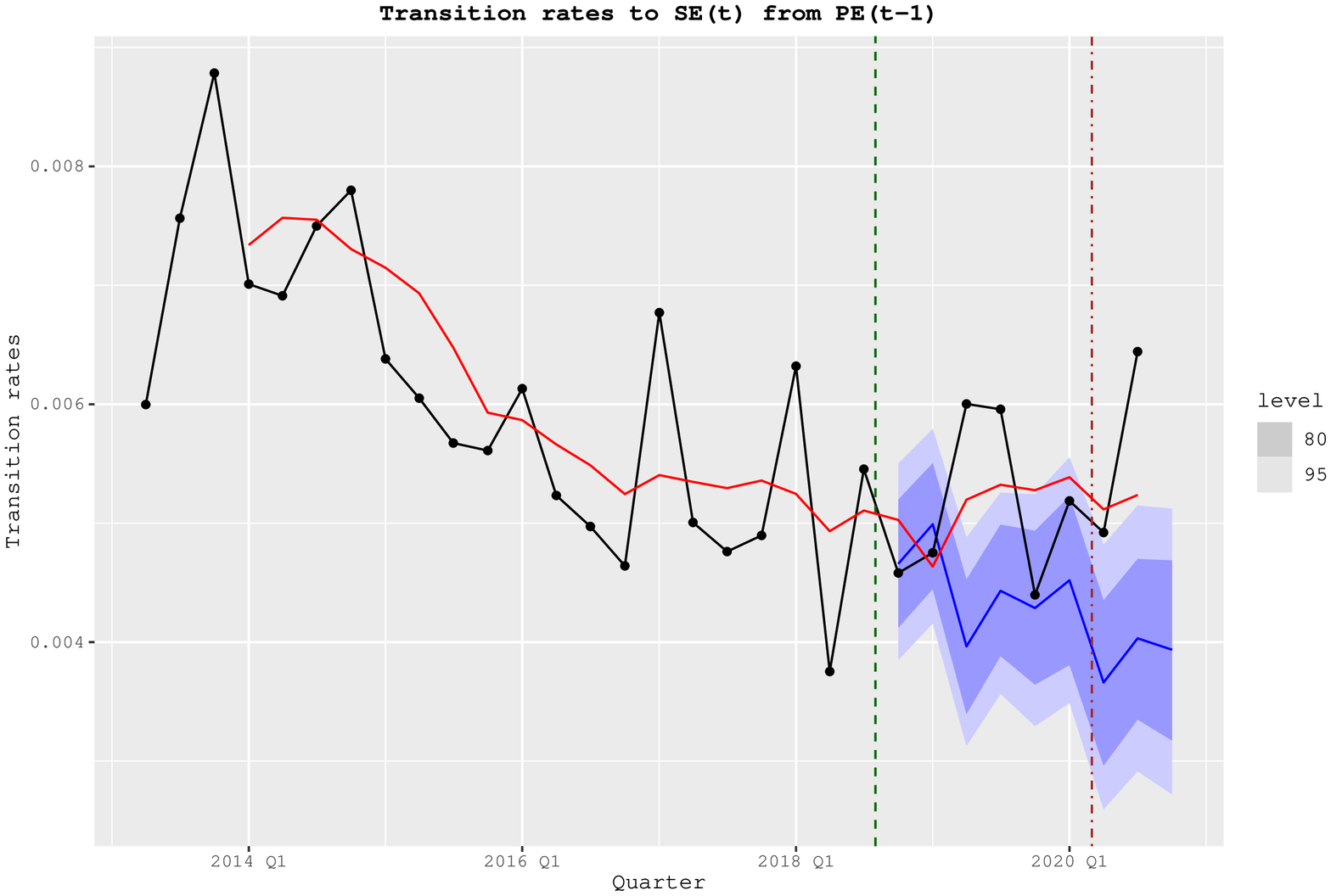}
		\caption{From PE to SE.}
		\label{tot_unem}
	\end{subfigure}
	\begin{subfigure}[b]{0.25\textwidth}
		\centering
		\includegraphics[width=\linewidth]{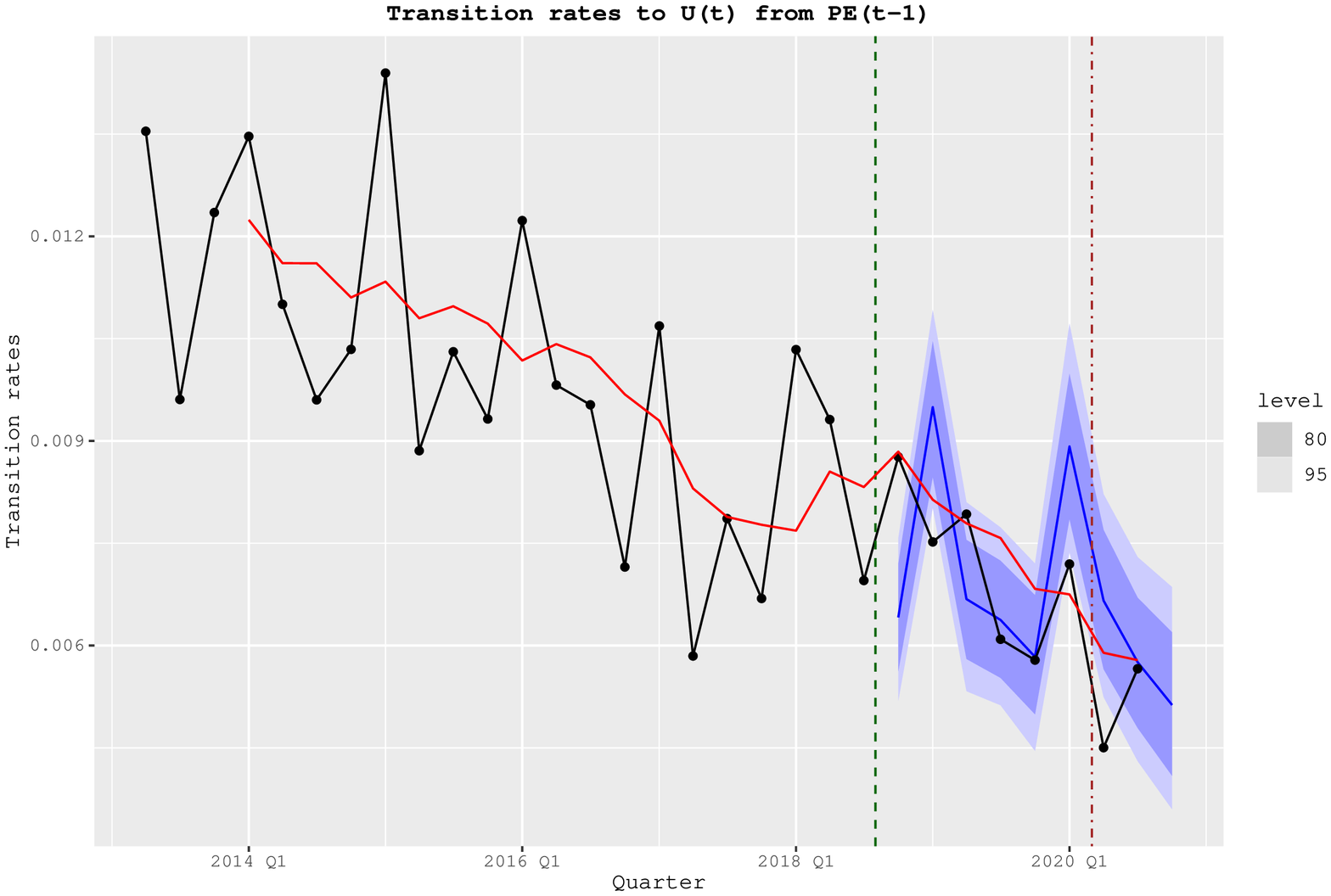}
		\caption{From PE to U.}
		\label{tot_unem}
	\end{subfigure}
	\begin{subfigure}[b]{0.25\textwidth}
		\centering
		\includegraphics[width=\linewidth]{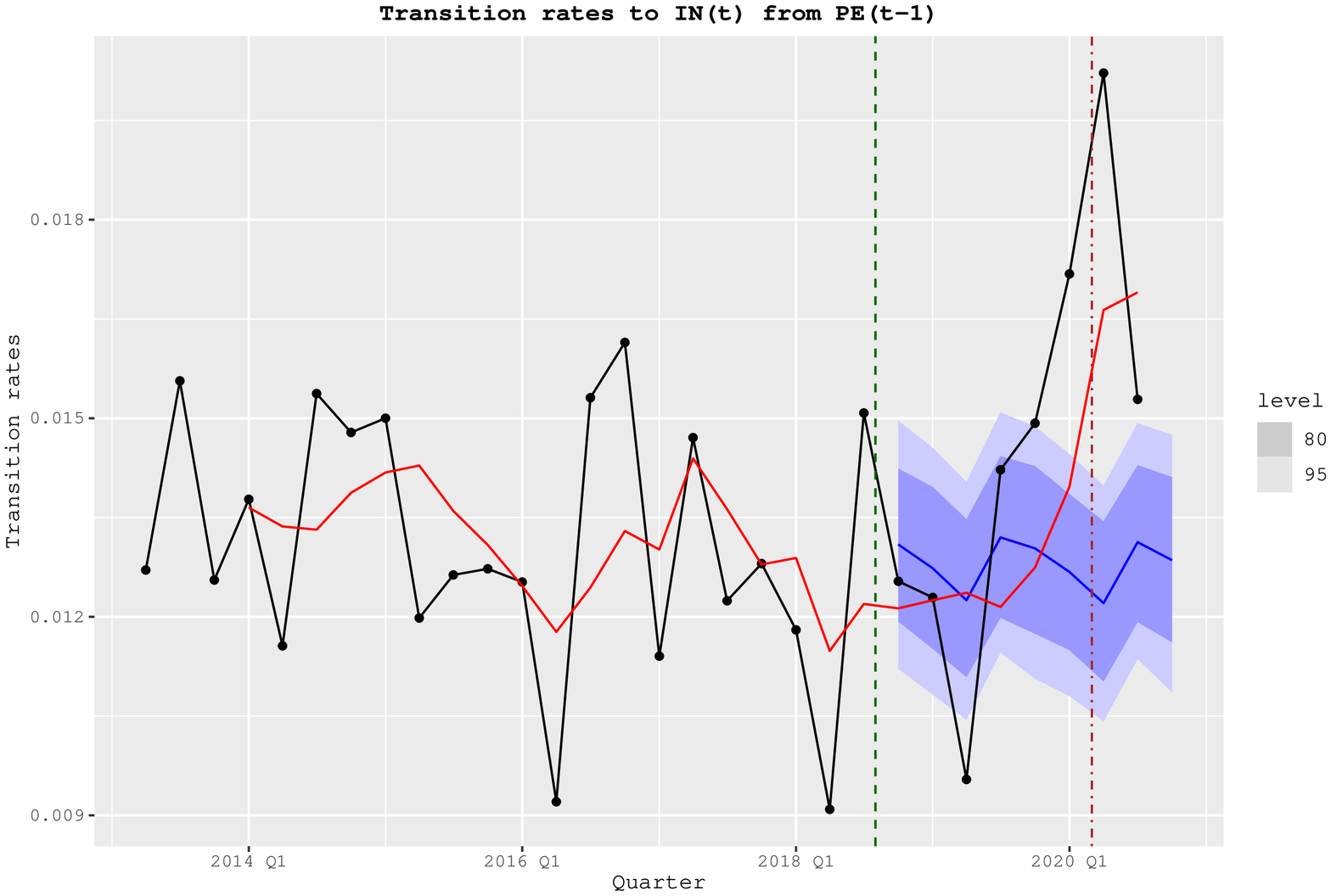}
		\caption{From PE to IN.}
		\label{tot_unem}
	\end{subfigure}

		\caption*{ \scriptsize{Note: the black line reports the observed share of individuals in each labour market state, while the blue line reports the counterfactual share, with standard errors (purple area) at 80\% and 95\%. The red line is the annual observed share calculated according to Equation (\ref{eq:annualQ}), given $\tau=4$. The vertical green line represents the \textit{Decreto Dignita'} reform implemented in August 2018. Finally, the vertical red dotted line in March 2020 represents the beginning of the Covid-19 lockdown.}}
\end{figure}
\end{landscape}

\begin{landscape}
	\begin{figure}[htbp]
			\centering
			\caption{Transition rates across labour market states (cont.d).}
			\label{fig:transIntensities2}
			\begin{subfigure}[b]{0.25\textwidth}
			\centering
			\includegraphics[width=\linewidth]{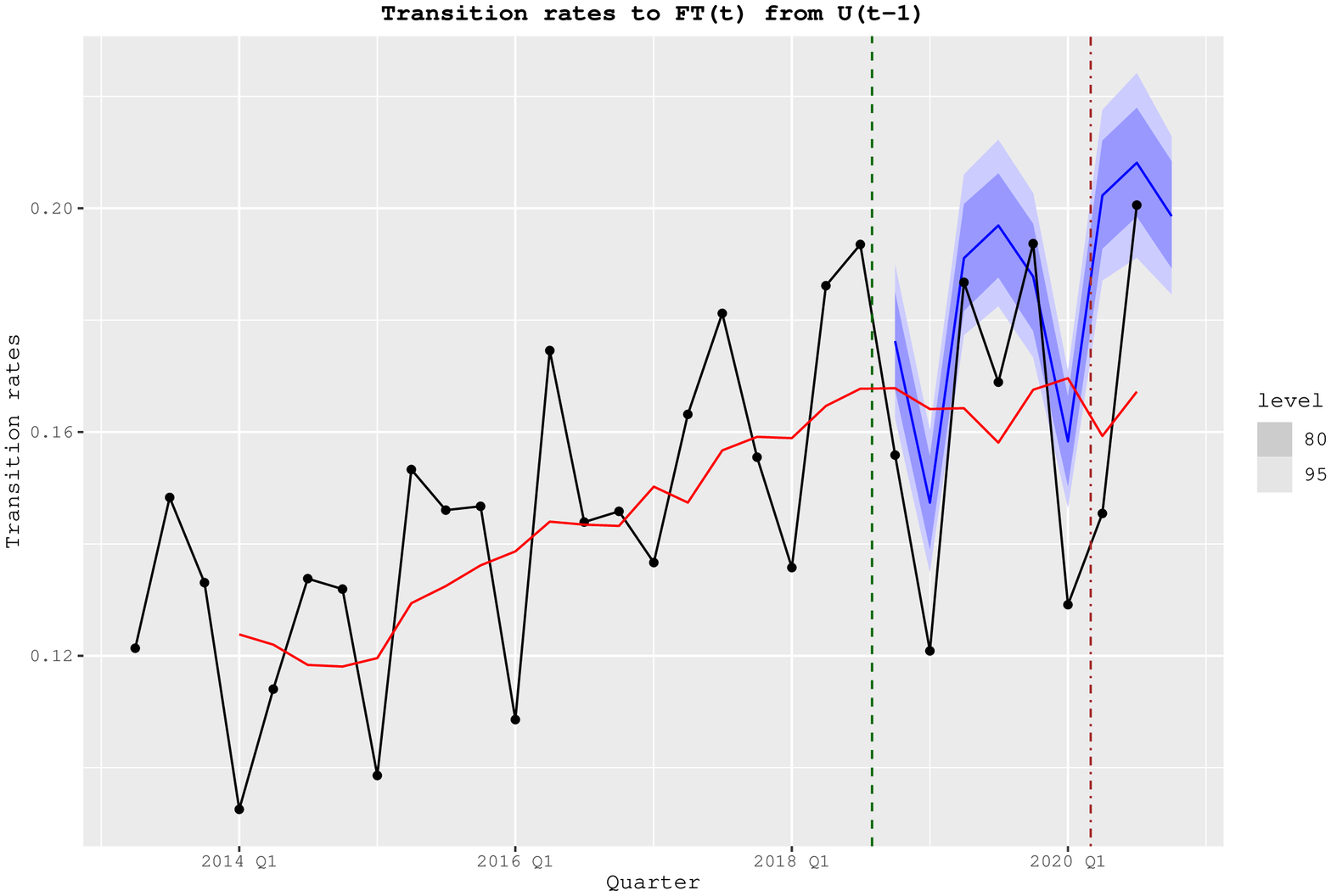}
			\caption{From U to FT.}
			\label{tot_unem}
		\end{subfigure}
		\begin{subfigure}[b]{0.25\textwidth}
			\centering
			\includegraphics[width=\linewidth]{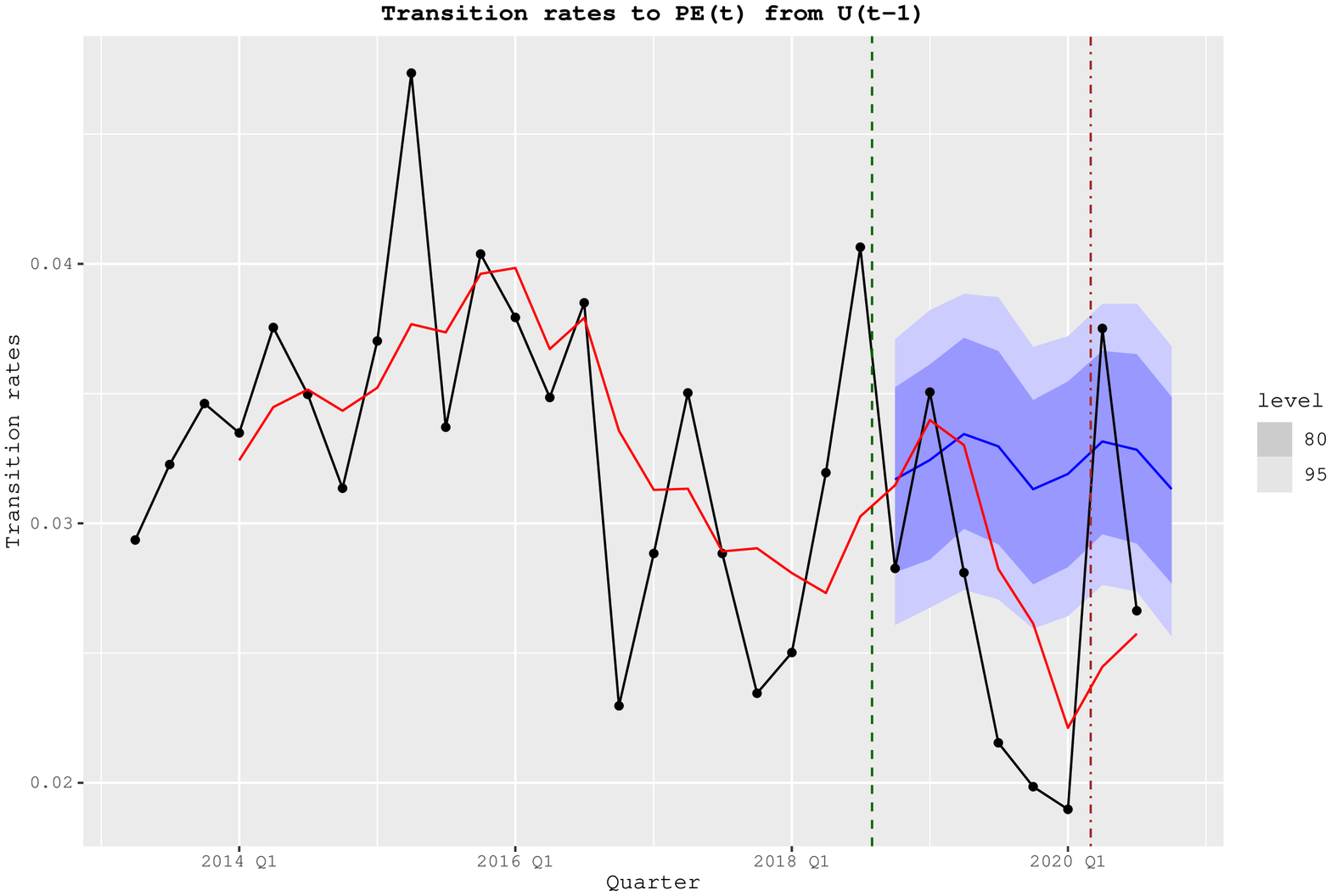}
			\caption{From U to PE.}
			\label{tot_unem}
		\end{subfigure}
		\begin{subfigure}[b]{0.25\textwidth}
			\centering
			\includegraphics[width=\linewidth]{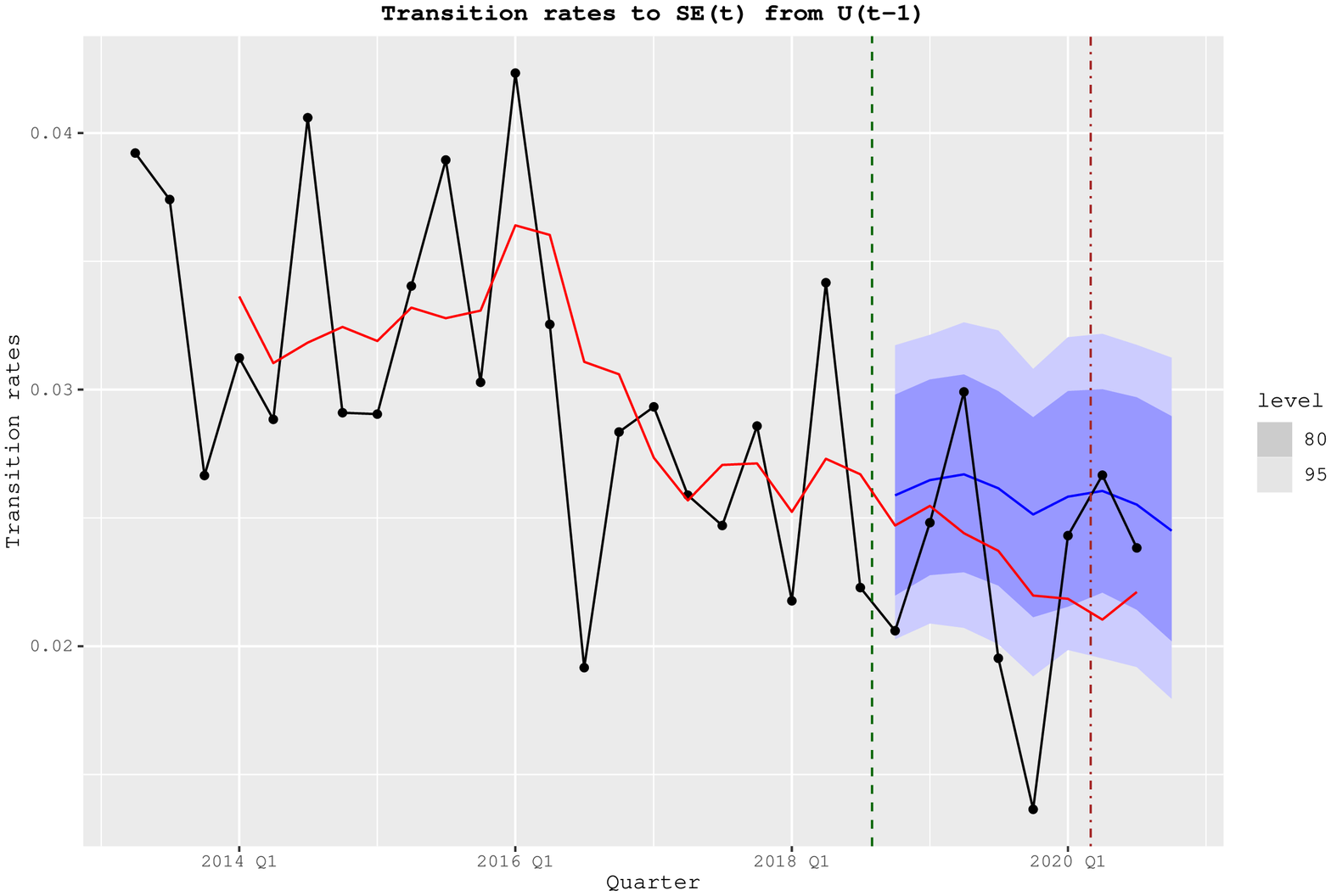}
			\caption{From U to SE.}
			\label{tot_unem}
		\end{subfigure}
		\begin{subfigure}[b]{0.25\textwidth}
			\centering
			\includegraphics[width=\linewidth]{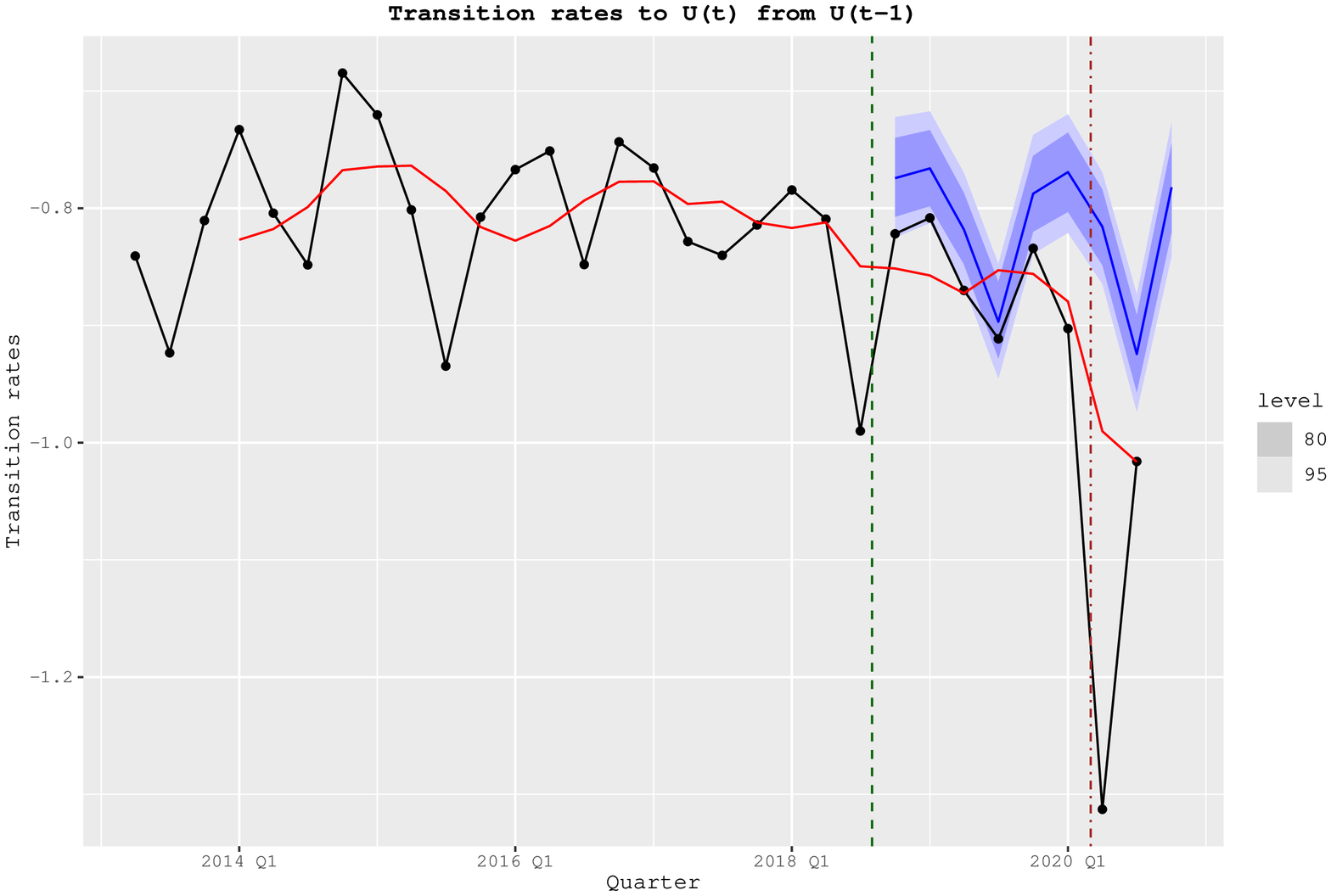}
			\caption{From U to U.}
			\label{tot_unem}
		\end{subfigure}
		\begin{subfigure}[b]{0.25\textwidth}
			\centering
			\includegraphics[width=\linewidth]{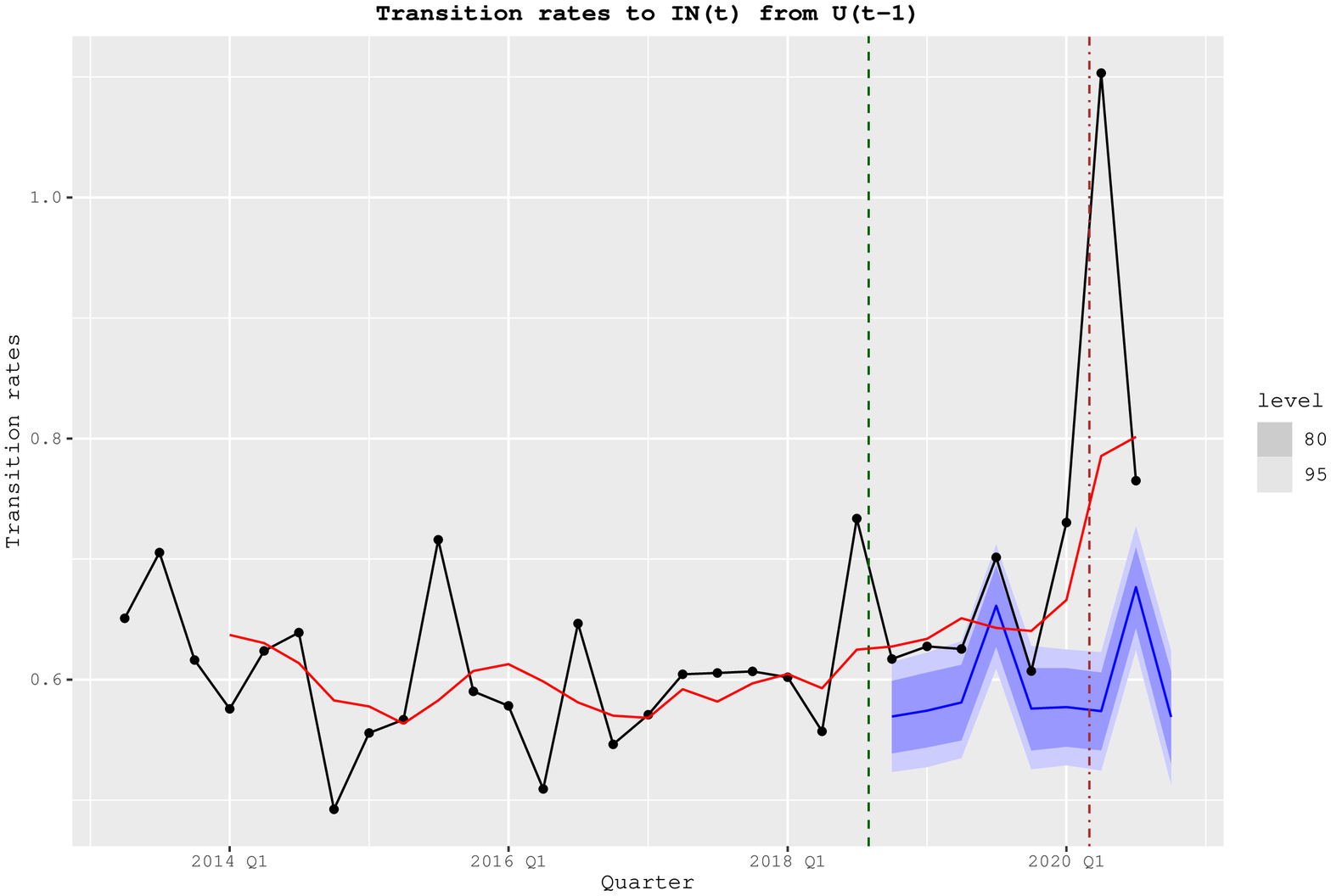}
			\caption{From U to IN.}
			\label{tot_unem}
		\end{subfigure}\\
		\begin{subfigure}[b]{0.25\textwidth}
			\centering
			\includegraphics[width=\linewidth]{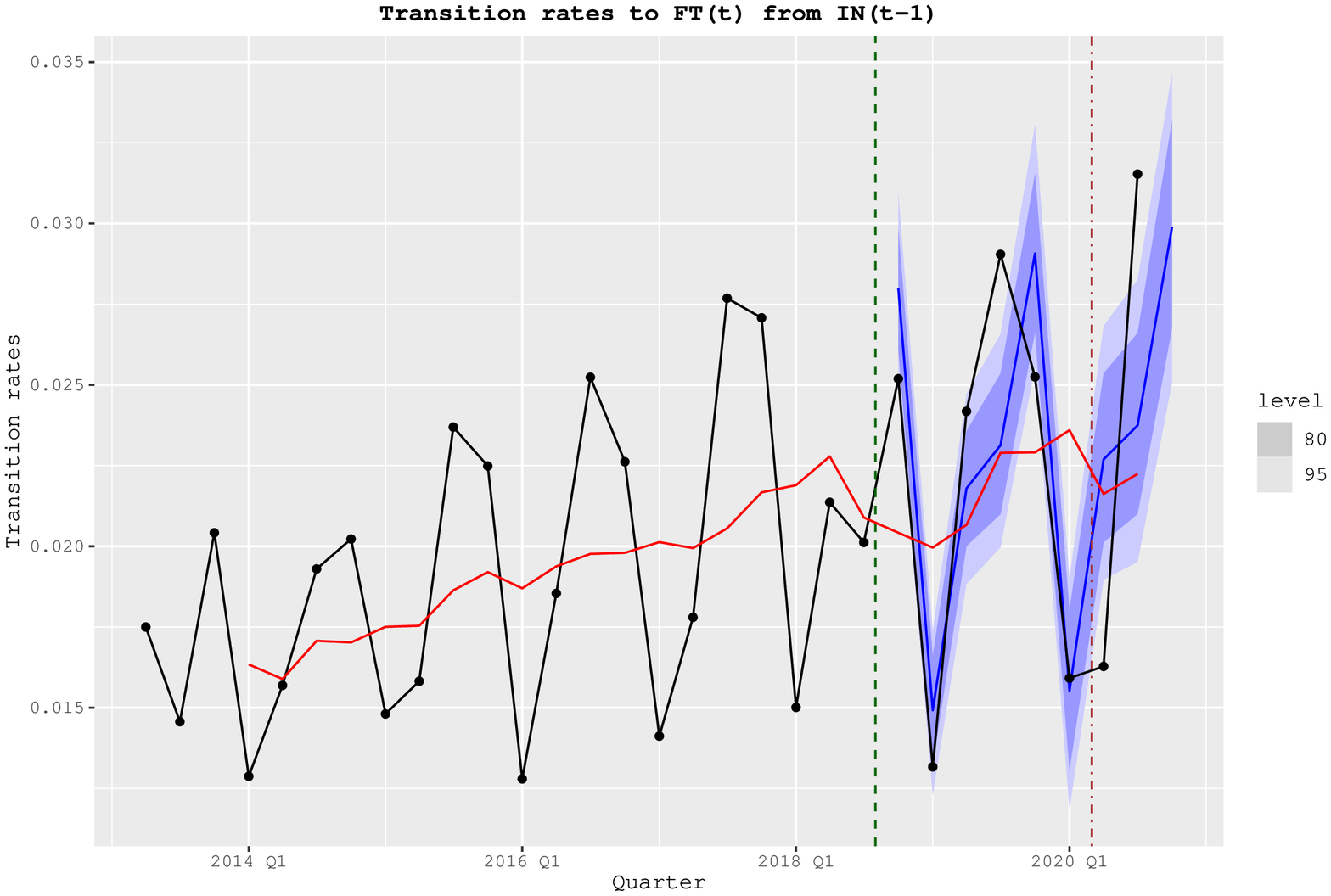}
			\caption{From IN to FT.}
			\label{tot_unem}
		\end{subfigure}
		\begin{subfigure}[b]{0.25\textwidth}
			\centering
			\includegraphics[width=\linewidth]{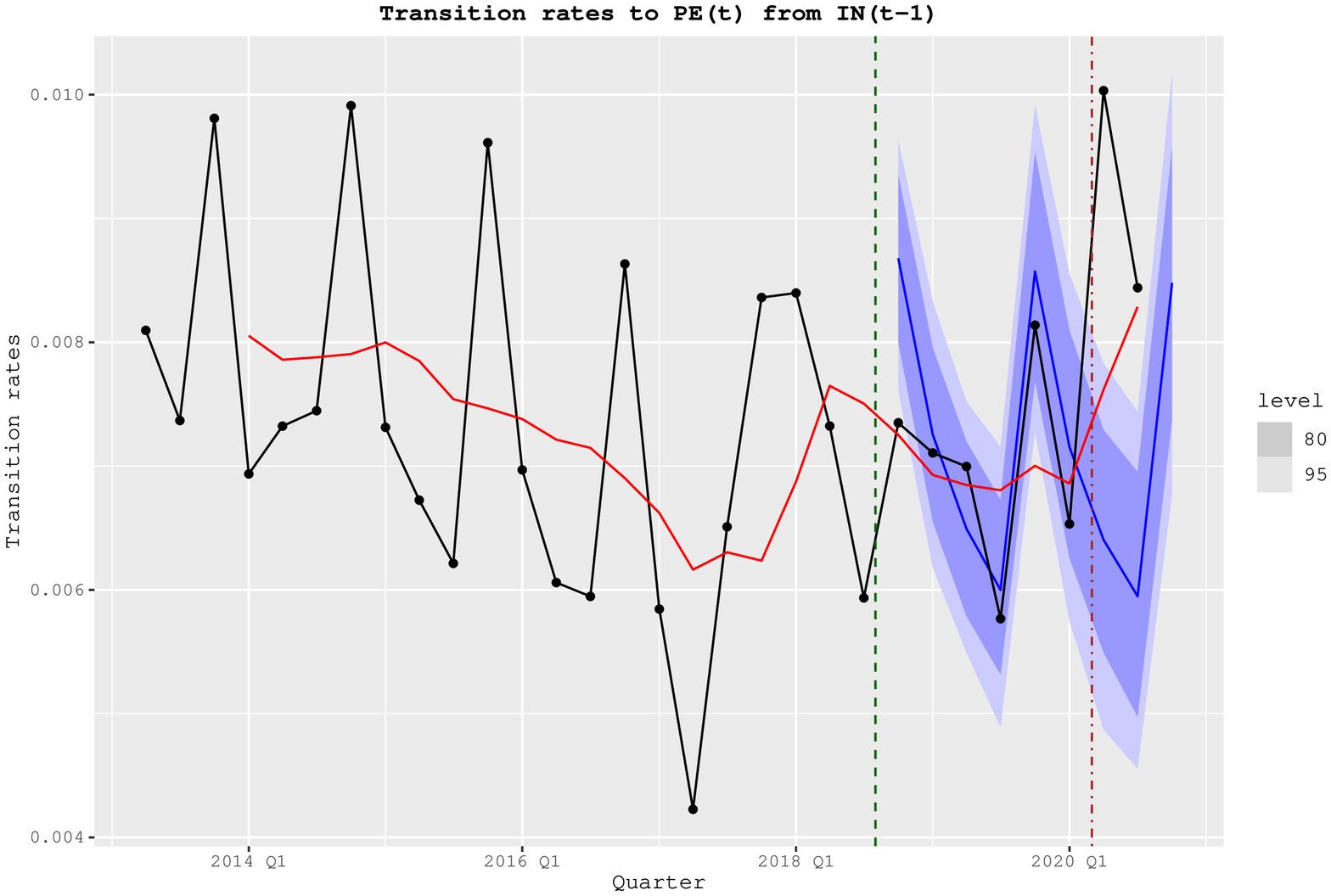}
			\caption{From IN to PE.}
			\label{tot_unem}
		\end{subfigure}
		\begin{subfigure}[b]{0.25\textwidth}
			\centering
			\includegraphics[width=\linewidth]{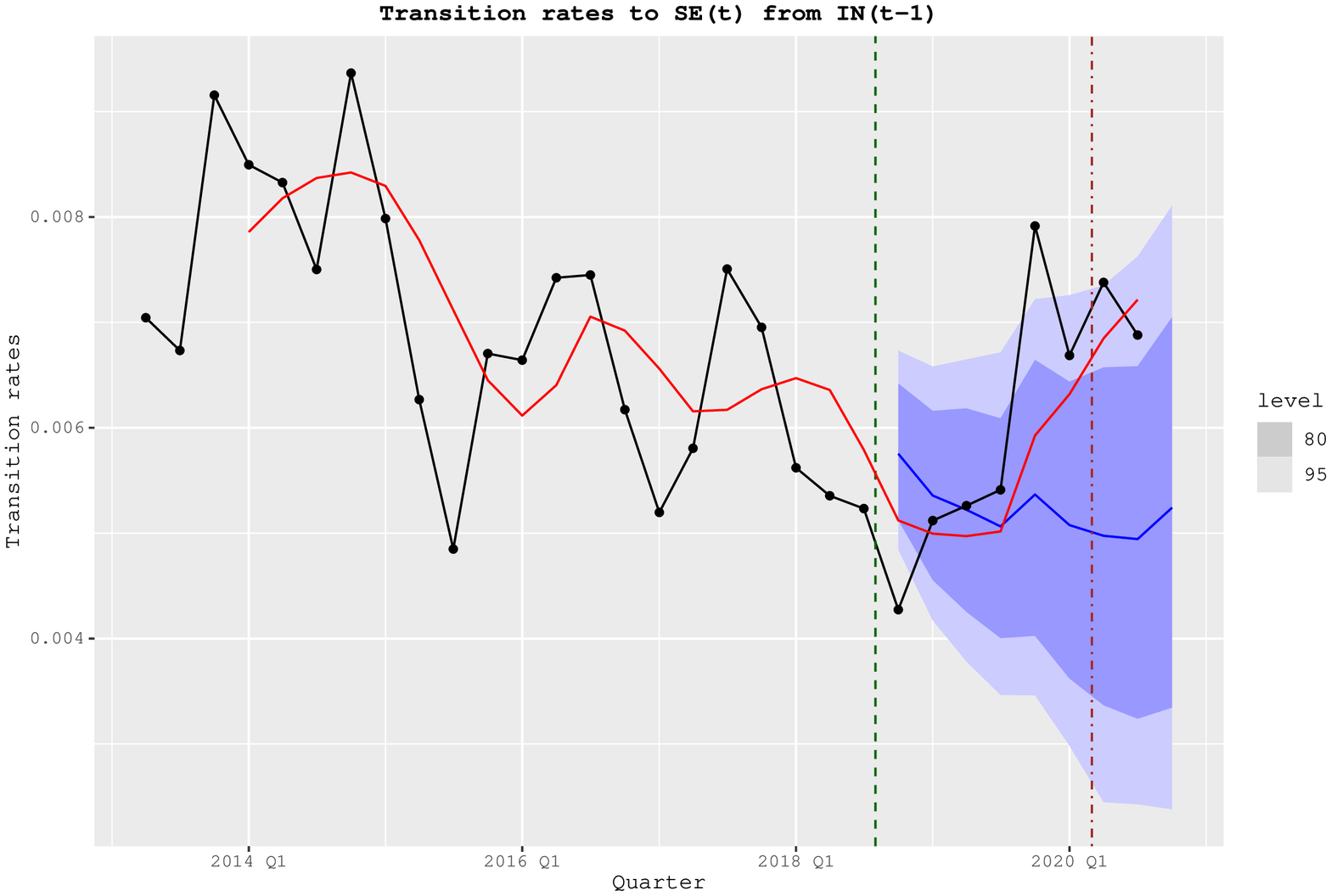}
			\caption{From IN to SE.}
			\label{tot_unem}
		\end{subfigure}
		\begin{subfigure}[b]{0.25\textwidth}
			\centering
			\includegraphics[width=\linewidth]{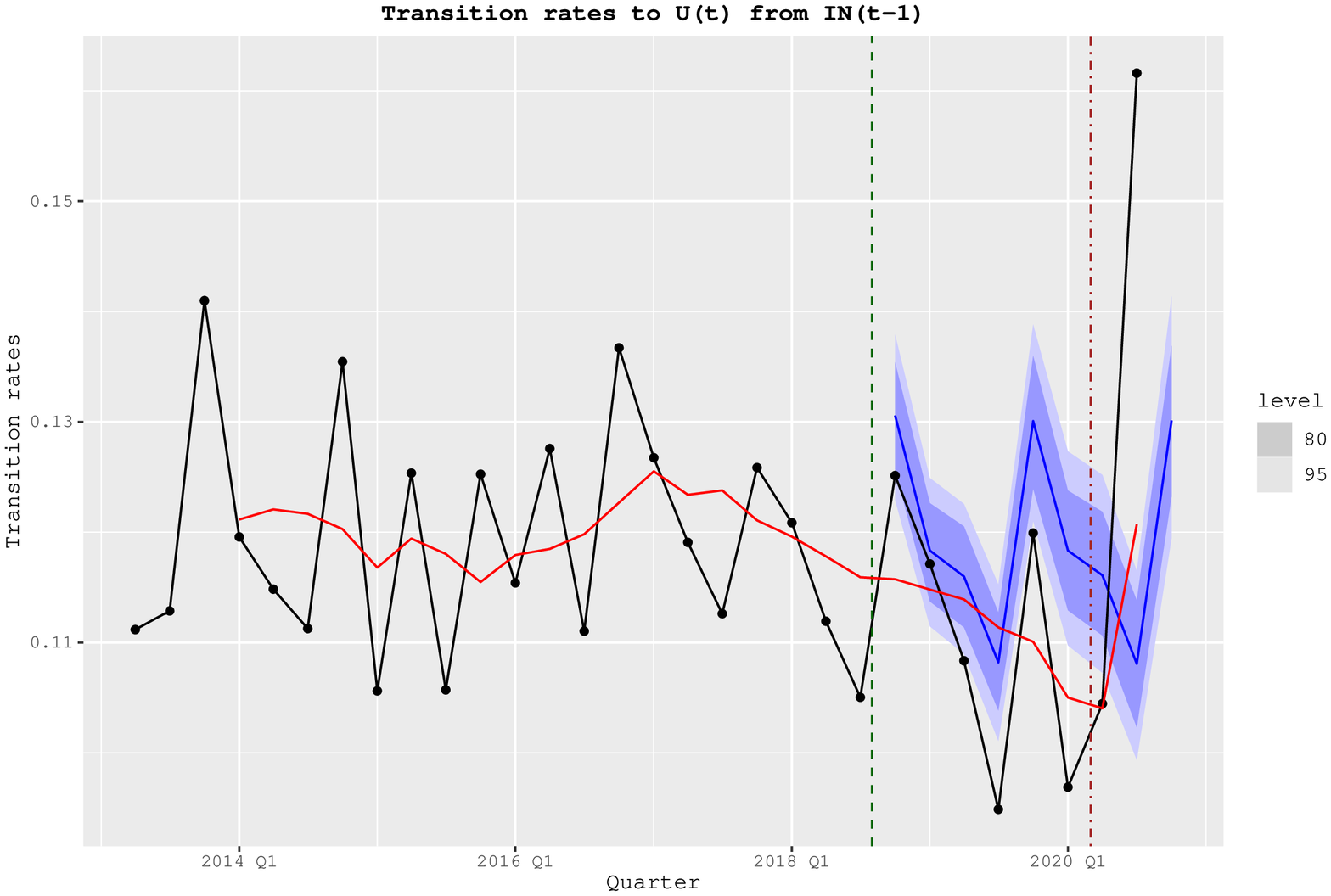}
			\caption{From IN to U.}
			\label{tot_unem}
		\end{subfigure}
		\begin{subfigure}[b]{0.25\textwidth}
			\centering
			\includegraphics[width=\linewidth]{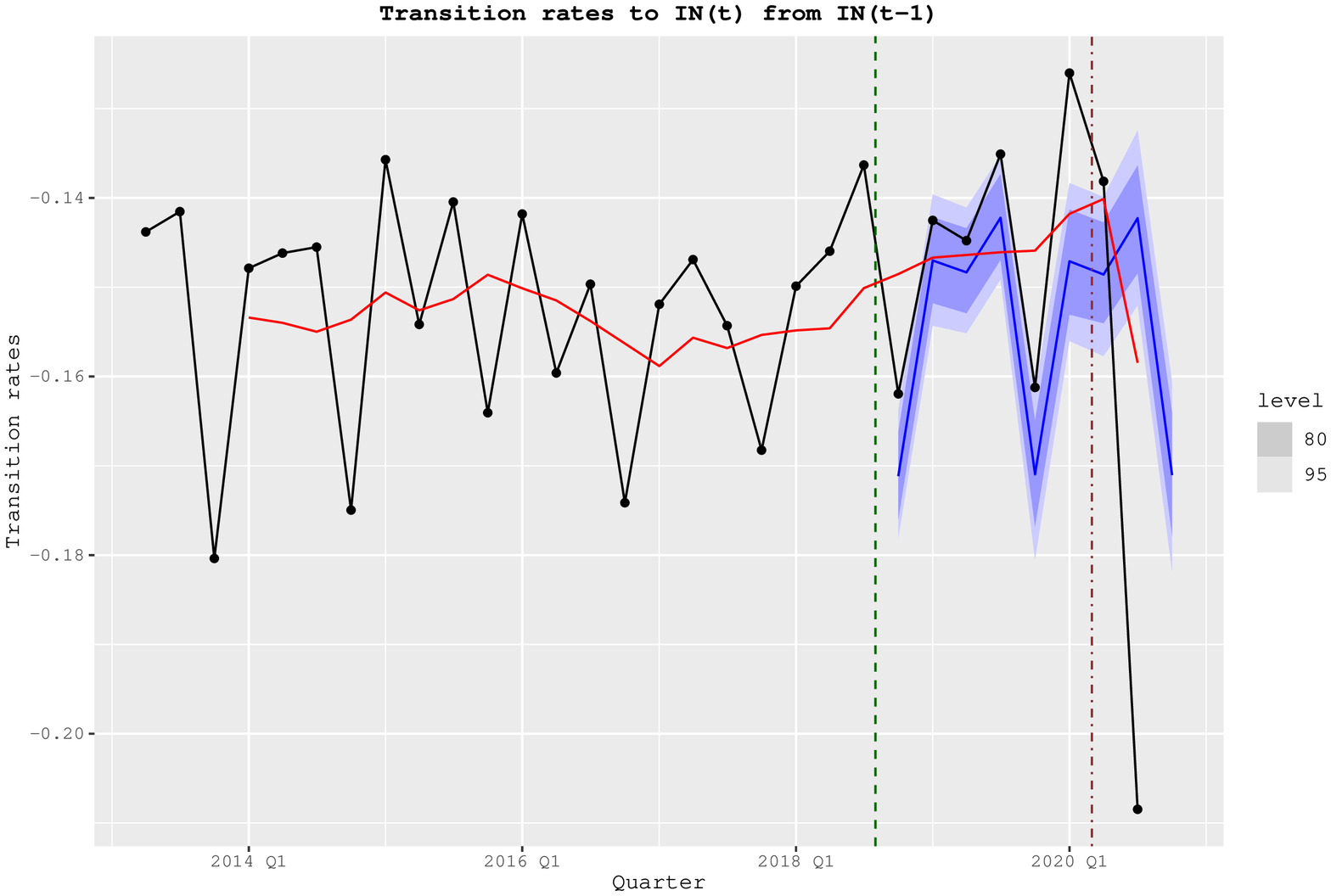}
			\caption{From IN to IN.}
			\label{tot_unem}
		\end{subfigure}

	\caption*{ \scriptsize{Note: the black line reports the observed share of individuals in each labour market state, while the blue line reports the counterfactual share, with standard errors (purple area) at 80\% and 95\%. The red line is the annual observed share calculated according to Equation (\ref{eq:annualQ}), given $\tau=4$. The vertical green line represents the \textit{Decreto Dignita'} reform implemented in August 2018. Finally, the vertical red dotted line in March 2020 represents the beginning of the Covid-19 lockdown.}}
\end{figure}
\end{landscape}

 To dig deeper into the underlying dynamics, we document how the transition rates across the five labour market states considered have evolved in each quarter  from 2013 (quarter I) to 2020 (quarter I). Specifically, the transition rates from self-employed, temporary employed, and permanent employed towards all other states are reported in Figure \ref{fig:transIntensities1}, while the transition rates from unemployed and inactive towards all other states are reported in Figure \ref{fig:transIntensities2}. The black line reports the observed transition rates, the red line reports the annual observed transition rates, while the blue line reports the forecasted transition rates, i.e., the counterfactuals without reform, together with their standard errors.
 
 We notice that the transition rates from self-employed to temporary employed and from self-employed to unemployed have significantly decreased compare to the counterfactual scenario of no reform. We also observe that the transition rates from temporary to temporary contract have declined significantly after the reform. The counterfactual transition rates instead show that in the absence of the reform we would have observed an increase in those transitions. We also observe that the transition rates from temporary to permanent employed have significantly increased after the reform, while the counterfactual transition rates show that in the absence of the reform we would have observed a flat trajectory. We do not detect additional significant changes in transition rates  from temporary employed to unemployed or inactive, although there are signs of increased transitions from temporary employed to self-employed. When we focus on transition rates from permanent employed, we observe increased transitions toward temporary employed, self-employed and inactive, although with some lags compared to the implementation of the reform. We also observe decreased transitions from permanent to permanent employed. Quite interesting are the dynamics of the transition rates from unemployed to other states. We notice decreased transition rates from unemployed to temporary employed, permanent employed and unemployed, while we observe increased rates towards inactive. Finally, we detect a decrease in the outflow from inactive towards towards unemployed, while a significant increase in the transition rates towards self-employed (with some lags) and inactive. 

In summary, although we do not claim any causal relationship between the reform and the transition rates, we can state that after the reform we observe a reduced flow of individuals transiting from temporary to temporary employed (which was the aim of the reform), an increased flow of individuals upgrading from temporary to permanent employed, but also a higher flow of individuals moving from unemployed to inactive and a higher persistence in the inactive state.
	
\section{Discussion and concluding remarks \label{sec:concludingRemarks}}

In this paper, we propose a general methodology to estimate transition rates between labour market states, which represent the core on which the search and matching theoretical framework is built. Specifically, our approach suggests a new method for the estimation of transition rates in continuous time when the data observed are in discrete time, without restrictions on the number of labour market states considered. Moreover, with this methodology we are able to perform inference using bootstrap and to forecast labour market dynamics.

We explore two applications of our methodology useful to learn about labour market dynamics using longitudinal labour force data for Italy. Specifically, first we compute the contribution of inflows and outflows in the labour market shares' volatility  and second, we evaluate how the transition rates between five labour market states and the equilibrium shares have changed after the implementation of a reform, which significantly modified the temporary contract regulation.
We find that the transitions from and to inactivity are the two most important factor contributing to the volatility of the unemployment rate, suggesting a week attachment of individuals to the labour force.
In terms of policy evaluation, although we find that after the reform, the shares of individuals in the five labour market states considered have changed, it is the analysis of the evolution of the transition rates which really informs about the dynamics of the labour market and its trajectory. Specifically, when looking at the transition rates we can identify movements of individuals across states, which are not picked up by the changes in the shares. For instance, by simply looking at the changes in the shares, we observe an increase in temporary employment and a decrease in unemployment, however, we cannot infer anything about the underlying dynamics, as this outcome can be the result of many different alternative combinations of transition rates \citep{Blanchard_Portugal_2001}. By analysing the transition rates, instead, we observe that there are many underlying movements of workers, such as an increase in the transitions from temporary to permanent employment, a decrease in the transition from temporary employment to unemployment, but also for instance an increase in the transitions from unemployment to inactivity. While the former two effects were set as goals of the reform, the latter may be an important unintended  consequence. The analysis of these transition rates are therefore paramount when designing labour market policies, as they provide important information about the direction the labour market is heading to. Moreover, the equilibrium shares are informative about the expected evolution of the labour market shares, as a consequence of the changed transition rates. Although we do not intend to establish causality, as a theoretical framework would be needed, our suggested methodology should provide a different approach to the estimation of instantaneous transition rates for policy evaluation.

\newpage

\bibliographystyle{chicago}
\bibliography{references}

\begin{thebibliography}{}

\bibitem[\protect\citeauthoryear{Anderson and Goodman}{Anderson and
  Goodman}{1957}]{anderson1957statistical}
Anderson, T.~W. and L.~A. Goodman (1957).
\newblock Statistical inference about markov chains.
\newblock {\em The Annals of Mathematical Statistics\/}, 89--110.

\bibitem[\protect\citeauthoryear{Barnichon and Garda}{Barnichon and
  Garda}{2016}]{barnichon2016forecasting}
Barnichon, R. and P.~Garda (2016).
\newblock Forecasting unemployment across countries: The ins and outs.
\newblock {\em European Economic Review\/}~{\em 84}, 165--183.

\bibitem[\protect\citeauthoryear{Barnichon, Nekarda, HATZIUS, STEHN, and
  PETRONGOLO}{Barnichon et~al.}{2012}]{barnichon2012ins}
Barnichon, R., C.~J. Nekarda, J.~HATZIUS, S.~J. STEHN, and B.~PETRONGOLO
  (2012).
\newblock The ins and outs of forecasting unemployment: Using labor force flows
  to forecast the labor market [with comments and discussion].
\newblock {\em Brookings Papers on Economic Activity\/}, 83--131.

\bibitem[\protect\citeauthoryear{Baussola and Mussida}{Baussola and
  Mussida}{2014}]{baussola2014transitions}
Baussola, M. and C.~Mussida (2014).
\newblock Transitions in the labour market: discouragement effect and
  individual characteristics.
\newblock {\em Labour\/}~{\em 28\/}(2), 209--232.

\bibitem[\protect\citeauthoryear{Blanchard and Portugal}{Blanchard and
  Portugal}{2001}]{Blanchard_Portugal_2001}
Blanchard, O. and P.~Portugal (2001, March).
\newblock What hides behind an unemployment rate: Comparing portuguese and u.s.
  labor markets.
\newblock {\em American Economic Review\/}~{\em 91\/}(1), 187--207.

\bibitem[\protect\citeauthoryear{Blanchard and Leigh}{Blanchard and
  Leigh}{2013}]{blanchard2013growth}
Blanchard, O.~J. and D.~Leigh (2013).
\newblock Growth forecast errors and fiscal multipliers.
\newblock {\em American Economic Review\/}~{\em 103\/}(3), 117--20.

\bibitem[\protect\citeauthoryear{Boeri and Garibaldi}{Boeri and
  Garibaldi}{2019}]{boeri2019tale}
Boeri, T. and P.~Garibaldi (2019).
\newblock A tale of comprehensive labor market reforms: Evidence from the
  italian jobs act.
\newblock {\em Labour Economics\/}~{\em 59}, 33--48.

\bibitem[\protect\citeauthoryear{Borowczyk-Martins and
  Lal{\'e}}{Borowczyk-Martins and Lal{\'e}}{2020}]{borowczyk2020ins}
Borowczyk-Martins, D. and E.~Lal{\'e} (2020).
\newblock The ins and outs of involuntary part-time employment.
\newblock {\em Labour Economics\/}~{\em 67}, 101940.

\bibitem[\protect\citeauthoryear{Clemen}{Clemen}{1989}]{clemen1989combining}
Clemen, R.~T. (1989).
\newblock Combining forecasts: A review and annotated bibliography.
\newblock {\em International journal of forecasting\/}~{\em 5\/}(4), 559--583.

\bibitem[\protect\citeauthoryear{Cox and Miller}{Cox and
  Miller}{1972}]{cox1977theory}
Cox, D.~R. and H.~D. Miller (1972).
\newblock {\em The theory of stochastic processes}.
\newblock Chapman and Hall Ltd.

\bibitem[\protect\citeauthoryear{Darby, Haltiwanger, and Plant}{Darby
  et~al.}{1986}]{darby1986ins}
Darby, M.~R., J.~Haltiwanger, and M.~W. Plant (1986).
\newblock The ins and outs of unemployment: The ins win.
\newblock {\em NBER working paper\/}~(w1997).

\bibitem[\protect\citeauthoryear{Di~Porto and Tealdi}{Di~Porto and
  Tealdi}{2019}]{di2019heterogeneous}
Di~Porto, E. and C.~Tealdi (2019).
\newblock Heterogeneous paths to stability.

\bibitem[\protect\citeauthoryear{Efron and Tibshirani}{Efron and
  Tibshirani}{1994}]{efron1994introduction}
Efron, B. and R.~J. Tibshirani (1994).
\newblock {\em An introduction to the bootstrap}.
\newblock CRC press.

\bibitem[\protect\citeauthoryear{Elsby, Hobijn, and {\c{S}}ahin}{Elsby
  et~al.}{2015}]{elsby2015importance}
Elsby, M.~W., B.~Hobijn, and A.~{\c{S}}ahin (2015).
\newblock On the importance of the participation margin for labor market
  fluctuations.
\newblock {\em Journal of Monetary Economics\/}~{\em 72}, 64--82.

\bibitem[\protect\citeauthoryear{Elsby, Michaels, and Solon}{Elsby
  et~al.}{2009}]{elsby2009ins}
Elsby, M.~W., R.~Michaels, and G.~Solon (2009).
\newblock The ins and outs of cyclical unemployment.
\newblock {\em American Economic Journal: Macroeconomics\/}~{\em 1\/}(1),
  84--110.

\bibitem[\protect\citeauthoryear{Fontaine, Galvez-Iniesta, Gomes, and
  Vila-Martin}{Fontaine et~al.}{2020}]{fontaine2020labour}
Fontaine, I., I.~Galvez-Iniesta, P.~Gomes, and D.~Vila-Martin (2020).
\newblock Labour market flows: Accounting for the public sector.
\newblock {\em Labour Economics\/}~{\em 62}, 101770.

\bibitem[\protect\citeauthoryear{Fujita and Ramey}{Fujita and
  Ramey}{2009}]{fujita2009cyclicality}
Fujita, S. and G.~Ramey (2009).
\newblock The cyclicality of separation and job finding rates.
\newblock {\em International Economic Review\/}~{\em 50\/}(2), 415--430.

\bibitem[\protect\citeauthoryear{Garibaldi and Mauro}{Garibaldi and
  Mauro}{2002}]{garibaldi2002anatomy}
Garibaldi, P. and P.~Mauro (2002).
\newblock Anatomy of employment growth.
\newblock {\em Economic policy\/}~{\em 17\/}(34), 67--114.

\bibitem[\protect\citeauthoryear{Garibaldi and Wasmer}{Garibaldi and
  Wasmer}{2005}]{garibaldi2005equilibrium}
Garibaldi, P. and E.~Wasmer (2005).
\newblock Equilibrium search unemployment, endogenous participation, and labor
  market flows.
\newblock {\em Journal of the European Economic Association\/}~{\em 3\/}(4),
  851--882.

\bibitem[\protect\citeauthoryear{Gomes}{Gomes}{2012}]{gomes2012labour}
Gomes, P. (2012).
\newblock Labour market flows: Facts from the {U}nited {K}ingdom.
\newblock {\em Labour Economics\/}~{\em 19\/}(2), 165--175.

\bibitem[\protect\citeauthoryear{Gomes}{Gomes}{2015}]{gomes2015importance}
Gomes, P. (2015).
\newblock The importance of frequency in estimating labour market transition
  rates.
\newblock {\em IZA Journal of Labor Economics\/}~{\em 4\/}(1), 1--10.

\bibitem[\protect\citeauthoryear{Hertweck and Sigrist}{Hertweck and
  Sigrist}{2015}]{hertweck2015ins}
Hertweck, M.~S. and O.~Sigrist (2015).
\newblock The ins and outs of german unemployment: a transatlantic perspective.
\newblock {\em Oxford Economic Papers\/}~{\em 67\/}(4), 1078--1095.

\bibitem[\protect\citeauthoryear{Hirsch, Smale, and Devaney}{Hirsch
  et~al.}{2012}]{hirsch2012differential}
Hirsch, M.~W., S.~Smale, and R.~L. Devaney (2012).
\newblock {\em Differential equations, dynamical systems, and an introduction
  to chaos}.
\newblock Academic press.

\bibitem[\protect\citeauthoryear{Hyndman and Athanasopoulos}{Hyndman and
  Athanasopoulos}{2021}]{hyndman2021forecasting}
Hyndman, R.~J. and G.~Athanasopoulos (2021).
\newblock {\em Forecasting: principles and practice}.
\newblock OTexts: Melbourne, Australia. OTexts.com/fpp3.

\bibitem[\protect\citeauthoryear{Israel, Rosenthal, and Wei}{Israel
  et~al.}{2001}]{israel2001finding}
Israel, R.~B., J.~S. Rosenthal, and J.~Z. Wei (2001).
\newblock Finding generators for markov chains via empirical transition
  matrices, with applications to credit ratings.
\newblock {\em Mathematical finance\/}~{\em 11\/}(2), 245--265.

\bibitem[\protect\citeauthoryear{Layard, Nickell, and Jackman}{Layard
  et~al.}{2005}]{layard2005unemployment}
Layard, R., S.~Nickell, and R.~Jackman (2005).
\newblock {\em Unemployment: macroeconomic performance and the labour market}.
\newblock Oxford University Press.

\bibitem[\protect\citeauthoryear{Mortensen et~al.}{Mortensen
  et~al.}{1970}]{mortensen1970theory}
Mortensen, D.~T. et~al. (1970).
\newblock A theory of wage and employment dynamics.
\newblock {\em Microeconomic foundations of employment and inflation
  theory\/}~{\em 219}.

\bibitem[\protect\citeauthoryear{Mortensen and Pissarides}{Mortensen and
  Pissarides}{1994}]{mortensen1994job}
Mortensen, D.~T. and C.~A. Pissarides (1994).
\newblock Job creation and job destruction in the theory of unemployment.
\newblock {\em The review of economic studies\/}~{\em 61\/}(3), 397--415.

\bibitem[\protect\citeauthoryear{OECD}{OECD}{2019}]{OECDreport2019}
OECD (2019).
\newblock {\em Recent trends in the Italian Labour Market}.

\bibitem[\protect\citeauthoryear{Pathak and Shi}{Pathak and
  Shi}{2020}]{pathak2020well}
Pathak, P.~A. and P.~Shi (2020).
\newblock How well do structural demand models work? {C}ounterfactual
  predictions in school choice.
\newblock {\em Journal of Econometrics\/}.

\bibitem[\protect\citeauthoryear{Perugini and Signorelli}{Perugini and
  Signorelli}{2007}]{perugini2007labour}
Perugini, C. and M.~Signorelli (2007).
\newblock Labour market performance differentials and dynamics in {EU}-15
  countries and regions.
\newblock {\em The European journal of comparative economics\/}~{\em 4\/}(2),
  209.

\bibitem[\protect\citeauthoryear{Petrongolo and Pissarides}{Petrongolo and
  Pissarides}{2008}]{petrongolo2008ins}
Petrongolo, B. and C.~A. Pissarides (2008).
\newblock The ins and outs of european unemployment.
\newblock {\em American economic review\/}~{\em 98\/}(2), 256--62.

\bibitem[\protect\citeauthoryear{Phelps}{Phelps}{1968}]{phelps1968money}
Phelps, E.~S. (1968).
\newblock Money-wage dynamics and labor-market equilibrium.
\newblock {\em Journal of political economy\/}~{\em 76\/}(4, Part 2), 678--711.

\bibitem[\protect\citeauthoryear{Pissarides}{Pissarides}{1986}]{pissarides1986unemployment}
Pissarides, C. (1986).
\newblock Unemployment and vacancies in {B}ritain.
\newblock {\em Economic policy\/}~{\em 1\/}(3), 499--541.

\bibitem[\protect\citeauthoryear{Pissarides}{Pissarides}{2000}]{pissarides2000equilibrium}
Pissarides, C.~A. (2000).
\newblock {\em Equilibrium unemployment theory}.
\newblock MIT press.

\bibitem[\protect\citeauthoryear{Raitano}{Raitano}{2018}]{Raitano2018}
Raitano, M. (2018).
\newblock {\em Italy: Para-subordinate workers and their social protection}.

\bibitem[\protect\citeauthoryear{Shimer}{Shimer}{2012}]{shimer2012reassessing}
Shimer, R. (2012).
\newblock Reassessing the ins and outs of unemployment.
\newblock {\em Review of Economic Dynamics\/}~{\em 15\/}(2), 127--148.

\bibitem[\protect\citeauthoryear{Silva and V{\'a}zquez-Grenno}{Silva and
  V{\'a}zquez-Grenno}{2013}]{silva2013ins}
Silva, J.~I. and J.~V{\'a}zquez-Grenno (2013).
\newblock The ins and outs of unemployment in a two-tier labor market.
\newblock {\em Labour Economics\/}~{\em 24}, 161--169.

\bibitem[\protect\citeauthoryear{Smith}{Smith}{2011}]{smith2011ins}
Smith, J.~C. (2011).
\newblock The ins and outs of {UK} unemployment.
\newblock {\em The Economic Journal\/}~{\em 121\/}(552), 402--444.

\bibitem[\protect\citeauthoryear{Valli}{Valli}{1970}]{valli1970programmazione}
Valli, V. (1970).
\newblock Programmazione e sindacati in italia.

\bibitem[\protect\citeauthoryear{Zahl}{Zahl}{1955}]{zahl1955markov}
Zahl, S. (1955).
\newblock A markov process model for follow-up studies.
\newblock {\em Human Biology\/}~{\em 27\/}(2), 90.

\end{thebibliography}

\newpage

\appendix

\begin{Large}
	\textbf{Appendix}
\end{Large}

\section{Bootstrap procedure \label{app:bootstrapProcedures}}

Given a sample of transitions $X$ of cardinality $N$, the bootstrap procedure is composed of three steps  \citep[Chapter 6]{efron1994introduction}:
\begin{enumerate}
\item Draw $B$ samples of cardinality $N$ by sampling with replacement from $X$;
\item For every bootstrapped sample $b$ estimate matrix $\mathbf{P}_b$ and the corresponding $\tilde{\mathbf{Q}}_b$;
\item Compute the standard errors of the transition rates $\tilde{q}_{ij}$, $\sigma_{q_{ij}}$ as:
\[
\sigma_{q_{ij}} = \sqrt{\sum_{b=1}^{B} \dfrac{ \left( \tilde{q}_{ij,b} - \overline{\tilde{q}}_{ij} \right)^2}{B}},
\]
where $ \tilde{q}_{ij,b}$ is the $(i,j)$ element of $\tilde{\mathbf{Q}}_b$ and $\overline{\tilde{q}}_{ij}$ is the average $(i,j)$ element of all the $B$ bootstraps.
\end{enumerate}

The test of zero difference between two transition rates and/or between two equilibrium labour market shares is based on the bootstrap procedure suggested in \citet[Chapter 16]{efron1994introduction}.

\section{Pairwise comparison of estimates \label{tab:pvaluesestimates}}

In this section we report the p-values of the pairwise comparisons of the elements of estimated $\mathbf{Q}$ matrices (Table \ref{app:PvaluesQelements}) and the p-values of the pairwise comparisons of the estimated equilibrium labour market shares in quarter II of 2018 and quarter IV of 2019 (Table \ref{app:PvaluesEqShares}). These p-values are calculated by the bootstrap procedure suggested in \citet[Chapter 16]{efron1994introduction}.
\begin{table}[htbp]
	\caption{P-values of pairwise comparisons of elements of estimated $\mathbf{Q}$ matrices in quarter II of 2018 and quarter IV of 2019. }
	\label{app:PvaluesQelements}
	\centering
	\begin{tabular}{r|rrrrr}
		\hline
		\hline
		& SE & FT & PE & U & IN \\ 
		\hline
		SE & 0.00 & 0.17 & 0.10 & 0.02 & 0.30 \\ 
		FT & 0.24 & 0.00 & 0.00 & 0.44 & 0.13 \\ 
		PE & 0.23 & 0.20 & 0.33 & 0.01 & 0.04 \\ 
		U & 0.04 & 0.37 & 0.38 & 0.01 & 0.00 \\ 
		IN & 0.26 & 0.47 & 0.20 & 0.01 & 0.01 \\ 
		\hline
		\hline
	\end{tabular}
	\caption*{\textit{Note}: P-values are calculated using bootstrap (1000 samplings).}
\end{table}

\begin{table}[htbp]
		\caption{P-values of the pairwise comparisons of estimated equilibrium labour market shares in quarter II of 2018 and quarter IV of 2019. }
		\label{app:PvaluesEqShares}
	\centering
	\begin{tabular}{rrrrr}
		\hline
		\hline
		SE & FT & PE & U & IN \\ 
		\hline
		0.205 & 0.000 & 0.025 & 0.000 & 0.232 \\
		\hline
		\hline
	\end{tabular}
\caption*{\textit{Note}: P-values are calculated using bootstrap (1000 samplings).}
\end{table}

\section{Counterfactual estimation through forecasts\label{app:counterfactualViaARIMA}}

Our strategy for the estimation of the counterfactual transition rates is inspired by \cite{blanchard2013growth} and based on forecasting.
The $f$-quarter ahead forecast of transition rate $q_{ij}$ in quarter $t$ can be expressed as:
\begin{equation}
q_{ij,t+f} = q_{ij,t+f|t} + \epsilon_{ij:t+f} ,
\end{equation}
where $q_{ij,t+f-1:t+f}$ is the \textit{observed} transition rate $(i,j)$ in quarter $(t+f)$, $q_{ij,t+f|t}$ is the \textit{forecasted} transition rate for the quarter $(t+f)$ calculated in quarter $t$ and $\epsilon_{ij:t+f}$ is the forecasting error. If the forecasting is computed exploiting any information available in period $t$, then the expected value of $\epsilon_{ij:t+f}$ is zero and $\epsilon_{ij:t+f}$ and $q_{ij,t+f|t}$ are orthogonal, i.e.:
\begin{equation}
E[q_{ij,t+f} - q_{ij,t+f|t}] = 0.
\end{equation}
Hence, any \textit{significant} divergence between $q_{ij,t+f}$ and $q_{ij,t+f|t}$ signals a novelty with respect to the information set available in period $t$ or, alternatively, $q_{ij,t+f|t}$ can be interpreted as a \textit{counterfactual}. The novelty in the labour market in quarter $t$ should appear as a significant gap between $q_{ij,t+f}$ and $q_{ij,t+f|t}$.

In estimating the forecast, we follow \citet{clemen1989combining} and \citet{hyndman2021forecasting}, and consider the combination of three different models of forecasting, i.e. AutoRegressive Integrated Moving Average (ARIMA), exponential smoothing (ETS) and linear regression (TSLR), allowing in all models for the presence of a trend and seasonality and using the modified AIC for model selection \citep[Chapters 8, 9 and 13]{hyndman2021forecasting}).

\end{document}